\documentclass[fleqn,usenatbib]{mnras}
\usepackage{newtxtext,newtxmath}

\usepackage[T1]{fontenc}
\usepackage{ae,aecompl}
\usepackage[normalem]{ulem}

\usepackage{graphicx}	
\usepackage{amsmath}	

\usepackage{amssymb}	
\usepackage[utf8]{inputenc}
\usepackage{lastpage}

\title[Bulges in MW/M31-like galaxies in TNG50]{High and low Sérsic index bulges in Milky Way- and M31-like galaxies: origin and connection to the bar with TNG50}
\author[I.D. Gargiulo et al.]{Ignacio D. Gargiulo$^{1,2}$\thanks{E-mail: gargiulo@fcaglp.unlp.edu.ar},
Antonela Monachesi$^{3,4}$, Facundo A. G\'omez$^{3,4}$, 
Dylan Nelson${^5}$, \newauthor{Annalisa Pillepich$^{6}$}, R\"udiger Pakmor$^{7}$,  R. J. J. Grand$^{7}$ , Francesca Fragkoudi$^{8}$ , 
  \newauthor{Lars Hernquist $^{9}$}, Mark Lovell$^{10}$, Federico Marinacci$^{11}$
\vspace{0.2cm}\\
\\
$^{1}$Instituto de Astrof\'isica de La Plata (CCT La Plata, CONICET, UNLP), Observatorio Astron\'omico, Paseo del Bosque
B1900FWA, La Plata, Argentina\\
$^{2}$Consejo Nacional de Investigaciones Cient\'ificas y T\'ecnicas (CONICET), Rivadavia 1917, Buenos Aires, Argentina\\
$^{3}$Instituto de Investigaci\'on Multidisciplinar en Ciencia y Tecnolog\'ia, Universidad de La Serena, Ra\'ul Bitr\'an 1305, La Serena, Chile\\
$^{4}$Departamento de Astronom\'ia, Universidad de La Serena, Av. Juan Cisternas 1200 Norte, La Serena, Chile\\
$^{5}$Institut für Theoretische Astrophysik, Zentrum für Astronomie, Universität Heidelberg, Albert-Ueberle-Str. 2, 69120 Heidelberg, Germany \\
$^{6}$Max-Planck-Institut f\"{u}r Astronomie, K\"{o}nigstuhl 17, 69117 Heidelberg, Germany.\\
$^{7}$Max-Planck-Institut f\"{u}r Astrophysik, Karl-Schwarzschild-Str. 1, D-85748, Garching, Germany\\
$^{8}$European Southern Observatory, Karl-Schwarzschild-Str. 2, 85748 Garching-bei-München, Germany \\
$^{9}$ Institute for Theory and Computation, Harvard-Smithsonian Center for Astrophysics, 60 Garden St., MS-51, Cambridge, MA 02138, USA \\
$^{10}$Center for Astrophysics and Cosmology, Science Institute, University of Iceland, Dunhaga 5, 107 Reykjavík, Iceland\\
$^{11}$Department of Physics and Astronomy “Augusto Righi”, Università di Bologna, Via Gobetti 93/2, 40129 Bologna, Italy}

\date{Accepted XXX. Received YYY; in original form ZZZ}

\pubyear{2021}

\begin{document}
\label{firstpage}
\pagerange{\pageref{firstpage}--\pageref{lastpage}}
\maketitle

\begin{abstract}

\noindent We study bulge formation in MW/M31-like galaxies in a $\Lambda$-cold dark matter scenario, focusing on the origin of high- and low-Sersic index bulges. For this purpose we use TNG50, a simulation of the IllustrisTNG project that combines a resolution of $\sim 8 \times 10^4 {\rm M_{\odot}}$ in stellar particles with a cosmological volume 52 {\rm cMpc} in extent. We parametrize bulge surface brightness profiles by the S\'ersic index and  the bulge-to-total (B/T) ratio obtained from two-component photometric decompositions. In our sample of 287 MW/M31-like simulated galaxies, $17.1\%$ of photometric bulges exhibit high-S\'ersic indices and $82.9\%$ show low-S\'ersic indices. We study the impact that the environment, mergers and bars have in shaping the surface brightness profiles. We explore two different definitions for local environment and find no correlation between bulge properties and the environment where they reside. Simulated galaxies with higher S\'ersic indices show, on average, a higher fraction of ex--situ stars in their kinematically selected bulges.  For this bulge population the last significant merger (total mass ratio $m_{\rm sat}/m_{\rm host} > 0.1$) occurs, on average, at later times. However, a substantial fraction of low-S\'ersic index bulges also experience a late significant merger. 
 We find that bars play an important role in the development of the different types of photometric bulges. We show that the fraction of simulated galaxies with strong bars is smaller for the high- than for the low-S\'ersic index population, reaching differences of $20\%$ at $z > 1$. Simulated galaxies with high fractions of ex--situ stars in the bulge do not develop strong bars. Conversely, simulated galaxies with long-lived strong bars have bulges with ex--situ fractions, $f_{\rm ex-situ} < 0.2$. 
 

\end{abstract}


\section{Introduction}
\label{sec:intro}

The galactic bulge has been, arguably, the least studied galactic component with cosmological simulations evolved within the current $\Lambda-$cold dark matter paradigm \citep[LCDM, see][and references therein]{Peeblesbook2020}. One of the main historical reasons behind this was the lack of computational resources needed to resolve the interaction of the many particles that co-exist within the volume of the galactic central region.  
Additionally, bulges are not easy to define and, as a result, theoretical works have been focused on the study of central spheroidal regions \citep[e.g.][]{Tissera2018, Tacchella2019}. 
The ``photometric bulge'' is, perhaps, a less ambiguous definition of a bulge \citep[see][for a succinct overview on bulge definitions]{Gadotti2012}. This observational definition refers to the excess of starlight in the central regions of a disk galaxy with respect to the exponential profile that is used to represent the surface brightness of the disk component. Indeed, it is a common practice to photometrically decompose the surface brightness profile of a disk galaxy by fitting two smooth functional components \citep{Freeman1970, Kent1985, Andredakis1995}. An exponential profile represents the disc and a S\'ersic profile \citep{Sersic1963}  accounts for the central part of the galaxy. The S\'ersic function includes a free parameter, known as the S\'ersic index, that modulates the shape of the S\'ersic profile. A S\'ersic index $n=1$ results in an exponential profile, while $n=4$ is equivalent to the de Vaucouleurs profile \citep{DeVaucouleurs1948}, that is generally used to fit the surface brightness profiles of elliptical galaxies. A larger S\'ersic index value is indicative of a more concentrated light profile \citep[see][for a complete reference of the related mathematical expressions]{Graham2005}.

The photometric bulge definition has two main advantages. First, it is independent of the galaxy inclination (under the assumption of no dust), meaning that one can consistently compare  large samples of observed photometric bulges, irrespective of galaxy orientation. Second, it is objective and reproducible. However, this lack of ambiguity comes with the price of an excess of vagueness.  This definition of a photometric bulge in a single galaxy includes in this component all the light of the stars that are not part of the disc, which can be of very different origins. Moreover, the photometric-based definition is also prone to put in one single category stellar components with markedly different kinematics \citep{Du2020, Du2021}.

It is worth noting that fitting procedures that include more than two components, like  bars, rings, or nuclear clusters,  when present, are also common \citep[e.g.][]{Laurikainen2005, Weinzirl2009}. However, when only two components are fitted to 1-dimensional surface brightness distributions of a sample of disk galaxies in the local Universe, an interesting trend is found. Using high-resolution images from HST in the V-band \citep{FisherDrory2008} and Spitzer images in near-infrared  \citep{FisherDrory2010}, it was shown that the S\'ersic index distribution obtained from the two-component photometric decompositions shows a correlation with bulge morphological type.
Bulges morphologically classified as pseudo-bulges, those with {\it morphology reminiscent of disk galaxies, with inner spiral structure, rings, or bars}, infrequently show $n>2$. On the other hand, those classified as classical bulges,  {\it easily recognized as having morphologies very similar to E-type galaxies},  rarely show a S\'ersic index $n<2$.  For a complete and updated guide of bulge classification into classical and pseudo-bulges according to these authors we refer the reader to \citet{FisherDrory2016}. 

Although classification into these two distinct types of bulges is not always straightforward because of the existence of overlapping criteria in composite systems \citep[e.g.][]{Mendez-Abreu2014, Erwin2015}, there are clear indications of the existence of more than one photometric bulge formation channel, underlying the differences observed in the bulge surface brightness profiles. The physical origin of this apparent dichotomy is not well understood. 

It should be noted that classification into classical and pseudo-bulges is also carried out using 2-dimensional surface brightness image decomposition of galaxies and invoking other criteria. For example \cite{Gadotti2009} classifies bulges according the position of galaxies in the mean surface brightness - effective radius ($<\mu_e>-r_{\rm eff}$) diagram  known as the Kormendy
relation \citep{Kormendy1977} and \cite{Luo2020} use the $\Delta\Sigma_1$ parameter, that measures the relative central stellar-mass surface density within the inner ${\rm kpc}$ of galaxies. Although the same terminology of pseudo- and classical bulge is used in these lines of work, the classification criteria differ from those of \citet{FisherDrory2016} and the dichotomy with bulge-type and S\'ersic index does not arise with clarity. 

Galaxies of the mass of our Galaxy, the Milky Way (MW) and our neighbour Andromeda (M31), are particularly interesting targets for studying the origin of different types of photometric bulges.  The bulge of the MW  seems to be a prototypical case formed exclusively via secular evolution \citep[see e.g.][]{Shen2010, Fragkoudi2020}, except for a small fraction of stars \citep{Kunder2016, Kunder2020}. M31, for its part, shows a more massive bulge \citep{Saglia2010} with a classical or composite bulge morphology \citep{Mould2013, Blana-Diaz2016, Blana-Diaz2018}. Beyond the local group, MW/M31-sized galaxies in the nearby Universe show a great diversity in their bulge properties \citep{Bell2017}. A successful model of galaxy formation within the LCDM paradigm should be able to describe this observed diversity.

From a theoretical perspective, galaxy formation models establish that two broad main drivers of photometric bulge formation are mergers \citep[][]{Toomre1977, Hopkins2009, BrooksChristensen2015} and secular evolution \citep[see, e.g.,][]{KormendyKennicutt2004, Athanassoula2005}. A third proposed channel of bulge formation is the infall of giant clumps of de-stabilized gas into the central regions of the galaxy, due to dynamical friction, in the early epochs of galaxy formation \citep{Elmegreen1995, Elmegreen2009, Dekel2009, Ceverino2010}. Historically, most theoretical works on bulge formation have focused on isolated simulations or tailored interactions with constrained orbital parameters. These have the advantage of being adequate to study in detail the effects of a given formation mechanism at the expense of losing perspective on the complex interactions and feedback loops between different physical processes. 
The study of bulge formation within a cosmological framework was done almost exclusively in the zoom-in regime, where a single DM halo is selected and the galaxy within it can be re-simulated at a higher resolution \citep{Okamoto2013,  Guedes2013, Tissera2018, Buck2018a, Gargiulo2019} The main disadvantage is the limited number of the available simulated galaxies.  Recently, for example, \citet[][]{Gargiulo2019} studied the photometric bulges of a sample of 30 high-resolution simulations from the Auriga project \citep{Grand2017} and found that all of them showed low S\'ersic indices and most of them had properties more akin to pseudo-bulges. They suggested that the galaxy sample may have an environmental bias due to the isolation criterion used to select the DM haloes in the parent simulation, where galaxies were later re-simulated.    

The advent of joint efforts to run cosmological simulations in large volumes, with particle and dynamical resolution proper of the re-simulation regime, such as the TNG50 simulation of the IllustrisTNG project \citep{Pillepich2019, Nelson2019b} as well as recent projects such as New-Horizon \citep{Dubois2020}, represents a renewed opportunity to study in detail the subject of bulge formation, as done recently, for example, by \citet{Du2021}. It is now possible to explore the complexity associated to the evolution in a cosmological context in a large sample of galaxies that is exempt of an environmental selection bias.  

In this work, we will make use of the TNG50 simulation
\citep[][TNG50 from now on]{Pillepich2019, Nelson2019a}, the cosmological box 
with highest resolution of the three IllustrisTNG volumes. We study
the properties of photometric bulges in a sample of MW/M31-like galaxies and search for the drivers of their formation. We will focus on the origin of the
different bulge surface brightness profiles of MW/M31-like galaxies, characterized by the S\'ersic index and bulge-to-total ratio. We place emphasis on the effects of the environment, the role of mergers, and the influence of bars. 

This paper is organized as follows. In Section~\ref{sec:methods} we describe the methods and simulation used throughout this work, and define our sample of MW-like galaxies selected from the cosmological box. In Section~\ref{sec:environmental-dependence} we explore the influence of environment on the type of photometric bulge in our sample of simulated galaxies. In Section~\ref{sec:insitu-acc} we characterize the populations of stars born in--situ and those accreted in mergers (ex--situ) that populate the bulges at $z=0$. In Section~\ref{sec:roleofmergers} we explore the role of mergers in shaping the surface brightness profile in the simulated galaxies. In Section~\ref{sec:barinfluence} we study the impact of bars in the formation of photometric bulges. In Section~\ref{sec:discussion} we discuss our results in a broader context and finally, in Section~\ref{sec:conclusions} we summarize our results and present our conclusions.   

\section{Methods}
\label{sec:methods}

In this section we describe the simulation used throughout this work and the selection of the sample of simulated MW/M31-like galaxies. We characterize the two different methods to quantify the local density for each galaxy in our sample. We also present the methodology used to measure the bar strengths in simulated galaxies.  

\subsection{Simulation}
\label{sec:sims}

 The Illustris-{\it The Next Generation} project \citep[hereafter IllustrisTNG,][]{Pillepich2018a, Nelson2018, Naiman2018, Springel2018, Marinacci2018} is a collection of cosmological  magnetohydrodynamical simulations that comprise a set of cosmological boxes with different sizes and mass resolutions\footnote{https://www.tng-project.org}, and were carried-out with the  magneto-hydrodynamical moving-mesh code {\sc arepo} \citep{Springel2010}. Currently, all simulations, including TNG50, are publicly available \citep{Nelson2019b}.
 
 IllustrisTNG is the successor of the original Illustris project \citep{Vogelsberger2014a, Vogelsberger2014b, Genel2014, Nelson2015} and is based on a substantially updated physical model \citep{Weinberger2017, Pillepich2018a} . One of the main distinct aspects of the IllustrisTNG physical model with respect to the one used in Illustris is the treatment of the energetic feedback from supermassive black holes (BHs). In particular, the low accretion regime of BH feedback \citep[also known as "radio mode",][]{Croton2006} was strongly modified. The bubble radio feedback based on the model from \citet{Sijacki2007}  was replaced by small-scale, kinetic winds \citep{Weinberger2017}. Other important features are the inclusion of seed magnetic fields that evolve and are amplified during cosmic evolution \citep{Pakmor2014} and a more efficient numerical implementation in {\sc arepo}.  Altogether, the updated IllustrisTNG physical model helps to reduce discrepancies with observational constraints identified for the original Illustris simulations \citep[][Sec.~6]{Nelson2015}. A detailed analysis of the updated physical scheme used in IllustrisTNG and its impact on the galaxy population can be found in \citet{Pillepich2018b}.
 
In this work we make use of the highest resolution version of IllustrisTNG, the TNG50-1 run \citep[hereafter  TNG50,][]{Pillepich2019, Nelson2019a}. 
The main parameters of the TNG50 simulations are shown in Table \ref{table:sims}. The TNG50-1 simulation consists of a periodic volume of $L_{\rm box} = 35 h^{-1} \sim 50$ {\rm cMpc} in extent with a gas cell mass resolution of $8.5 \times 10^4$ M$_\odot$. This resolution is only  a factor of approximately two worse than that of the Auriga simulations \citep[][hereafter G2017]{Grand2017} in their level 4 resolution hierarchy. Thus, TNG50 allows us to study the inner structure of a large sample of MW/M31-like galaxies (see next section), evolving under the influence of the cosmological environment, with a numerical resolution close to those obtained in typical zoom-in re-simulations.

\begin{table}
	\centering
	\caption{Attributes of the simulation analyzed in this work. The properties shown in rows are: box side-length (co-moving units); number of initial gas cells and dark matter particles; mean baryon and dark matter particle mass resolution (solar masses); mean size of star--forming gas cells. For a complete description of the simulation see the IllustrisTNG project homepage.}
	
	\label{table:sims}
	\begin{tabular}{lcl} 
		\hline \hline
 Run & & {\bf TNG50-1} \\ \hline
 $L_{\rm box}$ & [\,Mpc] & 51.7$^3$  \\
 $N_{\rm DM, GAS}$ & - & 2160$^3$  \\
 $m_{\rm baryon}$ & [\,M$_\odot$\,] & $8.5\times 10^4$  \\
 $m_{\rm DM}$ & [\,M$_\odot$\,] & 4.6 $\times 10^5$ \\
 $ <r_{\rm gas,SF}>$ & [\,pc\,] & $\sim 100-150$ \\
 $\epsilon_{\rm DM,\star}$ & [\,pc\,] & 290 \\
 
		\hline
	\end{tabular}
\end{table}

\subsection{Sample of MW/M31-like galaxies}
\label{sec:mw-like-gal}

 We select our sample of MW/M31-like galaxies in TNG50 based on the procedure described in Pillepich et al. (in prep, P22 from now on) and used in \citet{Engler2020} and \citet{Pillepich2021b}.  Note however that in this work we relaxed their isolation criterion to properly account for possible environmental effects. We select MW/M31 analogs such that:

\begin{enumerate}
    
    \item They have stellar masses $M_{\star}$ in the range $[10^{10.5} - 10^{11.2}] \rm M_{\odot}$, where $M_{\star}$ is the sum of all stellar particles enclosed in a sphere of 30 kpc, centered at the most bound DM particle, at $z = 0$.\\
    
    \item  They have a disk morphology. To quantify this, we select galaxies with $s < 0.45$, where $s=c/a$ is the minor-to-major axis ratio of the stellar moment of inertia tensor, measured between one and two times the stellar half-mass radius. Additionally, twenty five galaxies with $s > 0.45$ in the adopted mass range that show a clear disk morphology by visually inspecting synthetic 3-band stellar-light images are added to the simulated galaxy sample. \\
    
\end{enumerate}

 A total of 287 MW-like galaxies following these criteria are found in TNG50. As stated above, in P21 a third criterion based on isolation is applied to this sample of galaxies: no other galaxy with stellar mass $M_{\star} > 10.5$ should lie within a distance of 500 {\rm kpc} of the corresponding simulated galaxy and the galaxy should have a host DM halo with $M_{200c} < 10^{13} {\rm M_{\odot}}$, where $M_{200c}$ is the sum of the mass of DM particles inside a sphere centered in the halo (identified with a Friend-of-Friends algorithm) within which the mean density is $200$ times the critical density of the Universe.  When the isolation criterion is relaxed, 89 galaxies are added to their sample of 198 simulated galaxies, summing up to the 287 galaxies of our sample. 

\begin{figure*}
\includegraphics[scale=.4]{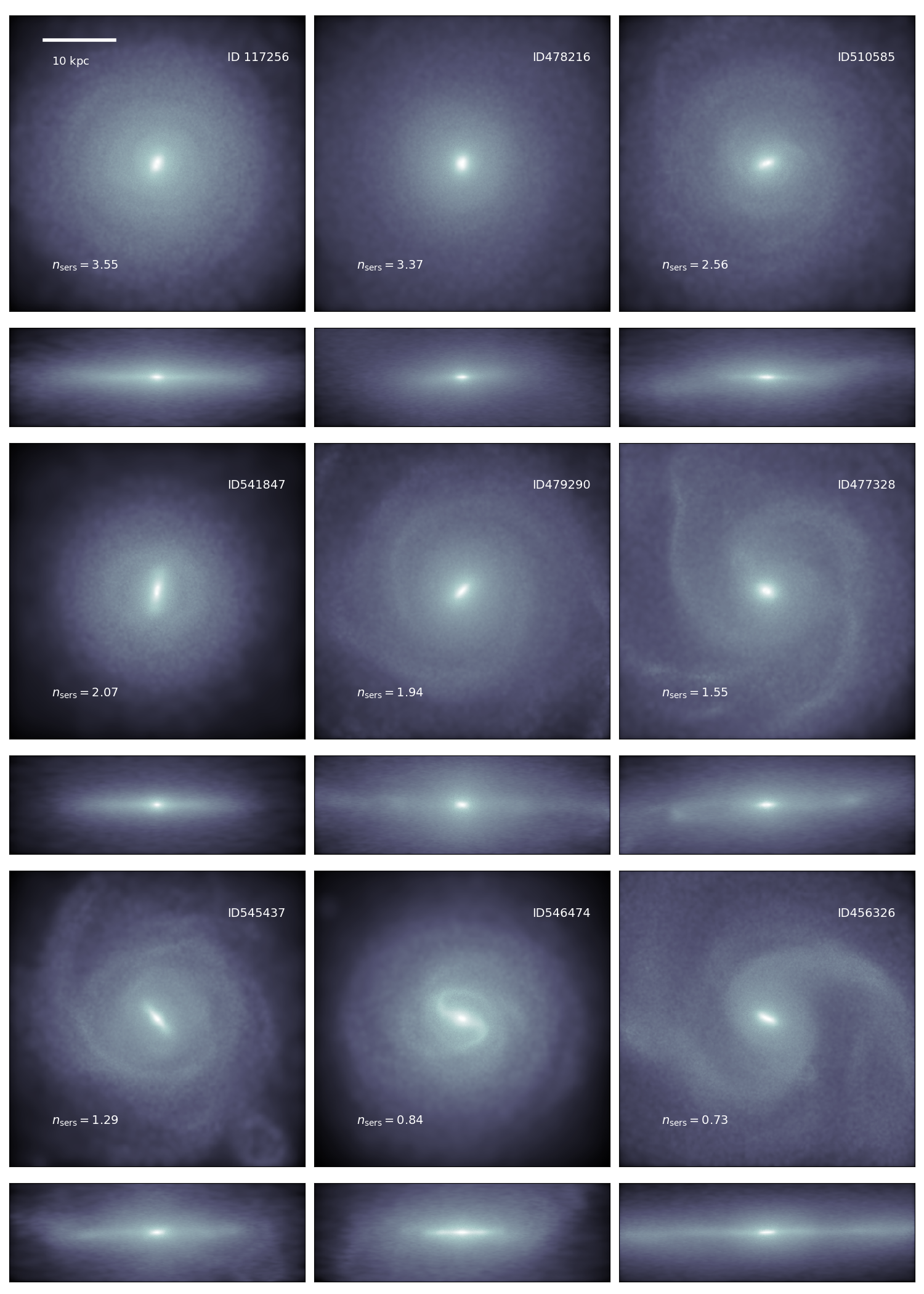}
\caption{Face-on and edge-on projections of mock images based on the stellar density of nine examples of simulated galaxies from the sample used in this work, composed of 287 TNG50 MW/M31-like galaxies at $z=0$. Galaxies are ordered from the top left corner to the right bottom corner by the S\'ersic index of the photometric bulge component. In each panel, the identifier of the galaxy in the simulation is shown in the top right corner, and the S\'ersic index in the bottom left corner. A scale of $10 {\rm kpc}$ is indicated by the white bar in the top left panel. Galaxies with higher S\'ersic bulges show featureless discs, and those with lower S\'ersic are prone to exhibit spiral arms and stronger features.}

\label{fig:galaxies}
\end{figure*}

\subsubsection{Two-component decomposition of galaxies: photometric bulges}
\label{sec:2components}

We perform a two-component decomposition of the surface brightness profiles (SBP) of the galaxies in our sample. We assume a smooth surface brightness distribution described by the sum of an exponential profile and a \citet{Sersic1968} function , 

\begin{equation}
I(r)=I_{\rm e,b}\exp\big\{ - b_{\rm n} \big[(r/r_{\rm eff})^{1/n} -1  \big] \big\} + I_{0,d}\exp\left[- (r/R_{\rm scale}) \right ]\, ,
\end{equation}
\label{eq:serexp}

\noindent where $r_{\rm eff}$ is the effective radius of the S\'ersic model, $n$ is the S\'ersic index, $I_{\rm e,b}$ is the intensity of the bulge component at $r_{\rm eff}$, $R_{\rm scale}$ is the disc scale length and $I_{\rm 0,d}$ is the central intensity of the disc component. The factor $b_{\rm n}$ is such that $\Gamma(2n) = 2\gamma (2n, b_{\rm n})$, where $\Gamma$ is the complete gamma function and $\gamma$ is the incomplete gamma function.

We follow \citet[][G19 hereafter]{Gargiulo2019} and measure the SBP from the face-on projection of each galaxy.
The SBP is computed after averaging the total luminosity in the V-band inside 500 {\rm pc} wide concentric annuli, from just outside the resolution limit and out to the optical radius of each galaxy. Luminosities are derived from the magnitudes of the stellar particles, which are treated as single stellar populations after formation using population synthesis models \citep{BruzualCharlot2003}. The resolution limit is defined as three times the minimum allowed softening length for gas cells ( $3 \times   \epsilon_{\rm gas,min} = 222 {\rm pc} $). 
The optical radius is defined as the radius at which the surface brightness in the B band drops below $\mu_B = 25~ {\rm mag~arcsec^{-2}}$.  

We have to bear in mind that a relatively large fraction of the simulated galaxies of our sample contain bars at $z=0$ (from $\sim 30\%$ to $\sim 55\%$, depending on the bar strength  threshold used, see Sec. ~\ref{sec:barinfluence}). We do not fit an extra component to the surface brightness profiles to account for this component. Although we exclude the points that show evidence for an excess of light due to a bar when it is identifiable in the SBP from the fitting procedure, contamination from the bar will be present in each of the components of the fit. If the bar is short and with low ellipticity, the majority of the bar light will be absorbed by the S\'ersic component. In the case of long bars with a close-to-exponential profile, the majority of their light will be part of the exponential profile. We deliberately choose to fit only two-component models 
to the SBPs. This is because the two-peaked distribution of S\'ersic indices reported by \citet{FisherDrory2008, FisherDrory2010, FisherDrory2011}, was derived using two-component photometric decompositions of SBPs of galaxies. However, adding a third or more components to the SBPs would also yield interesting and complementary insights of the physical origin of photometric bulges \citep[see, e.g.,][]{Blazquez-Calero2020}  and we defer this to a future work.

\begin{figure}
\includegraphics[scale=0.425]{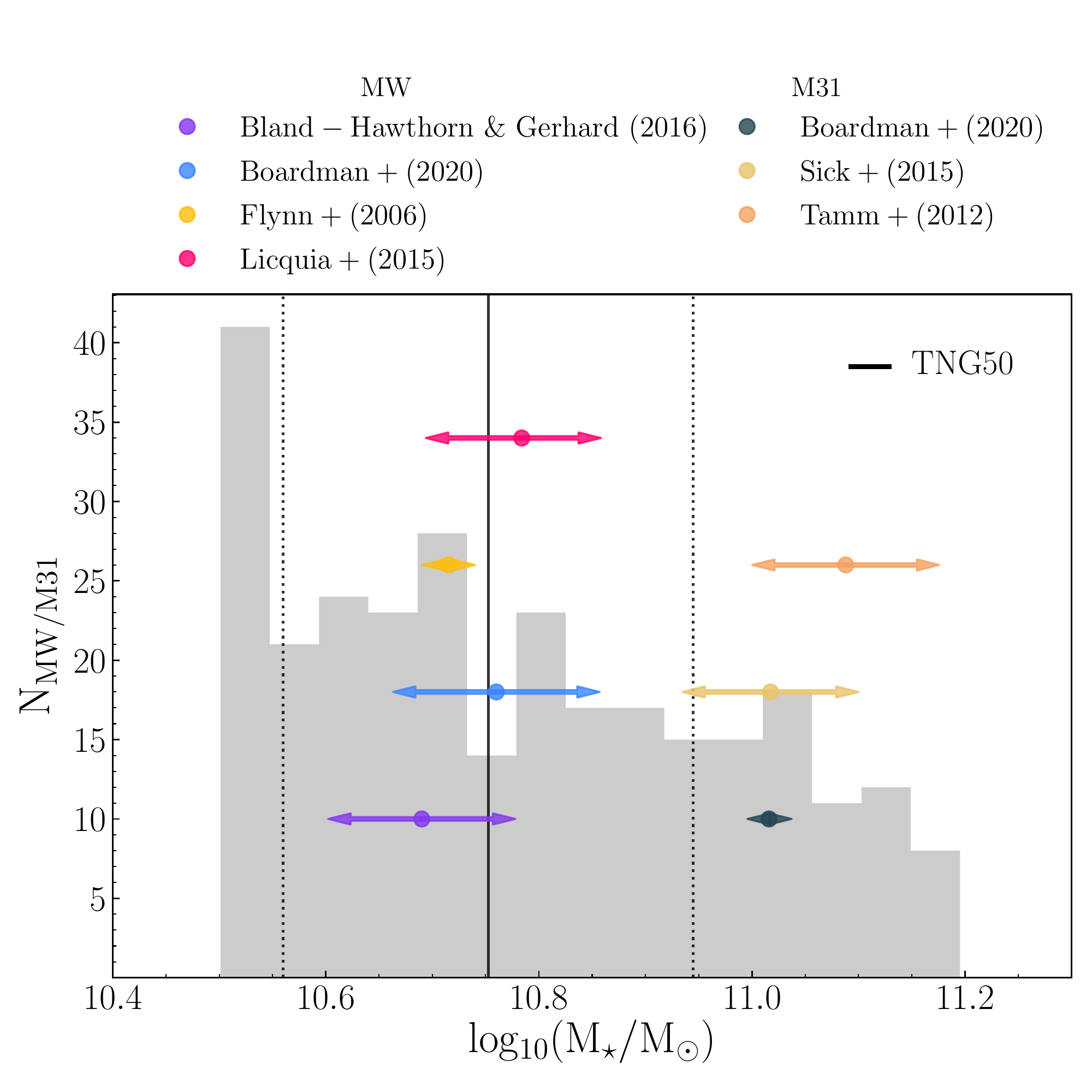}
\caption{Stellar mass distribution of the sample of 287 TNG50 MW/M31-like galaxies described in Sec.~\ref{sec:mw-like-gal}. The solid black line indicates the median of the distribution and the dotted black lines show the interquartile range. The coloured circles with bidirectional arrows indicate the estimated ranges for the MW and M31 stellar mass, from different authors as indicated in the legend and cited in the text.}
\label{fig:massdistMW}
\end{figure}

\subsubsection{Kinematic bulges}
\label{sec:kinematicbulges}

To define the kinematic bulge, we select particles inside a spherical region of $2 \times r_{\rm eff}$ with circularities $|\epsilon| < 0.7$, where the circularity $|\epsilon|$ is defined as  $\epsilon = J_z / J(E)$ \citep{Abadi2003}. Here, $J_z$ is the angular momentum component perpendicular to the disk plane of a stellar particle with orbital energy $E$,  and $J(E)$ is the maximum possible angular momentum for the given {\it E}. Particles with circularities $|\epsilon| > 0.7$ are considered disk particles. As discussed by \citet{Peebles2020}, this particular circularity cut can include in the disc component a fraction of particles with orbits that significantly depart from circular. 
We note that this circularity cut is commonly assumed in the literature and, thus, it facilitates comparison with previous works. More stringent circularity cuts (e.g.$|\epsilon| > 0.8$) modify only slightly the in--situ/ex--situ particle fractions and our conclusions remain unmodified. The fixed spatial cut of $2 \times r_{\rm eff}$ is, in general,  close to the radius beyond which the exponential disk starts to dominate the light profile. We keep this fixed definition to compare with results presented in G19 for the Auriga simulations. 

\begin{figure*}
\includegraphics[scale=0.46]{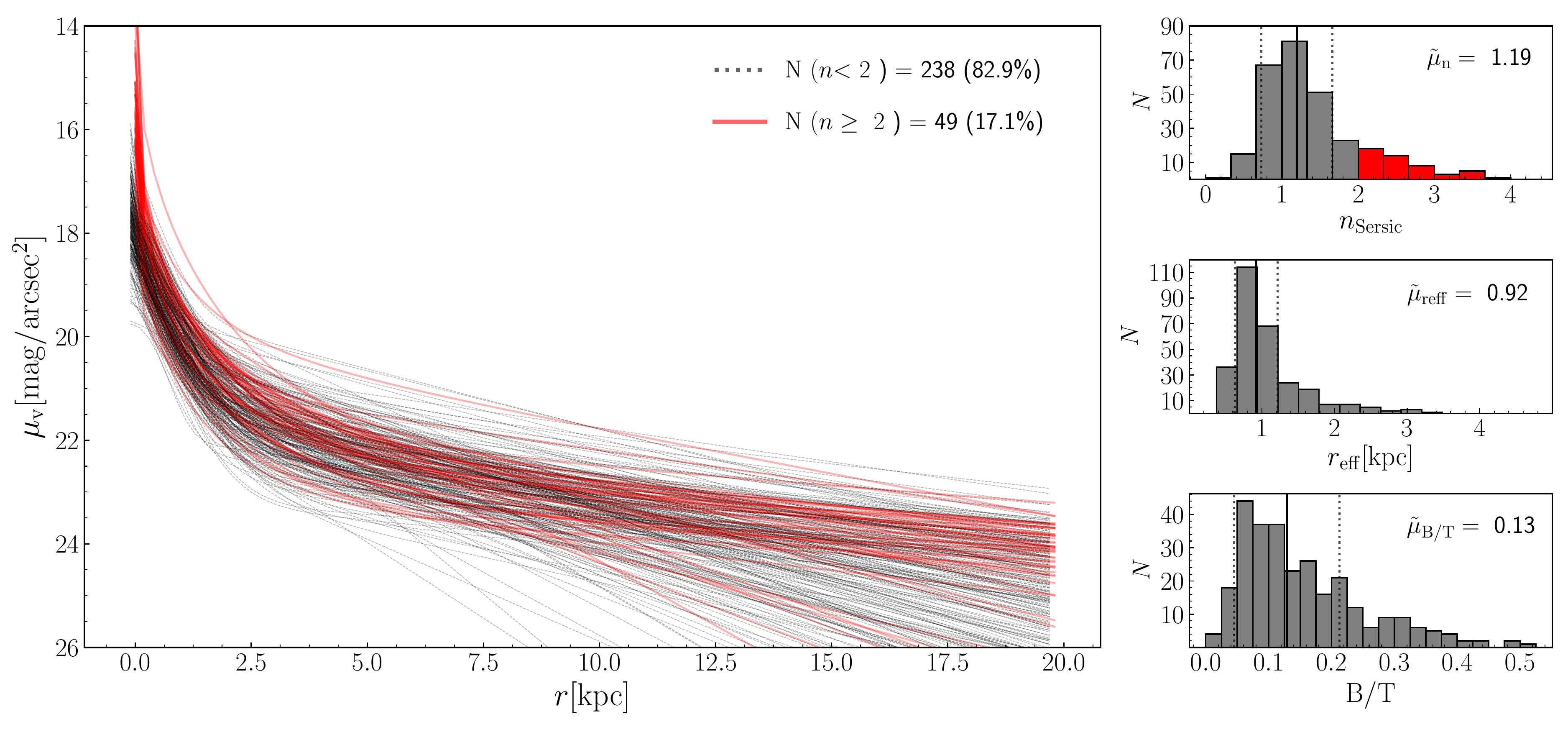}

\caption{{\it Left panel}: Best-fitting functions to the surface brightness profiles of all TNG50  MW/M31-like galaxies of our sample, composed of a S\'ersic profile plus an exponential profile (see details on the text). Highlighted with red are the fitting functions with high S\'ersic index ($17.1\%$ of the sample) that show a concentrated light distribution.  
{\it Right panels}: Distributions of S\'ersic index, effective radii and B/T from the top panel to the bottom panel, respectively. Median values of the distributions are indicated with a solid vertical line and the interquartile range is indicated with dotted vertical lines.}
\label{fig:profiles}
\end{figure*}

\subsubsection{Environment definitions}
\label{sec:def-environment}

Our first estimate of the local density of the environment is the overdensity parameter:

\begin{equation}
\label{eq:local_overdens}
1+\delta = \frac{P}{P_{\rm median}},
\end{equation}

\noindent where $P$ is the volumetric density of galaxies at the position of the {\it i}th MW/M31-like galaxy defined as:

\begin{equation}
P(\bf r\rm_{i})= \frac{3k}{4 \pi \sum_{j=1}^{k} d_{ij}^{3}}.
\end{equation}

\noindent Here ${\bf r}_{\rm i}$ is the position of the ${\it i}$th MW/M31-like galaxy, ${\rm d_{ij}}$ is the distance between the ${\it i}$th MW/M31-like galaxy and its ${\it j}$th neighbour with a mass above given mass cut, and ${\rm k}$ is the number of neighbours considered.  
$P_{\rm median}$ is the median volumetric density of galaxies with the adopted mass cut in the cosmological volume of the simulation.

A second approach to measure the local density of galaxies is to count neighbours in a fixed size region \citep[e.g.][]{BlantonMoustakas2009}. 
 We define a $738.2~{\rm kpc}$ sphere around each MW/M31-like galaxy in our sample 
 and count the number of neighbouring galaxies inside the sphere with total mass above a given mass cut. 

\subsubsection{Bar strength measurement}
\label{sec:bar-definitions}

We compute the bar strength for galaxies in our sample by means of Fourier mode analysis \citep[e.g.,][]{Grand2016}. 
We define equally spaced radial annuli in the face-on projections of the disc galaxies and compute the complex Fourier coefficients to quantify azimuthal patterns in the mass distribution with {\it n}-fold axisymmetry:

\begin{equation}
 a_{n} (R_j) = \sum^{N_{R}}_{i=1} m_{i} \, {cos}(n \, \theta_i ),
 \end{equation}
and
\begin{equation}
 b_{n} (R_j)= \sum^{N_{R}}_{i=1} m_{i} \, {sin}(n \, \theta_i ),
\end{equation}

\noindent where $a_n$ and $b_n$ are the real and imaginary components of the Fourier coefficients. The sum is over the {\it i}-th particle in the {\it j}-th annulus. Here, $m_i$ and $\theta_i$ are the mass and azimuthal angle of the $i$-th particle, respectively. 
We characterize the strength of the $n$-th Fourier mode by its amplitude, given by: 

\begin{equation}
B_n (R_j,t) = \sqrt{a_n(R_j,t)^2 + b_n(R_j,t)^2}.
\label{BFou}
\end{equation}
 The second mode, $n=2$, corresponds to a bisymmetric signal with a periodicity of $\pi$ radians, such as a double arm or a bar. In order to correctly quantify the bar strength one should determine where the bar ends and the spiral arms begin. To do so, we make use of the bar phase angle, which  is computed as:

\begin{equation}
\theta ' _2 = \frac{1}{2} {\rm atan2} (b_2,a_2).
\label{ang}
\end{equation}

The bar phase angle remains almost constant in consecutive radial annuli until the bar ends. Following \citet{Grand2016} we define the bar length, i.e. the bar semi-major axis, as the radius at which the difference in the phase angle between two consecutive radial bins is larger than 0.5. Finally, the mass-weighted mean of the amplitude of the $m=2$ Fourier mode within the bar region, i.e., the bar strength, is defined as:  

\begin{equation}
A_2 (t) = \frac{\sum _j   B_{2} (R_j,t) }{\sum _j  B_{0} (R_j,t)}.
\label{a2bar}
\end{equation}

\begin{figure}
\includegraphics[scale=0.55]{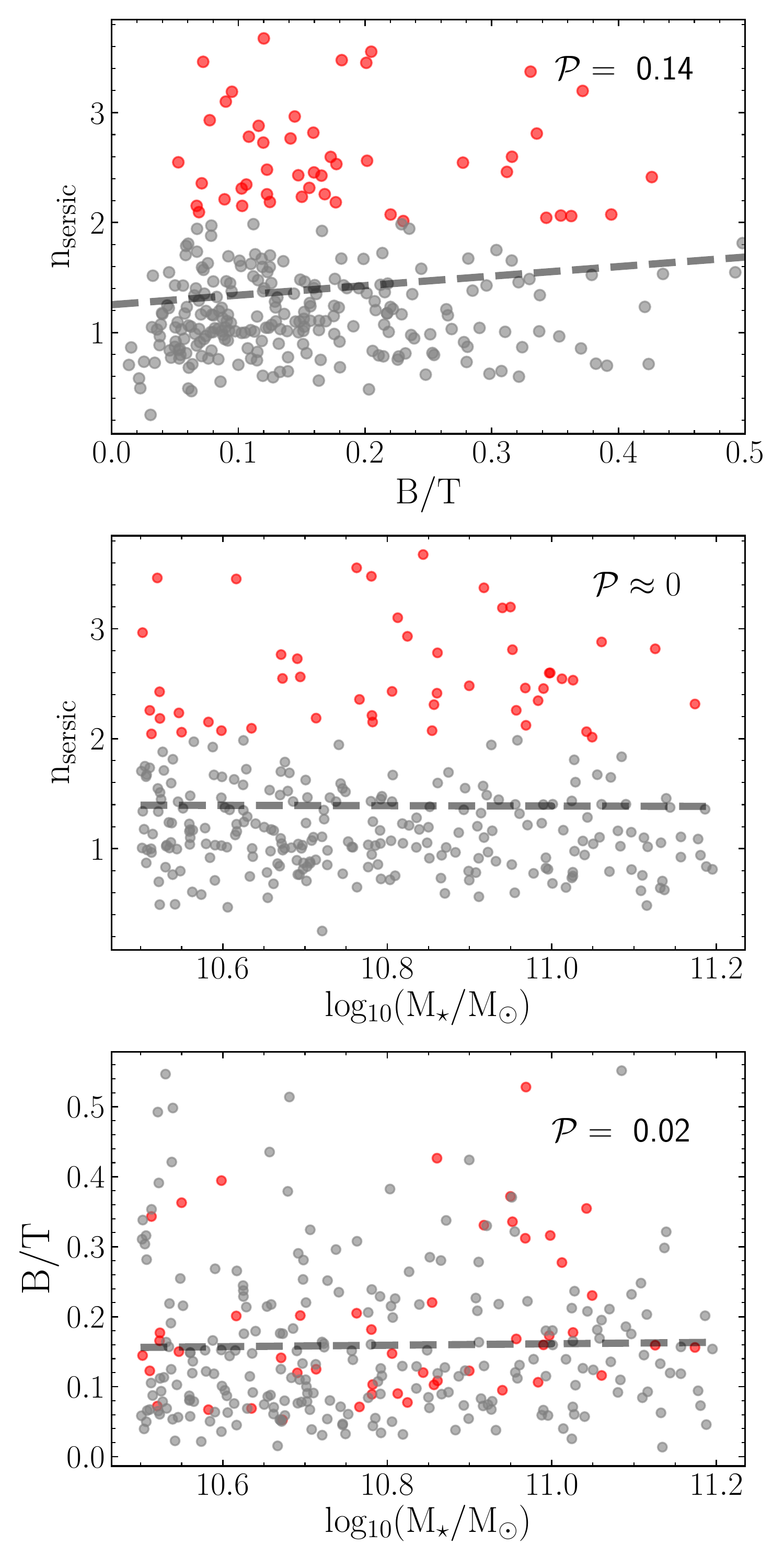}

\caption{{\it Upper panel: }S\'ersic index vs. {\rm B/T} for our sample of TNG50 MW/M31-like galaxies, derived from 2-component decompositions to the surface brightness profiles in the V-band. Simulated galaxies with S\'ersic indices above $n=2$ are highlighted in red.  {\it Middle panel:} S\'ersic index vs. stellar mass for our sample of MW/M31-like galaxies
{\it Lower panel:} S\'ersic index vs. stellar mass for our sample of MW/M31-like galaxies.  The dashed lines show a least-squares linear fit to the data in the three panels. The value of the pearson coefficient, that measures the degree of linear correlation, is indicated in the top right corner of each panel. S\'ersic index and {\rm B/T} are independent quantities, and neither of them is correlated with the stellar mass of the simulated galaxies.} 

\label{fig:BoT-sersic}
\end{figure}

\section{THE BULGES OF TNG50 MW/M31-like GALAXIES}

We begin by analysing the general properties of our sample of 287 MW/M31-like simulated galaxies. Fig.~\ref{fig:galaxies} shows a selection of 9 galaxies from our sample at $z=0$, ordered by their S\'ersic index (see Sec.~\ref{sec:2components}) from left to right and top to bottom. These examples highlight the differences of the concentrated central regions and featureless discs of  galaxies with high S\'ersic bulges close to the top left corner and the nearly bulgeless galaxies with clear spiral features towards the lower right corner.

Fig.~\ref{fig:massdistMW} shows the stellar mass distribution of our sample with the median and interquartile range indicated with  solid and dotted black vertical lines, respectively. The galaxy stellar mass is calculated summing the stellar masses of all particles inside a sphere of 30 {\rm kpc} radius and centered on the most bound DM particle.  The coloured points and arrows indicate the ranges of different stellar mass estimates for the the MW \citep{Flynn2006, LicquiaNewman2015, Bland-HawthornGerhard2016, Boardman2020} and M31 \citep{Tamm2012, Sick2015, Boardman2020}.

\begin{figure*}
\includegraphics[scale=0.45]{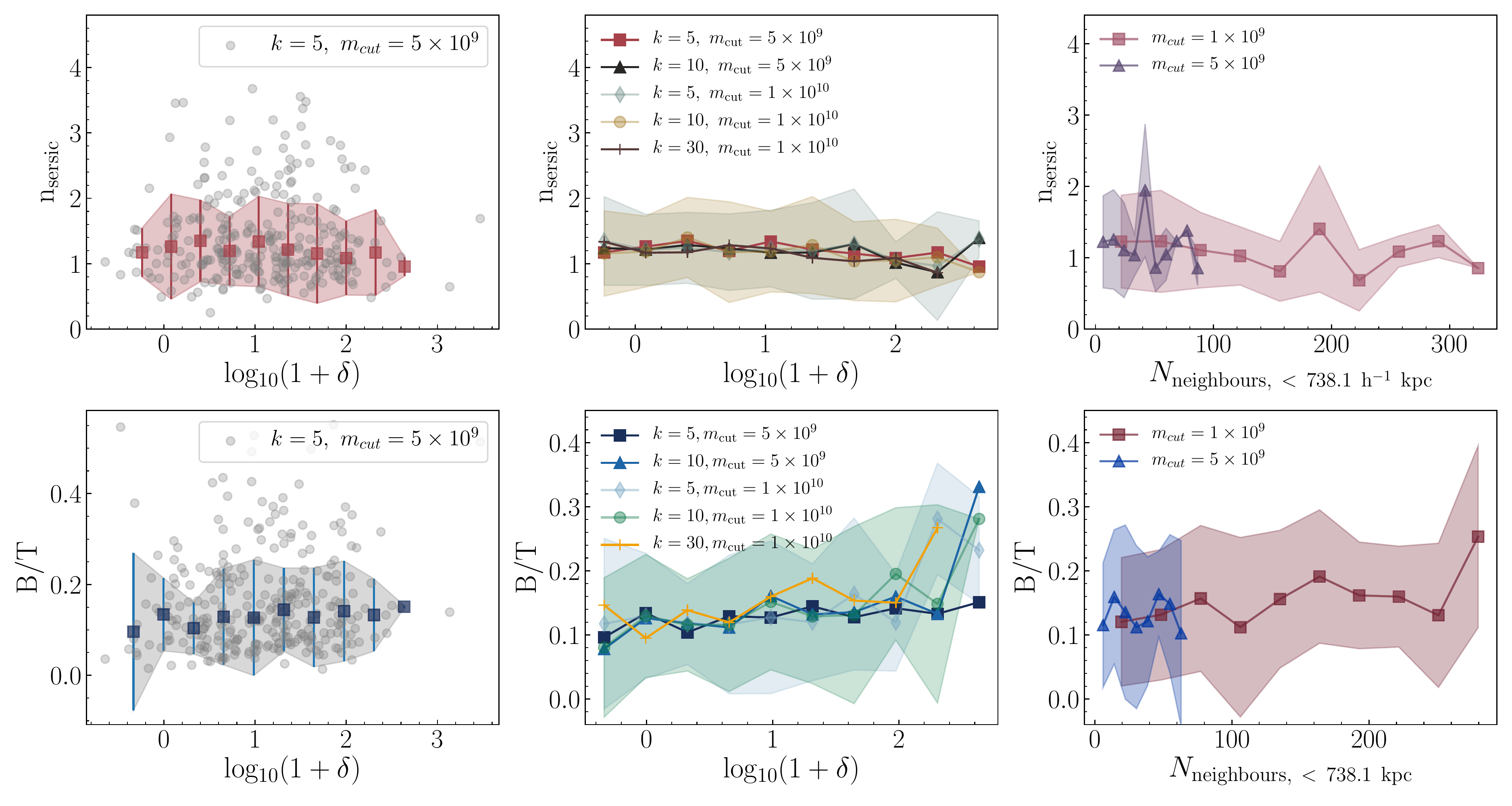}

\caption{{\it Top left}: S\'ersic index of bulges in our sample of TNG50 MW/M31-like galaxies, as a function of overdensity parameter (see Sec.~\ref{sec:def-environment} for the definition) considering $k=5$ and a stellar mass cut of $5 \times 10^9$ M$_{\odot}$. Dark red squares show the median values in overdensity bins and the corresponding shaded region and errorbars indicate the standard deviation in each bin. {\it Top middle:} Medians of S\'ersic indices in overdensity bins, for different choices of parameters $k$ and $m_{\rm cut}$ in solar masses, as indicated in the legend. We only show two shaded regions representing the standard deviation in each bin to avoid overcrowding in the plot.
{\it Top right:} S\'ersic index as a function of the number of neighbours inside a 738.1 {\rm kpc} sphere. {\it Bottom panels:} Same as top panels but for B/T values. No measurable dependence of S\'ersic index with environment is found and a slight increase of B/T values with overdensity of galaxies can be seen.  
}
\label{fig:sers-bot-environmental}
\end{figure*}

In the left panel of Fig.~\ref{fig:profiles} we show the best-fitting functions of the two-component decomposition (see Sec.~\ref{sec:2components}) for the whole sample of MW/M31-like galaxies. Highlighted with red lines are the best-fitting functions for galaxies with high S\'ersic indices.  We find that $17.1\%$ of the galaxies in our sample have S\'ersic indices $n > 2 $. In the following sections we will explore the physical process, or combination of processes, that drives the evolution of these types of concentrated bulges. We will contrast these results with the evolution of the more abundant systems with flatter SBP.  In the right panels of Fig.~\ref{fig:profiles}, we show the S\'ersic indices, the effective radii and the B/T distributions. All these quantities were derived from the fits to the surface brightness profiles, where ${\rm B/T}$ is the quotient of the integrals of the S\'ersic component and the complete fitted function. The medians are indicated with a black  solid line and the inter-quartile ranges are depicted with dotted lines.  The three distributions, with median values of $n_{\rm Sersic} = 1.19$ , $r_{\rm eff} = 0.92$ and  $\mu_{\rm B/T} = 0.13$, respectively, show a peak skewed towards lower values of and a tail to higher values.

We will see how the S\'ersic index
and {\rm B/T} are affected by different aspects of galaxy evolution, so it is important to show that both properties are independent and
that we can extract different insights from both of them. The upper panel of Fig.~\ref{fig:BoT-sersic} shows that, effectively, the S\'ersic index and {\rm B/T} are not correlated. Notice the low value of the Pearson coefficient that quantifies the grade of
correlation in a linear fit. A Pearson coefficient greater that $\mathcal{P} =0.5$ would imply a significant correlation.  Additionally, we show how the two quantities derived from SBPs behave as a function of the stellar mass of the simulated galaxies, to ensure that there is no underlying dependence of our results with stellar mass. The middle and lower panel of Fig.~\ref{fig:BoT-sersic} show the S\'ersic index and {\rm B/T} as a function of stellar mass, respectively.  Note that neither the Sersic index nor {\rm B/T} show a correlation with stellar mass, at least in the relatively narrow stellar mass range adopted to select our sample of simulated galaxies. A correlation between S\'ersic index and/or {\rm B/T} with mass could emerge if a wider mass range of simulated galaxies is considered, but such analysis is beyond the scope of this paper and is postponed to future work.  

\subsection{Does bulge type depend on environment?}
\label{sec:environmental-dependence}

We study the influence of environment on the properties of photometric bulges in our sample of simulated galaxies, using the definitions presented in Sec.~\ref{sec:def-environment}.   We focus on the S\'ersic index and B/T ratio derived as described in Sec.~\ref{sec:2components}.
As mentioned in Sec.~\ref{sec:intro}, G19 suggested that the prevalence of low S\'ersic bulges in the Auriga galaxies might be, in part, a consequence of the isolation criterion used to select the Auriga DM Haloes for re-simulation. This assumption can be tested with TNG50 in a broader picture, thanks to the larger diversity of environments surrounding our sample of simulated galaxies. 

 We begin by analyzing the dependence of S\'ersic index and B/T with the overdensity parameter. In the left panels of  Fig.~\ref{fig:sers-bot-environmental} we show the S\'ersic index and B/T as a function of the logarithm of the overdensity parameter, considering a number of  neighbours $k = 5$, and a mass cut $ m_{\rm cut} = 5 \times 10^9$ M$_{\odot}$ (only galaxies with total masses, i.e. the sum of all types of particles of the subhalo, larger than the cut, are counted). The squares show the median values of the Sersic index and B/T in bins of $\log_{10}(1+\delta)$.
The shaded areas show the standard deviations on each bin. For these particular values of $k$ and  $m_{\rm cut}$ we find no significant correlation between S\'ersic index, or B/T, with the local overdensity. The middle panel of this figure explores the impact that different values of $k$ and $m_{\rm cut}$ can have on this analysis. These panels show that, independently of the values chosen for these parameters, we find no significant correlation between $n_{\rm sersic}$ and $\log_{10}(1+\delta)$ and only a very mild correlation with B/T. 

As a second test, we measure the local density of galaxies as the number of neighbours inside a sphere of fixed size (see Sec.~ \ref{sec:def-environment}). The rightmost panels of Fig.~\ref{fig:sers-bot-environmental} show the dependence of the S\'ersic index and B/T  on this local density. We find a result similar to the previous one: the S\'ersic indices do not correlate with environment as characterized by the number of galaxies that can be found within a sphere of 738.2 {\rm kpc} radii, while B/T shows a mild correlation towards a higher number of neighbours. This is also independent of the $m_{\rm cut}$ adopted. 

Our results strongly point towards a lack of environmental dependence of bulge type in MW/M31-like galaxies. The lack of environmental dependence on these two particular properties of SBPs of galaxies is not surprising. Structural properties of galaxies are shown to be rather independent of the abundance of neighbouring galaxies and more dependent on intrinsic properties like their mass \citep[e.g.]{BlantonMoustakas2009, PeeblesNusser2010}. We acknowledge that, even with the TNG50 simulations, the environmental dependence diagnostics considered in this work may suffer from low number statistics, especially towards larger densities. A larger cosmological
volume would be desirable to densely populate all the overdensity bins. Moreover, although TNG50 stands as one the largest efforts to simulate a large cosmological volume with a mass and spatial resolutions comparable with those achieved in the realm of the zoom-in technique, our results may still be influenced by cosmic variance. 

\section{In--situ and ex--situ components}
\label{sec:insitu-acc}

In this section we analyze the origin of stellar particles in the central regions of our sample of MW/M31-like galaxies, where photometric bulges arise, dividing them into {\it in--situ} and {\it ex--situ} particles. 
In--situ particles are defined as those formed from condensation of the gas that belongs to the host galaxy, while accreted particles are those formed within the potential wells of galaxies that are later accreted onto the main host.
This grouping of star particles has been previously used several times in the study of stellar haloes to help understand the kinematics and chemical abundances distribution of their stars \citep[see e.g.][]{Tissera2012, Tissera2014, Pillepich2015, Monachesi2016, Monachesi2019}. It has also  proven useful to decode the origin of stellar populations in bulges and discs and evaluate how mergers contribute to their formation \citep[][G19]{Guedes2013, Gomez2017, Gargiulo2017, Fragkoudi2020}. 
The catalog of in--situ and ex--situ particles of the galaxies analyzed here was constructed following the considerations in \citet{Rodriguez-Gomez2016}.

For this section, to analyze the in--situ and ex--situ components of the bulge, we isolate the stellar particles in the inner regions of the galaxies that are not part of the disc, using kinematic information (see Sec.\ref{sec:kinematicbulges}). As shown by \citet{Abadi2003}, bulge components selected by a kinematic decomposition show an increase in surface brightness towards the inner regions, similar to a  photometric bulge. Note, however, that it is not possible to unequivocally isolate the particles that constitute a photometric bulge using a kinematic decomposition \citep[see][]{Abadi2003, Du2020}. This is because the properties of photometric bulges are obtained by decomposing the light distribution in the inner galactic region with two overlapping profiles representing the disc and the bulge. Nevertheless, this type of analysis can give important insights about the physical mechanisms responsible for the formation of different kinds of photometric bulges.

\begin{figure}
\includegraphics[scale=0.53]{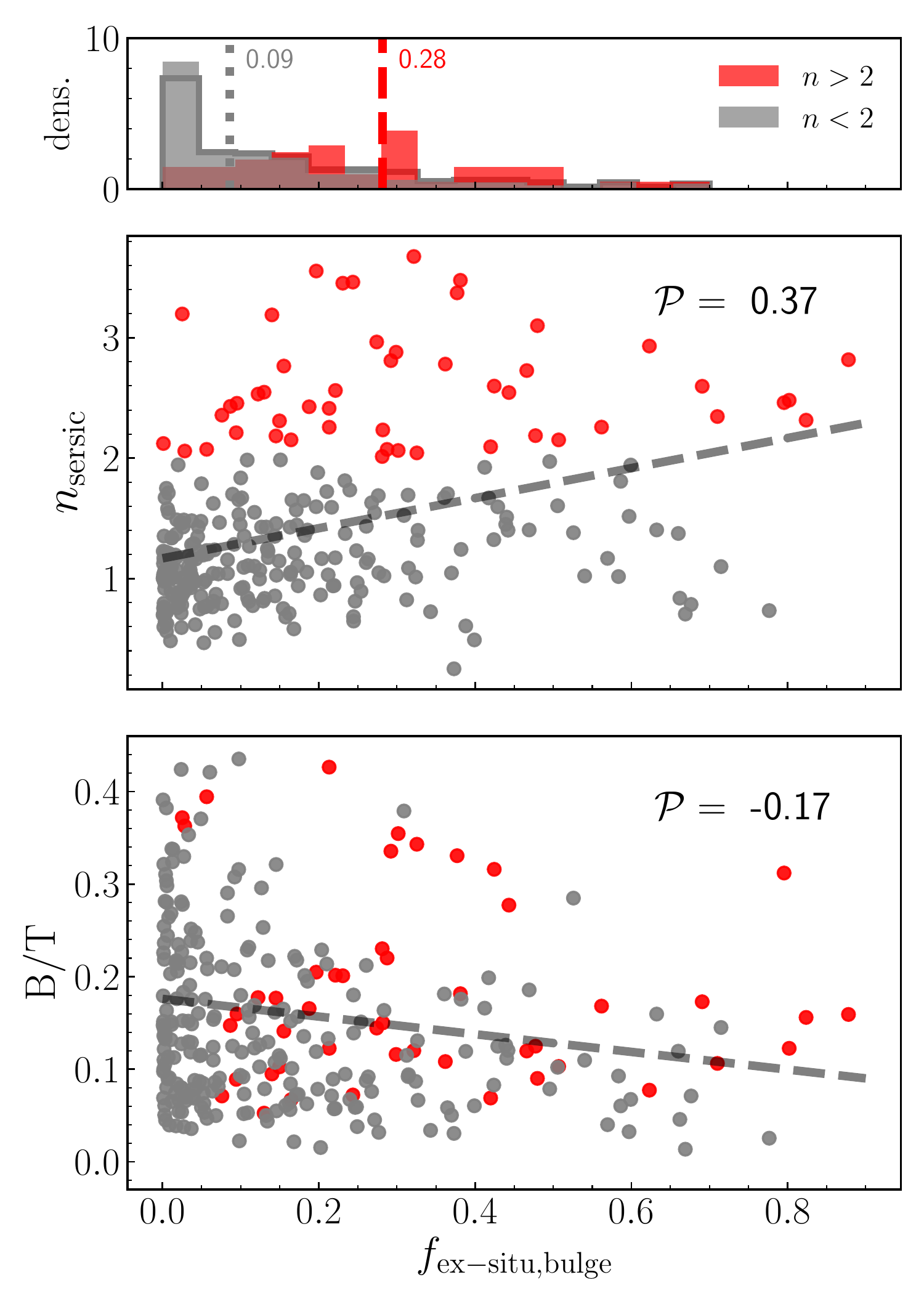}

\caption{{\it Top panel:} Distributions of the ex--situ fractions of stars in kinematic bulges of TNG50 MW/M31-like galaxies  with high- and low-S\'ersic indices in red and grey, respectively. Median values are indicated with grey dotted and red dashed lines. The median of the ex--situ fraction distribution of high-S\'ersic bulges is significantly larger than the one for low-S\'ersic bulges.  
{\it Middle panel:} S\'ersic index as a function of ex--situ fraction of stars in kinematic bulges. Galaxies with high-S\'ersic index are highlighted in red. The dashed grey line is the best-fitted linear function using a least-squared method. The Pearson coefficient of the linear adjustment is shown in the legend. The S\'ersic index and the fraction of ex--situ stars in the kinematic bulges are linearly uncorrelated , according to this metric, in our sample of MW/M31-like galaxies. {\it Bottom panel:} {\rm B/T} as a function of ex--situ bulge fraction. There is a null correlation between the ex--situ fraction of stellar particles in kinematic bulges, and the excess of light in the central regions of our sample of simulated galaxies.}

\label{fig:sers-bot-accreted-fractionsTNG50}
\end{figure}

The top panel of Fig.~\ref{fig:sers-bot-accreted-fractionsTNG50} shows the kinematic bulge ex--situ mass fraction for galaxies with high ($n \geq 2$) and low ($n <2$) S\'ersic index bulges, in red and gray colors, respectively. We can see that high-S\'ersic index bulges typically have larger fractions of ex--situ stars in their kinematic counterpart than low-S\'ersic bulges. This is shown by the median values of the distributions, $\tilde{\mu}= 0.09$ and $\tilde{\mu}= 0.28$ for low and high S\'ersic bulges, respectively, which are highlighted with vertical dashed and dotted lines. However, there is no significant linear correlation between S\'ersic index and $f_{\rm ex-situ, bulge}$. This is shown in the middle panel of this figure, and quantified by the Pearson coefficient. We note here that the accumulation of galaxies with low ex--situ fractions and low S\'ersic index is not dominated by the numerous low mass galaxies in our sample.   In the bottom panel of this figure we show the {\rm B/T} values as a function of the fraction of ex--situ stars ($f_{\rm ex-situ, bulge}$) within the kinematic bulges and find that there is, as well, no correlation. 
The lack of a correlation of the {\rm B/T} ratio with
the ex--situ fraction is evident, with a Pearson coefficient value of $-0.18$. 
These results indicate that the prominence and concentration of bulges are not
trivially related to mergers. Although, it is important to emphasize here that the ex--situ
fraction is not an exact proxy of bulge growth via mergers, since all the stars formed
during a starburst triggered by a merger would be considered here as originating in--situ. The
role of mergers in the formation of different kinds of photometric bulges will be the
subject of Section~\ref{sec:roleofmergers}. 
Despite the lack of linear correlation between the ex--situ fraction of stars in the kinematic bulges and the structural parameters of the photometric bulges, an interesting result is that the majority of the kinematic bulges from TNG50 are formed in--situ, given that the ex--situ fractions are low for most cases.  This is in qualitative agreement with the results found in G19 using the Auriga simulations, which motivates us to compare our results in a more quantitative fashion. 

\subsection{Comparison with the Auriga simulations}

 G19 studied the fraction of ex--situ particles in the central regions of the Auriga simulations and found that the stellar particles of kinematic bulges were formed mostly in--situ.  The Auriga simulations are a suite of high resolution re-simulations of galaxies in MW-sized haloes \citep{Grand2017}, run with the  same magneto-hydrodynamic code as the TNG50 simulation, AREPO. However, the physical models used in the simulations differ in the implementation of the AGN feedback and other aspects (see Sec.~\ref{sec:sims}). Moroever, TNG50 was run within a moderately large cosmological box. As a result, we have a MW/M31-like simulated galaxies sample that is $\sim 7$ times bigger than the one used in G19. It is thus worthwile to compare the results of this work with those found in G19 using the Auriga simulations. We now wish to test in a more quantitative way whether the dominance of in--situ bulges mentioned in the last section is robust and sensible to changes in the physical model of the TNG50 simulation and to the much larger and more statistically representative sample of galaxies.
 
\begin{figure}
\includegraphics[scale=0.53]{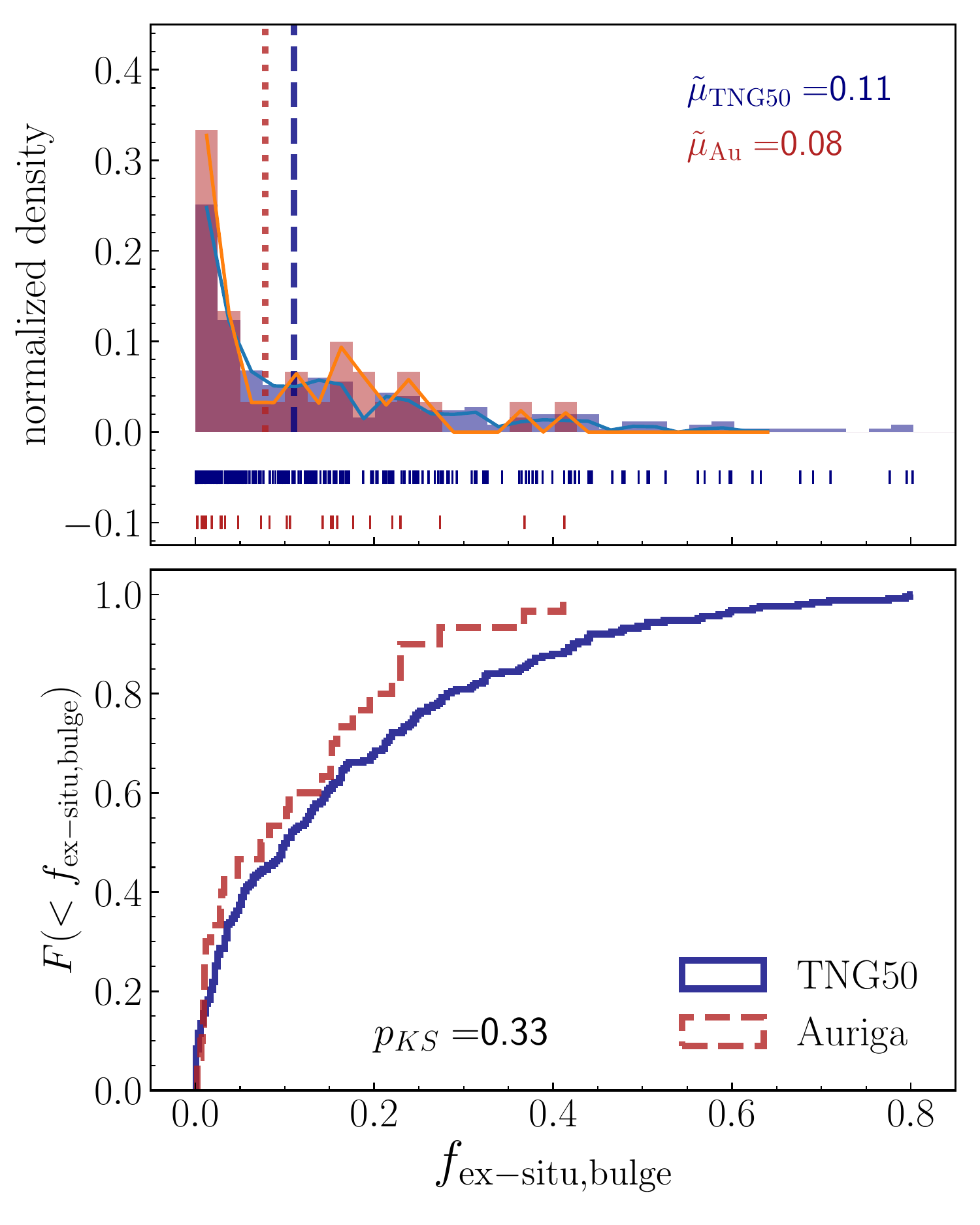}

\caption{{\it Top panel:} Normalized density of ex--situ fractions in bulges of a sample of 30 MW--mass galaxies of the Auriga simulations (red), compared with ex--situ fractions of a subsample of TNG50 MW/M31-like galaxies (blue), selected to be in the same mass range.  Solid lines are kernel density estimations of the distribution for both simulations. The red dotted and blue dashed line indicate the median values as indicated in the legend. Blue and red short lines in the bottom of the figure show the individual values in the distribution. {\it Bottom panel:} Cumulative distributions of the ex--situ bulge fractions for both simulations. The p-value of a two-sample Kolmogorov-Smirnov statistical test is also 
shown. Both simulations show a similar distribution of  accreted stellar fraction in kinematic bulges, but statistically differ due to differences in the physical model.}
\label{fig:accreted-fractions-Au-TNG50}
\end{figure}

The top panel of Fig.~\ref{fig:accreted-fractions-Au-TNG50} shows the 
  density distributions of ex--situ stellar fraction inside bulges of MW/M31-like galaxies in TNG50 and the MW-mass galaxies from the Auriga simulations considered in G19. We take into account only 251 galaxies from our sample of TNG50 simulated galaxies that lie in the same stellar mass range as the Auriga sample, $M_{\star}~\epsilon~[2.75, 10.97] \times 10^{10} {\rm M_{\odot}}$.  
  The kernel density estimations are shown with red and blue lines, for the Auriga and TNG50 simulated galaxies, respectively. The ex--situ fractions distributions are similar, with a relative excess of bulges with $f_{\rm ex-situ,bulge} < 0.1$ in the Auriga simulations and the presence of galaxies with very high ($f_{\rm ex-situ,bulge} > 0.4$) ex--situ bulge fractions only in the TNG50 sample. To quantify  the  differences between the distributions we show, in the bottom panel, the cumulative
 distributions of ex--situ fractions for both samples of simulated galaxies. A two-sample Kolmogorov--Smirnov statistical test yields a p-value of $p_{\rm KS} = 0.33$, not high enough to discard the null hypothesis, i.e, that both samples
 come from different distributions.  Our results show that the prevalence of in--situ formed stars in the modelled kinematic bulges, first reported in G19, is confirmed by TNG50. However, statistically, the distributions differ slighly. The low number of galaxies in the Auriga sample relative to the TNG50 sample may play a role, as well as the different median masses of both samples, which are $6.07\times10^{10} {\rm M_{\odot}}$ and $5.09\times10^{10} {\rm M_{\odot}}$ for the Auriga and the restricted TNG50 sample, respectively . But, given that the mass range of galaxies was constrained to be the same in both samples, it is fair to assume that the differences in the distributions come, at least partially, from the differences in the galaxy formation models. The AGN model implemented in TNG50 (see Sec.~\ref{sec:sims}) might prevent to some degree the formation of stars in the central regions \citep[see also][]{Nelson2021}, and elevate the fraction of ex--situ stars with respect to the Auriga simulations. 
 Another noteworthy fact is that in the Auriga sample G19 did not find any high-S\'ersic bulge, differently from the restricted TNG50 sample, where $17.5\%$ of the galaxies host a high-S\'ersic bulge. The low number of galaxies in the Auriga sample, with respect to the restricted TNG50 sample, prevents us to draw further conclusions on this fact.

\begin{figure}
\includegraphics[scale=0.53]{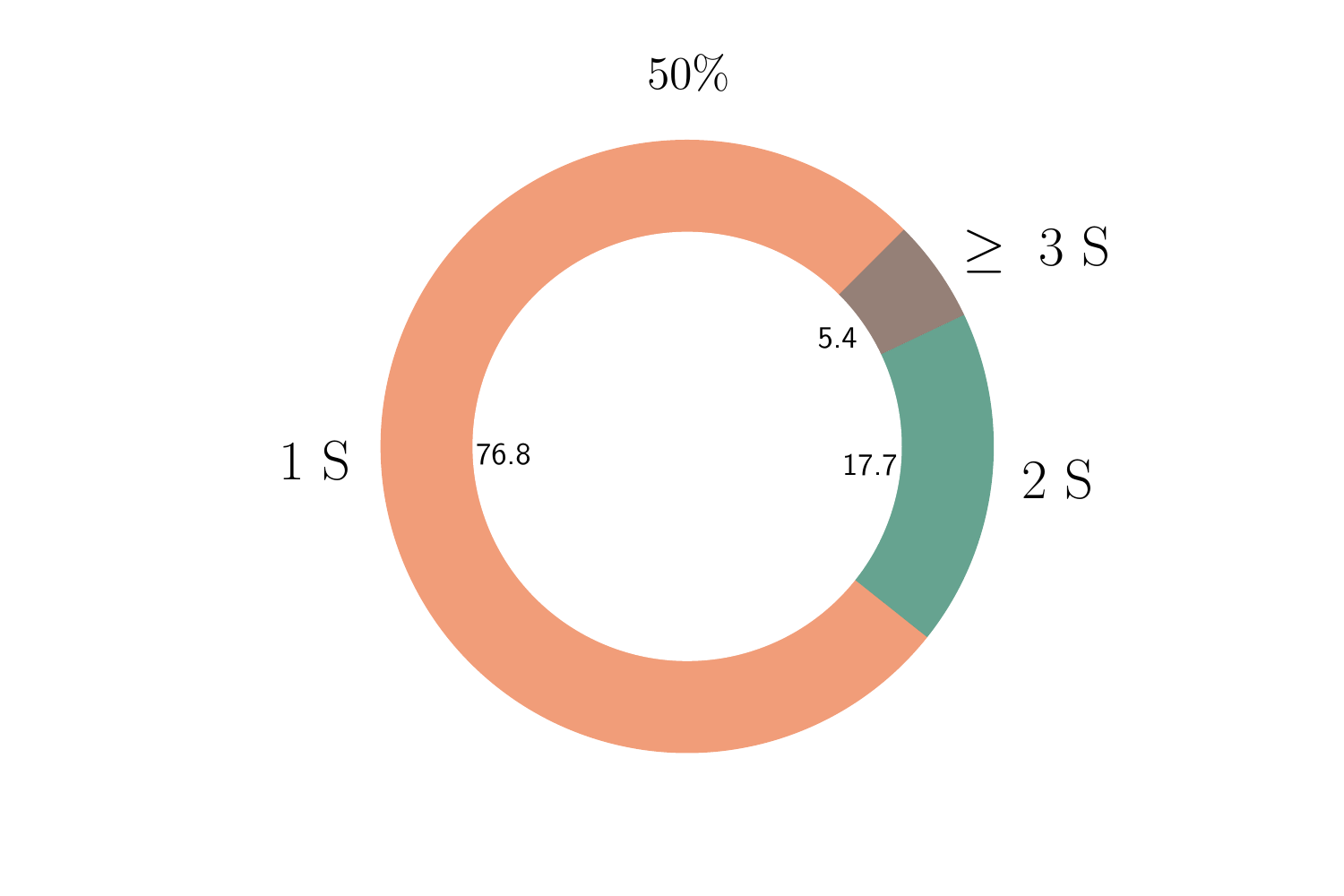}
\includegraphics[scale=0.53]{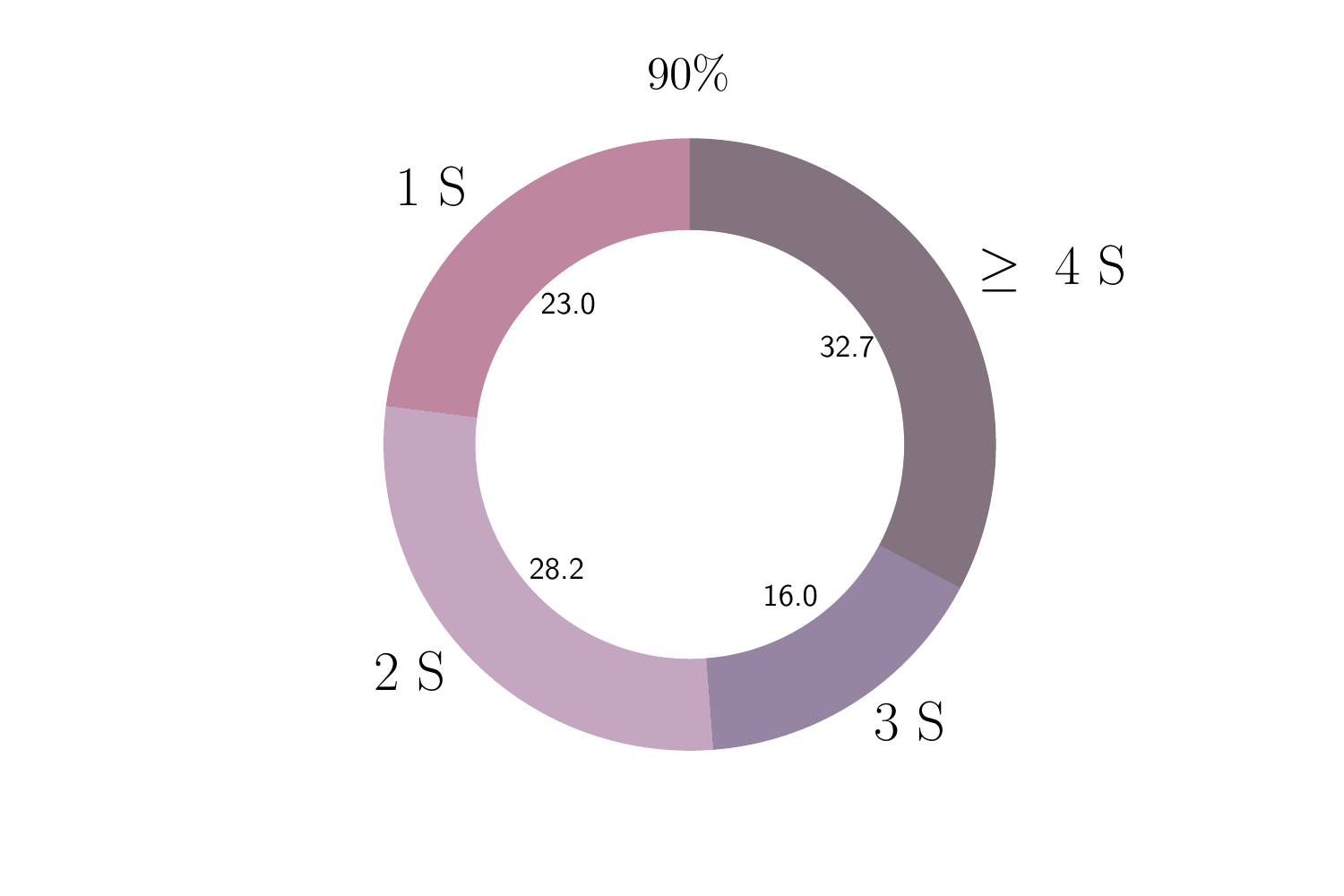}

\caption{Fractions of simulated TNG50 MW/M31-like galaxies in our sample that need the indicated number of satellites to build-up $50\%$ (top chart) and $90\%$ (bottom chart) of the total ex--situ stellar mass of kinematically selected bulges. A single satellite is enough to explain the majority of the accreted mass budget in bulges. Only a few of them are enough to sum almost the total of the accreted stars. }

\label{fig:90-of-acc}
\end{figure}

 An interesting aspect regarding the ex--situ component of the kinematic bulges, and also studied by G19, relates to the number of  satellites that contributed to the majority of their ex--situ stellar component. Fig.~\ref{fig:90-of-acc} shows donut charts displaying the number of satellites needed to add-up $50\%$ and $90\%$ of the total ex--situ stellar mass in the kinematically selected bulges of our TNG50 MW/M31-like sample.  We find that $\sim 76.8\%$ of the galaxies have bulges in which half of the ex--situ component was contributed by a single accretion event. For the remaining, we find that in $17.7\%$ and $5.4\%$ of the galaxies half of the ex--situ component was brought to the bulge by two and three or more satellites, respectively. To build up $90\%$ of the ex--situ mass of the kinematic bulge we find that less than four satellites are sufficient in $67.2\%$ of the cases, with a median value of two satellites. These results are in line with those obtained in G19 with the Auriga simulations, where only a few satellites (a median of 3 satellites) account for $90\%$ of the ex--situ component of kinematically selected bulges and, in most cases (with a median value of one satellite), the stellar particles of one single satellite dominate the
ex--situ component (i.e., this satellite accounts for more than $50\%$ of the total ex--situ stellar
particles in the kinematic bulge).
 
\section{Role of mergers}
\label{sec:roleofmergers}

\begin{figure}
\includegraphics[scale=0.42]{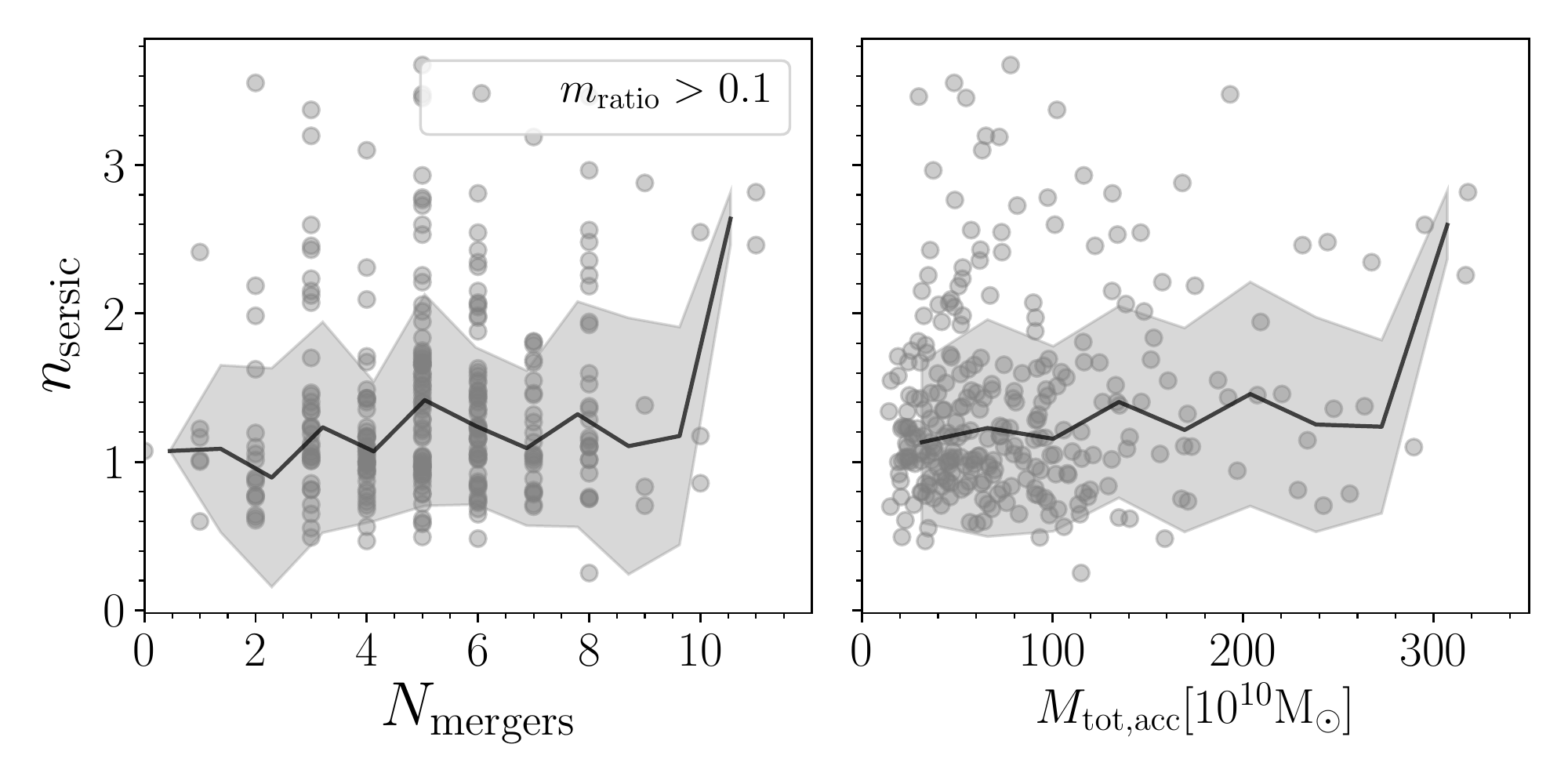}

\caption{{\it Left panel:} S\'ersic index as a function of the number of mergers experienced by each TNG50 MW/M31-like galaxy with merger ratio $m_{\rm ratio} > 0.1$. {\it Right panel:} S\'ersic index as a function of the total mass accreted in mergers with merger ratio $m_{\rm ratio} > 0.1$. The black line represents the median value per bin and the shaded region the interquartile range. The number of mergers and total accreted mass in mergers in galaxies have no measurable effect in the S\'ersic index of bulges }

\label{fig:nummerger}
\end{figure}

In this section we further explore the relation between bulge properties and merger events. In the left panel of Fig.~\ref{fig:nummerger} we show the S\'ersic index as a function of the total number of significant mergers, $N_{\rm mergers}$, experienced by our TNG50 sample of MW/M31-like galaxies. Here we only consider mergers with mass ratios $m_{\rm ratio} = m_{\rm tot,sat}/m_{\rm tot,host} > 0.1$, since redshift $z=12$ up to present day, where  $m_{\rm tot,sat}$ and $m_{\rm tot,host}$ are the total mass of the merging satellite and the host galaxy, respectively. The black line represents the median value per bin and the shaded region the interquartile range. 
Note that the  $N_{\rm mergers}$ does not account for the diversity in the merging history of galaxies, since mergers with different mass ratios, and at different times, are counted in this variable. In the right panel of Fig.~\ref{fig:nummerger} we now show the total accreted mass (i.e. DM, gas and stars) from those significant mergers (i.e. $m_{\rm ratio}> 0.1$)  by each MW/M31-like galaxy as a function of S\'ersic index.  Although the median values in both diagrams show an increase towards larger number of mergers and total accreted mass, the number of simulated galaxies in the last bins is too low, and most galaxies with high S\'ersic index show a moderate number of significant mergers ($3-8$) and total accreted mass ($20-150 \times 10^{10} {\rm M_{\odot}}$). All in all, we find no significant correlation between merging history measured in this way and S\'ersic index. 

The timing of mergers may play a role in shaping the surface brightness profile of a galaxy. A late merger is more likely to induce recognizable perturbations on the host's present-day stellar kinematics than an early event since secular processes have less time to act and reconfigure the galactic phase-space distribution. Moreover, physical conditions, such as the availability of  gas supply to form stars in a burst, are different at different stages of a galaxy's evolution. As a result, although the build-up of the light profile of galaxies is a complex and cumulative process, either the last significant accretion event, or the most massive merger a galaxy has experienced, are more likely to leave a visible imprint in the resulting light profile at $z=0$ than other events.

\begin{figure}
\includegraphics[scale=0.49]{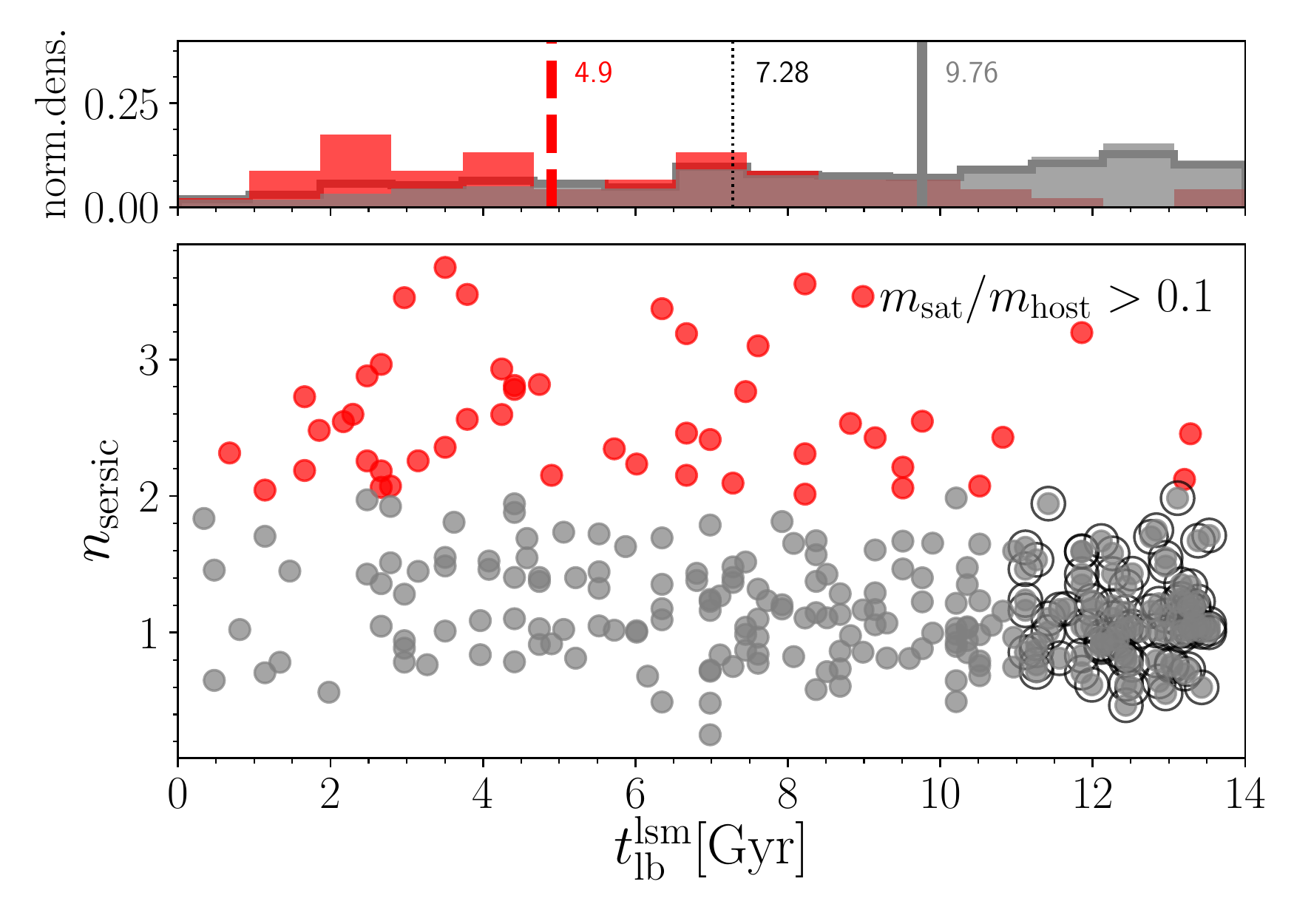}

\caption{{\it Top panel:} Distribution of  lookback time of the last merger with $m_{\rm tot,sat}/m_{\rm tot,host} > 0.1 $, $t_{\rm lb}^{\rm lsm}$, for galaxies with high-S\'ersic and low-S\'ersic photometric bulges, in red and grey, respectively. Median values of the distributions are shown alongside the filled grey line and a dashed red line. A dotted black line indicates the median value of the $t_{\rm lb}^{\rm lsm}$ distribution of simulated galaxies hosting low-S\'ersic bulges when extremely early last mergers ($t_{\rm lb}^{\rm lsm}$ > 11 Gyrs) are not considered. {\it Bottom panel:} S\'ersic index as a function of $t_{\rm lb}^{\rm lsm}$. Grey indicates simulated galaxies with low-S\'ersic photometric bulges and red indicates those with high-S\'ersic bulges. Highlighted with black circles are galaxies with low-S\'ersic bulges that experienced the last significant merger more than 11 Gyrs ago. Galaxies hosting a high-S\'ersic bulge experience the last significant merger at substantial later times, on average.}

\label{fig:lastconsmer}
\end{figure}

We will now focus on the characteristics of the last significant merger experienced by each galaxy. The bottom panel of Fig.~\ref{fig:lastconsmer} shows the S\'ersic index of the photometric bulges as a function of look-back time ($t_{\rm lb}$) of the last significant merger,  $t_{\rm lb}^{\rm lsm}$ (i.e., with merger ratio $m_{\rm sat}/m_{\rm host} > 0.1$).
Here, the time of the merger is defined as the look-back time of the snapshot at which the satellite (or secondary progenitor), is no longer identified by the {\rm subfind} algorithm. Our goal is to better identify the instant of time when the merger could have a stronger effect on the surface brightness profile of the host. The mass ratio, however, is measured at the time when the satellite reaches its maximum stellar mass, because a substantial amount of loose particles that belonged to the satellite are assigned to the central galaxy by the halo finder, which results in a significant underestimation of the merger ratio. This, in turn, adds a resolution dependent effect, since in simulations with better resolution a satellite takes longer to merge \citep[see Sec.~5.2 in][]{Rodriguez-Gomez2015}. 

On average, galaxies hosting high-S\'ersic photometric bulges experienced the last significant merger at later times than galaxies with low-S\'ersic bulges. The distribution of $t_{\rm lb}^{\rm lsm}$ is shown in the top panel. The low- and high-S\'ersic index population have median $\tilde{t}_{\rm lb}^{\rm lsm} = 9.764$ and $\tilde{t}_{\rm lb}^{\rm lsm} = 4.9 $ Gyr  and are indicated with a filled grey line and red dashed line, respectively. Note, however, that a significant group of simulated galaxies with low-S\'ersic bulges, highlighted in Fig.~\ref{fig:lastconsmer} with black circles,  did not experience a merger with $m_{\rm tot,sat}/m_{\rm tot,host} > 0.1 $ during the last 11 Gyr. These galaxies have undergone a very quiet merger history over their entire history. If we do not consider this group of galaxies, the low-S\'ersic distribution shows a median value of $t_{\rm lb}^{\rm lsm} = 7.28$. This result indicates that there is a link between the occurrence of a late merger and the concentration of the surface brightness profile.  Although there is no linear correlation between the S\'ersic index of photometric bulges and the time of the last significant merger suffered by the simulated MW/M31-like galaxies, the median value of $t_{\rm lb}^{\rm lsm}$  is higher for galaxies with high S\'ersic index bulges. A similar analysis taking into account the most massive merger of each simulated galaxy showed no correlation with this particular event and the S\'ersic index of photometric bulges, with similar distributions of $t_{\rm lb}$ of the most massive merger for low- and high-S\'ersic bulges.   

\begin{figure}
\includegraphics[scale=0.45]{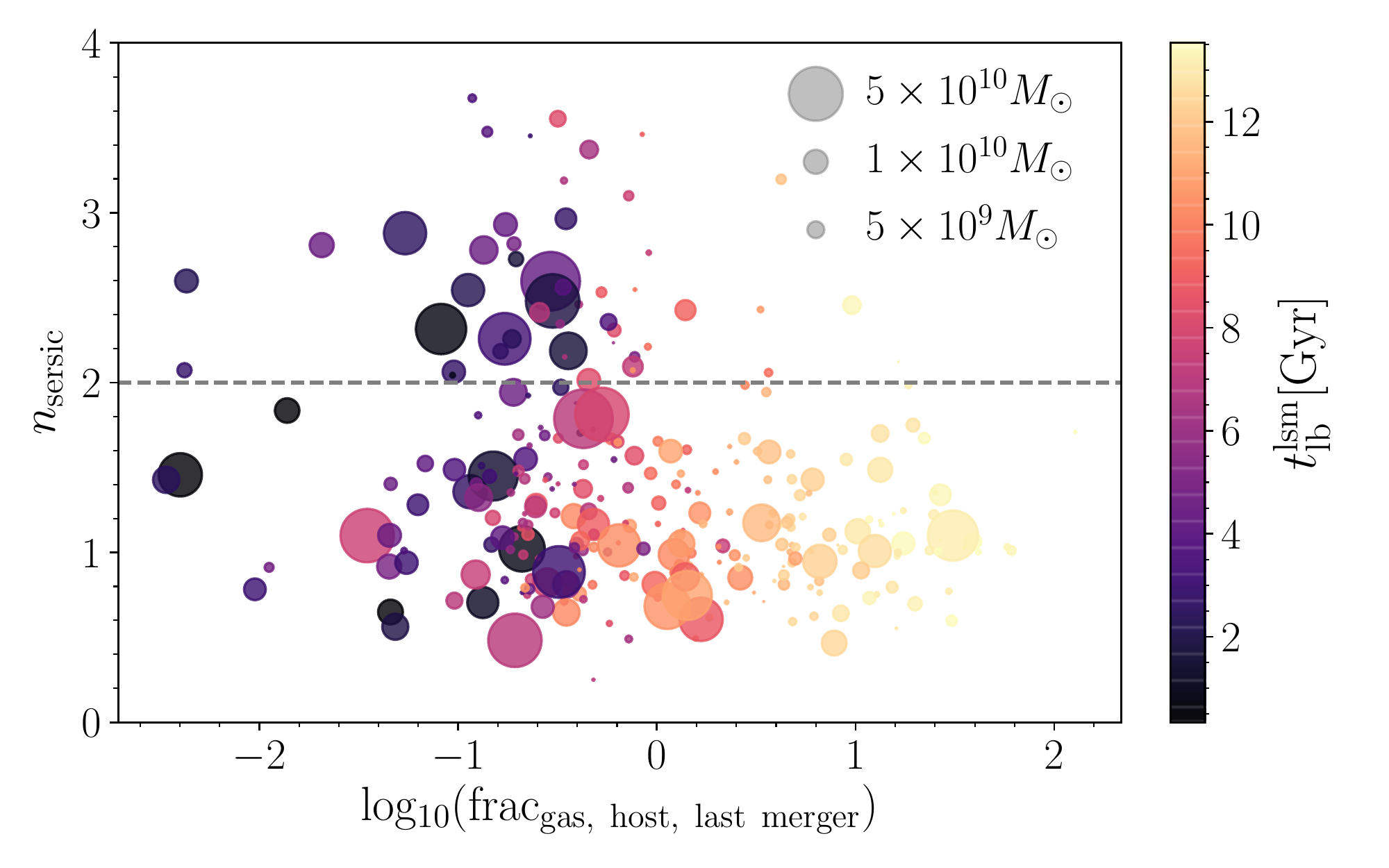}

\caption{S\'ersic index as a function of the logarithm of the host gas fraction at the time of the last significant merger ($m_{\rm tot,sat}/m_{\rm tot,host} > 0.1 $). The color coding indicates the lookback time of the last merger and marker sizes are indicative of the stellar mass of the accreted satellite in the last merger. Gas--poor mergers in the late Universe are not exclusive of galaxies with high-S\'ersic bulges. Early gas-rich mergers are exclusive of galaxies with low S\'ersic bulges.}

\label{fig:sersic-fraccold-lbt}
\end{figure}

Another noteworthy property of galaxies that may affect the outcome
of mergers is the gas fraction at the moment of the interaction.  We define the gas fraction, $f_{\rm gas}$, as the ratio between the mass in cold gas and the stellar mass of the host galaxy at the time of peak mass of the satellite (the same instant of time that we use to define the merger mass ratio). To measure the amount of cold gas we sum the mass of gas in cells defined as {\it star forming} in the simulation, i.e., the gas cells that are above the star formation threshold for the \citet{SpringelHernquist2003} ISM model. We show in Fig.~\ref{fig:sersic-fraccold-lbt} the S\'ersic index as a function of the logarithm of the gas fraction, at the time of the last considerable merger ($m_{\rm tot,sat}/m_{\rm tot,host} > 0.1$). The color-coding indicates the $t_{\rm lb}^{\rm lsm}$ of this last merger and the size of the circle is proportional to the stellar mass of the accreted satellite. The gas fraction clearly correlates with the $t_{\rm lb}^{\rm lsm}$, as is expected because of the exhaustion of the gas supply of galaxies with time. We can see that a group of galaxies with low S\'ersic index with no significant mergers through most of their history accreted low mass satellites at early times in a very wet merger, that is, with a high fraction of cold gas available, larger than unity. High S\'ersic bulges had predominantly mergers with gas fraction between unity and a tenth, but there is no clear correlation between the gas fraction and the concentration of bulges. Many low-S\'ersic bulges show the same pattern as high-S\'ersic bulges.  Looking at the mass of the satellite galaxy reflected in the sizes of circles we can tell that there is also no correlation with the behaviour of galaxies with low and high-S\'ersic bulges, except for the fact that there is a large group of low S\'ersic galaxies that do show a last merger with low mass satellites, with high gas fractions that occurred at early times.      

The lack of clear correlations with the properties of mergers and the S\'ersic index of bulges does not necessarily mean that mergers have a low impact on the formation of bulges. The fact that the ex--situ fraction in kinematic bulges is higher more frequently in high S\'ersic bulges (see Fig.~\ref{fig:sers-bot-accreted-fractionsTNG50}) is the first indication that mergers play a role.
Mergers are ubiquitous in the current paradigm of galaxy formation and the fact that some large galaxies in the local group and the MW galaxy itself show no signatures of a merger-built bulge component (although see \citealt{Kunder2016}) has led to the idea that there is a tension of the paradigm with observations \citep[e.g.][]{Shen2010}. In G19 it was found that despite the rich merger histories of simulated MW--mass galaxies, kinematically selected bulges were formed mostly {\it in--situ} and show properties more akin to pseudo-bulges.  Here we see that on one side, there are MW/M31-like simulated galaxies that do not suffer a significant merger since the very early epoch of their formation. On the other side, mergers evidently do not always reach and affect the central region of the galaxy. In the following Section we explore other relevant physical mechanisms that could play a significant role in the formation and evolution of bulges, modulating the effects associated with merger events.

\section{Influence of bars}
\label{sec:barinfluence}

In this section, we analyze the influence of bars on the shape of the bulge surface
brightness profile, characterized by the S\'ersic index and {\rm B/T}.  

\subsection{Bar demographics}
\label{sec:bardemo}

\begin{figure}
\includegraphics[scale=0.52]{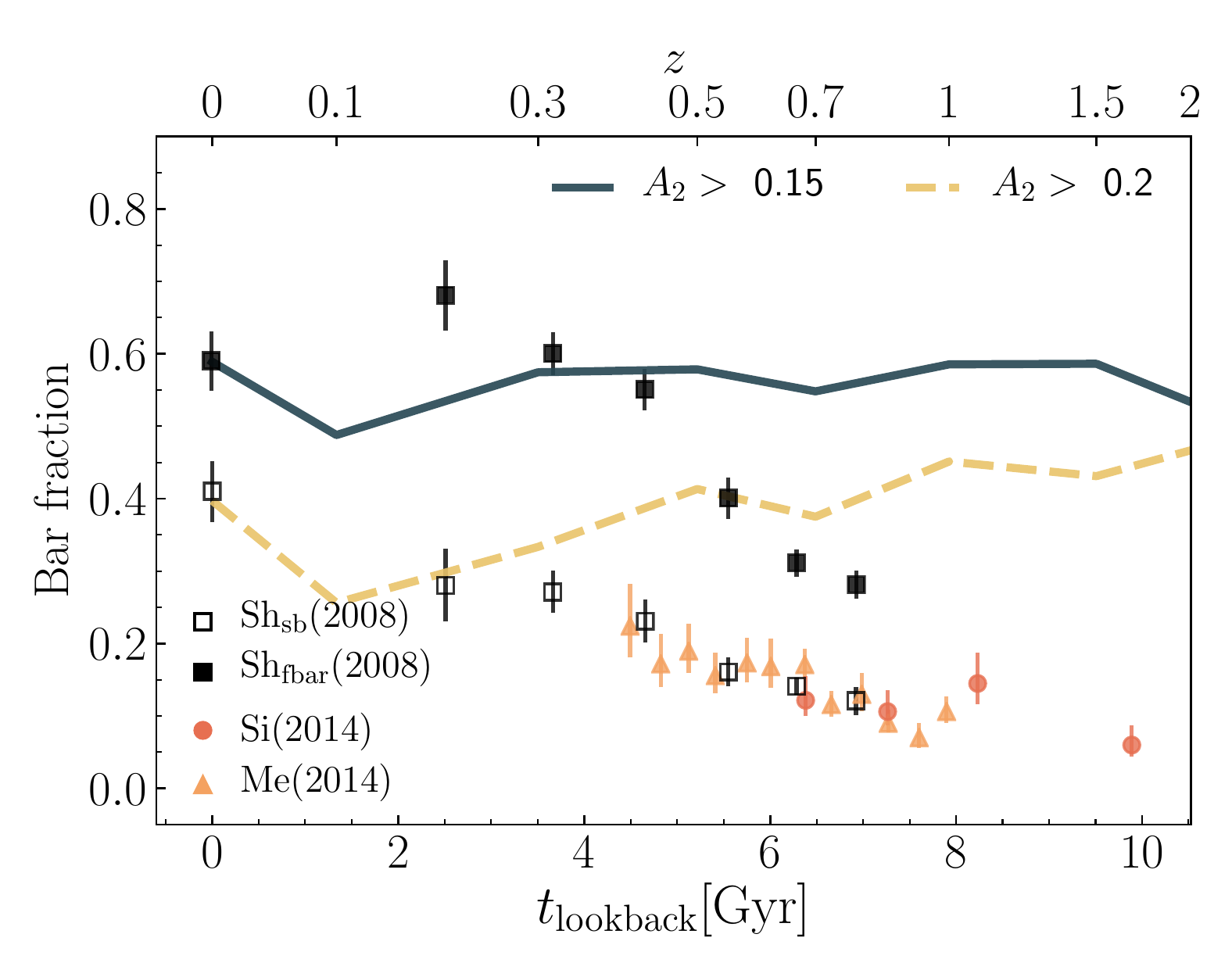}

\caption{Bar fraction as a function of lookback time. Bar fractions
  computed for the simulated sample of TNG50 MW/M31-like galaxies are subject to a minimum V-band luminosity threshold (see text) at $z=0$, together with the direct progenitor of each at $z>0$. Two different
  thresholds of bar strengths to classify a galaxy as barred are shown, as indicated in the legends.
  Observational estimates of bar fractions from \citet{Sheth2008} are shown with filled black squares and empty black squares. Empty squares correspond to strong bars. Dark Orange circles are from \citet{Simmons2014} and orange triangles are from \citet{Melvin2014}. Bar fractions in the progenitors of our sample of MW/M31-like galaxies do not decrease with redshift. No direct comparison of observations and simulations is suggested, as the galaxy samples selected -- as well as the methods for identifying bars -- differ between the two.}

\label{fig:barfrac-sheth}
\end{figure}

We first check the evolution of the bar fraction in our sample of MW/M31-like galaxies.
Fig.~\ref{fig:barfrac-sheth} shows the fraction of barred galaxies in our TNG50 sample as a function of $t_{\rm lb}$, together with different observational estimates of the bar fraction in $\sim L_{\star}$ galaxies as a function of $t_{\rm lb}$. \citet{Sheth2008} measured bar fractions in the COSMOS survey \citep{Scoville2007}. The existence of a bar is determined by the ellipse fitting method \citep{MenendezDelmestre2007}, which consists in following the excess in the ellipticity profile up to a sudden change in the position angle profile of the fitted ellipses. They report two different estimates of bar fraction based on the ellipticity of the isophotes. One is simply referred to as the bar fraction ($f_{\rm bar}$), and takes into account all bar detections. The other is estimated considering only bars with ellipticities greater than 0.4, and named the strong bar fraction (sb). The total bar fraction is shown with black filled squares, and the strong bar fraction is shown with empty black squares. Orange triangles show the bar fraction estimates of galaxies in the COSMOS survey, but based on the visual classification of the Galaxy Zoo project from \citet{Melvin2014}. Dark orange circles shows the observational data from \citet{Simmons2014} who estimated the fraction of galaxies with strong bar features in the Galaxy Zoo \citep{Lintott2008} using galaxies from the CANDELS survey \citep{Koekemoer2011} up to $z=2$. 
Observational estimates rely on the selection of galaxies at different redshifts that are brighter than the  ${\rm M_V^{\star}}$ region of the rest-frame luminosity function at higher redshifts. The ${\rm M_V^{\star }}$ parameter as a function of redshift was estimated in \citet{Marchesini2012}, by fitting a Schechter function to the rest-frame luminosity functions in the V band at redshifts $z = [0.55, 0.90, 1.30 ,1.8, 2.4 , 3., 3.64]$. Later on, a parametrized curve is fitted to the observed 
${\rm M_V^{\star }}$ values as a function of redshift. We follow the progenitors of simulated galaxies in our sample at $z=0$ using the merger trees from the simulation. By means of the parametrized curve fitted to the observed 
${\rm M_V^{\star }}$ we impose this magnitude threshold to our sample of galaxies at each redshift.  
Observations also establish different selection criteria of strong barred galaxies than simulations.  To take this into account, we consider two different thresholds in the amplitude of the second Fourier mode $A_{2}$ to consider a galaxy as barred. 
One of the caveats of the comparison is that we quantify bar strength and length based on the stellar mass distribution, whereas observationally this depends of the photometric band used for the analysis. Moreover, the definition of barred galaxies in an observational sample is based on visual inspection or the ellipse fitting method.
Overall, the methods to estimate bar fractions of galaxies in observations and simulations and the sample selection differ considerably, and both present limitations, so a direct comparison between observed estimates of bar fractions and those obtained in our simulated galaxies is not intended.
However, we find it useful to show the observational estimates to guide the reader in the following argument. The observed fractions of barred galaxies show a decrease after $z\sim 0.4$, in all cases. The bar fraction of the progenitors of galaxies in our sample do not decrease toward higher redshifts.  This apparent excess of barred galaxies in our sample at high redshifts might contribute to the excess of low-S\'ersic bulges that we see in our sample of MW/M31-like galaxies compared with observations, as we later discuss in Sec.~\ref{sec:discussion}. Nevertheless, we caution once again that a better and more quantitatively meaningful comparison between simulations and observations is required to clearly assess the degree of this apparent discrepancy. For example \citet{Rosas-Guevara2021} study the evolution of the bar fraction of galaxies in TNG50 and faced the same limitations. When considering only bars larger than 2 {\rm kpc} in TNG50, which is a typical resolution limit in images used to measure bars in observations at high redshift, the bar fractions show a decrease with redshift that is qualitatively similar to what observers find.  

\begin{figure*}
\includegraphics[scale=0.39]{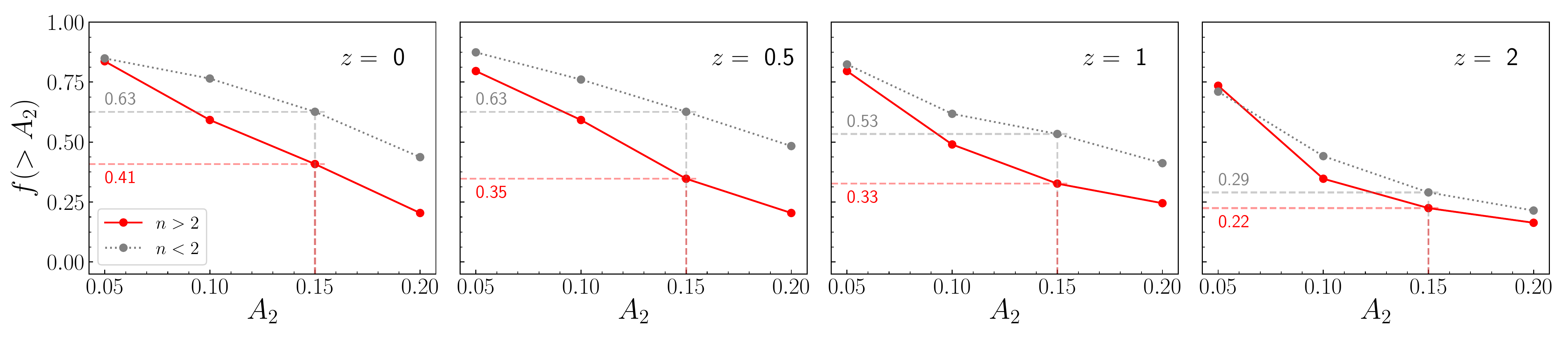}
\includegraphics[scale=0.39]{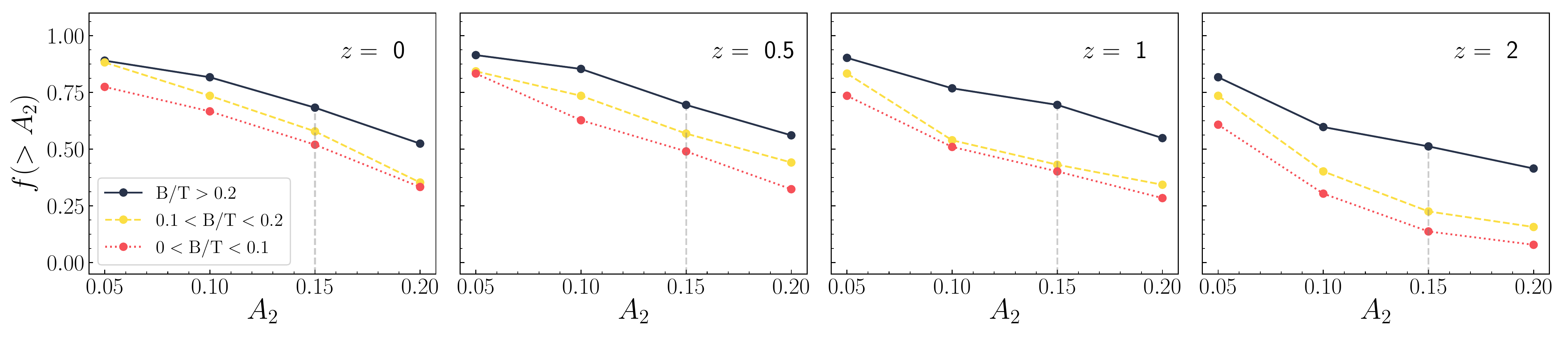}
\caption{{\it Top panels:} Cumulative fraction of bar strengths of TNG50 MW/M31-like simulated galaxies with high- and low-S\'ersic indices in their bulges at z=0 are shown as red solid lines and dotted grey lines, respectively, at four different redshifts, as indicated in the keys. An excess of bars in the progenitors of galaxies with low-S\'ersic bulges at $z=0$ is present at all redshifts since $z=2$.
{\it Bottom panels:} Cumulative fraction of bar strengths in TNG50 MW/M31-like simulated galaxies in three intervals of {\rm B/T} as indicated in the keys. Progenitors of galaxies with larger {\rm B/T} at $z=0$ show a higher bar fraction at all cosmic epochs, highlighting the influence of bars in the formation of photometric bulges.}
\label{fig:barfrac-z}
\end{figure*}

\subsection{Bar evolution}
\label{sec:barevo}

We now concentrate on the evolution of the bar strengths to investigate the effect that bars may have on the morphology and formation of photometric bulges. The top panels of Fig.~\ref{fig:barfrac-z} show the cumulative fraction of galaxies with bars stronger than a given $A_2$ value. Galaxies with high- and low-S\'ersic bulges are shown with red and grey lines, respectively. Different panels show results at  different redshifts. As shown in the top leftmost panel, at $z=0$ there is a clear excess of bars with $A_2 > 0.1$ in simulated galaxies with low S\'ersic index. At $z = 0 $, $\ 63\%$ of simulated galaxies with low-S\'ersic index have bars stronger than $A_2 = 0.15$ and only $\ 41\%$ of those with high-S\'ersic index show bars with that magnitude. As we move to higher redshifts, the relative difference in the bar fraction regime (i.e. $A_2 > 0.15$) remains large, reaching a maximum difference of $0.28$  at $z=0.5$. At $z=2$, bar fractions drop significantly for both samples. This indeed suggests a correlation between bulge prominence and the prevalence of bars. In the bottom panels of Fig.~\ref{fig:barfrac-z} we now divide the sample of galaxies based on the parameter  {\rm B/T}. The red, yellow and black lines show the cumulative number of galaxies as a function of $A_2$ for galaxies with $0.0 <{\rm B/T} < 0.1$, $0.1 <{\rm B/T} < 0.2$ and ${\rm B/T} > 0.2$, respectively. We can see that across all redshifts, and for all bar strengths considered, galaxies with larger {\rm B/T} values show a greater fraction of bars than less luminous bulges. 

The results found here are
qualitatively in agreement with those shown by \citet{Weinzirl2009}, who found
that $\ 65\%$ of their sample of bright spirals with low S\'ersic index are barred, whilst $\ 38\%$ of them with high S\'ersic index ($2 < n < 4$) have bars. However, a one-to-one comparison is not possible for two main reasons. First, their galaxy sample spans a wider range in mass and, second, they include a bar component in their 2-dimensional surface brightness decomposition.  
The addition of a bar component is probably the main reason for a disagreement in the trend found in our results of bar fraction for different ranges of {\rm B/T}; \citet{Weinzirl2009} found that galaxies with lower ${\rm B/T}$ are most
likely barred. Galaxies with ${\rm B/T} < 0.2$ have a high bar fraction of $\sim 68\%$, and those with $0.2 <{\rm B/T} < 0.4$ and ${\rm B/T} > 0.4$ have bar fractions of $\sim 42\%$ and $\sim 17\%$, respectively. The trend exhibited by their results is opposite
to what we find in our simulated galaxies, where a bar component in the surface brightness profile decomposition is not included. When a bar is considered as a separated entity, a significant fraction of the light associated with the bulge in a two-component fit would be associated to the bar component. This effect can be readily seen comparing the results of G19, with those of \citet{Blazquez-Calero2020}, who analyzed the same sample of the Auriga simulation suite and applied two-component light decompositions and three-component light decompositions (including a bar component), respectively. The {\rm B/T} ratios derived by G19 are consistently larger than those obtained by \citet{Blazquez-Calero2020}.  The implications of these results are discussed in Sec.~\ref{sec:discussion}.   

\subsection{On the bulge-bar connection}
\label{sec:bulgebarconn}

Bars are not always persistent features. Is it fair to assume that the duration of a bar feature would determine the influence of this component on the evolution of a galaxy and, in particular, its photometric bulge. To quantify the period of time during which a given galaxy contained a significant bar we proceed as follows. First we define a bar strength threshold $A_{\rm thresh}$. We then search for all those snapshots in the simulation where $A_2 > A_{\rm thresh}$. At every snapshot, $S_i$, where $A_2 > A_{\rm thresh}$, and due to the relatively poor time resolution, we assume that the bar has been above $A_{\rm thresh}$ for a period equal to $[t_{\rm lb}(S_{i-1}) - t_{\rm lb}(S_{i+1})]/2$. Finally the total time a bar has had an amplitude above $A_{\rm thresh}$  since $t_{\rm lb} < 10~ {\rm Gyrs}$ is computed as

\begin{equation}
t_{(>A_{\rm thresh})} = \frac{1}{2} \sum_{i_{>A_{\rm thresh}}}~\big([t_{\rm lb}(S_{i-1}) - t_{\rm lb}(S_{i+1})]~\big),
\label{eq:tA2}
\end{equation}

\noindent where $i_{>A_{\rm thresh}}$ counts over the snapshots where the amplitude of the second Fourier mode is larger than $A_{\rm thresh}$.

\begin{figure}
\includegraphics[scale=0.6]{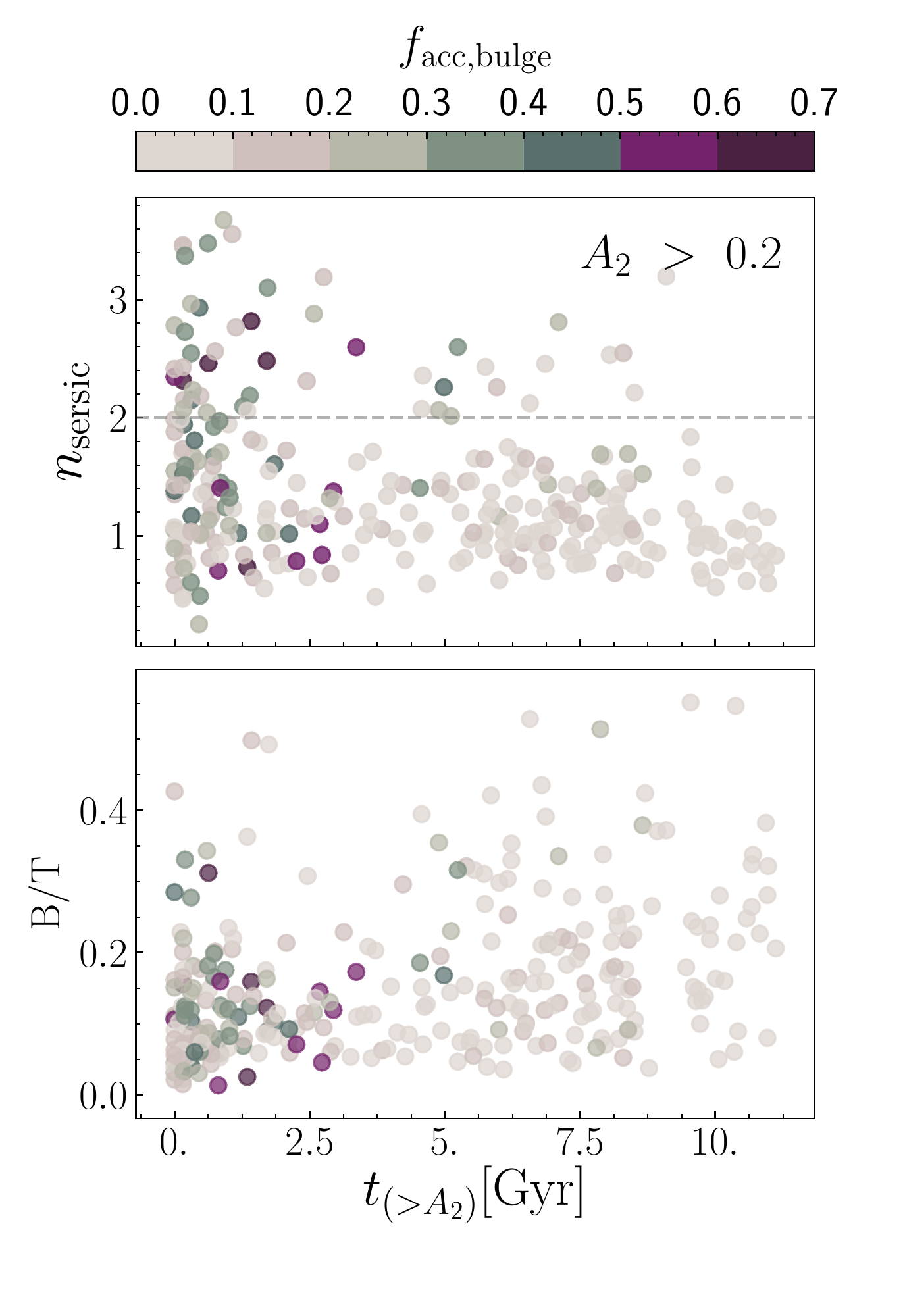}

\caption{{\it Top panel:} S\'ersic index as a function of the duration of the bar, color-coded with the fraction of ex--situ stars in the bulge, for our sample of simulated TNG50 MW/M31-like galaxies.  {\it Bottom panel:} B/T as a function of the duration of the bar with the same color-coding as in the top panel. High-S\'ersic bulges usually  develop in galaxies that do not host a bar for a long time.}

\label{fig:time-over-A2}
\end{figure}

In the top panel of Fig.~\ref{fig:time-over-A2} we show the dependence of S\'ersic index of our simulated galaxies with the period that a system hosted a bar with $A_2 > 0.2$. The color coding indicates the fraction of ex--situ stars in each bulge. We find no correlation between S\'ersic index and the duration of a strong bar (top panel). However, it is interesting to see how different regions in this diagram are populated. Galaxies with high-S\'ersic index bulges typically do not develop strong bars during most of their evolution, as was already hinted in Fig.~\ref{fig:barfrac-z}. There are a dozen galaxies with high-S\'ersic bulges that developed a bar during more than 3 {\rm Gyrs} in their history, some of them with S\'ersic indices close to the threshold imposed to divide them into low and high-S\'ersic bulges. Most of the bulges with long-lived bars ($t_{>A_2} > 3$ {\rm Gyrs}) show either low or negligible ex--situ fractions in the kinematic bulge, as well as low S\'ersic bulges. 
Galaxies with short-lived bars, on the other hand, show both low- and high-S\'ersic index bulges. Among those, the galaxies with high-S\'ersic index bulges commonly show high ex--situ fractions, as already shown in Fig.~\ref{fig:sersic-fraccold-lbt}.

The bottom panel of  Fig.~\ref{fig:time-over-A2} shows the {\rm B/T} ratio as a function of strong bar duration with the same color coding as in the top panel. There is no clear correlation between the photometric {\rm B/T} ratio and the duration of the bar, but two broad main branches can be seen in this panel, departing from the locus of points with low {\rm B/T} and short-lived bars, i.e ${\rm B/T} < 0.1$ and $t_{(>A_2)} < 3~ {\rm Gyrs}$. On one hand, we see galaxies with long-lived strong bars that in some cases show significant values of {\rm B/T}, larger than ${\rm B/T = 0.15}$. These likely grew their photometric bulges through bar-related processes. On the other hand, there are galaxies with short lived bars that also present large values of {\rm B/T}. Some of them show higher ex--situ fractions in their kinematic bulge, thus, in this case, also mergers likely contributed to grow their photometric bulges.

\section{Discussion}
\label{sec:discussion}

We have studied the formation of bulges in a sample of galaxies of MW/M31-like galaxies drawn from TNG50. In this section, we discuss our findings in general and in relation to the current understanding about the role that mergers and bars can play in the formation and evolution of bulges.

By means of one-dimensional, two component photometric decompositions of the surface brightness profiles of our sample of MW/M31-like galaxies we find that $17.1\%$ have S\'ersic indices $n_{\rm sersic} > 2$, i.e., very concentrated bulges components,  consistent with the structure expected for classical bulges.
An interesting question to address would be if this fraction of classical bulges is consistent with the fraction of concentrated photometric bulges estimated from observations. \citet{FisherDrory2011} compiled a set of $\approx 100$ observations of galaxies in a local $11$ {\rm Mpc} volume, obtained with a consistent methodology, and produced a census of bulge type as a function of the stellar mass of galaxies. They found that, for galaxies with stellar masses in the range defined by our sample (${\rm log_{10}}({\rm M_{\star}}) \in [10.5, 11.2]$), $\sim 50\%$ of bulges can be classified as classical, which have S\'ersic index $n > 2$ almost in all cases. This proportion is higher than the $17.1\%$ of concentrated bulges found in the present study.  Note however that the observed sample of galaxies has a low number in the mass range considered in our work, and also include elliptical galaxies in the classical bulge classification; the latter are excluded from our analysis.
Another possibility is that the TNG50 simulation is underproducing high Sérsic index bulges, when compared with MW-mass galaxies in the local Universe. However, it is not possible to draw a definitive conclusion from the analysis presented in this work. A more precise answer to this question must be addressed in a new study, in which the same galaxy selection criteria are applied for both the observed and simulated sample of galaxies.

It has been pointed out in previous works that the overabundance of galaxies with low-S\'ersic bulges can be a resolution issue, via two main channels \citep{Martig2012}: \citet{Bois2010} showed that dynamical resolutions near 32 {\rm pc} are needed to correctly remove angular momentum in mergers and produce slowly rotating ellipticals. Although the TNG simulations are successful in producing slowly rotating ellipticals \citep[e.g.][]{Pulsoni2020}, it is fair to speculate that a higher resolution might still be needed to correctly redistribute angular momentum in the central regions of disc galaxies, thus producing more concentrated bulges supported by anisotropy. Additionally, using zoom-in simulations, \citet{SparreSpringel2016} showed that simulations with higher resolutions than Illustris (similar to TNG100 and hence 16 times worse in particle mass than TNG50) are more efficient to produce starburst in galaxy mergers. With TNG50 we show that bursts of star formation are prominent during the phases of gas-rich mergers of MW/M31-like galaxies (Still et al, in prep), yet it is unclear to what degree this is converged and particularly so for the question of bulge formation.

Second, the lack of accuracy in the follow-up of the fragmentation of gas in the early evolution of the galaxy might suppress a bulge formation channel, through formation of gas clouds that sink into the central parts of a galaxy due to dynamical friction \citep{Dekel2009, Ceverino2010, Perez2013}. This process is thought to contribute mostly to form concentrated bulges. 

The resolution analysis shown in Appendix~\ref{sec:appendix-res} suggests that convergence on  the SBP of these late type simulated galaxies is only starting to be reasonably achieved at the resolution level of TNG50-1. This is in good agreement with the results presented in \citet{Grand2021}, where they studied this problem considering two higher resolution levels, all in the context of the Auriga project. They show that surface density profiles of MW-mass galaxies are starting to be well converged at resolution levels of $5.4 \times 10^4 {\rm M_{\odot}}$ in the baryonic component, which is comparable to the TNG50-1 resolution.  It is worth highlighting, however, that lower resolution levels than that associated with TNG50-1 will suffer from this issue. For example, \citet{Du2020} analyzed the TNG100-1 simulation, with a mass resolution 16 times worst than the one considered here, and found only a few galaxies with concentrated bulges.

Nonetheless, simulations like TNG50-1 studied in this work, with a minimum gravitational softening of $\sim 300$ physical {\rm pc} for the stars, are a major step in understanding the physical processes that contribute to form the diversity of bulges found in MW/M31-like galaxies in the local Universe. \\

Another issue worthy of attention and recently discussed by \citet{Peebles2020} is the relative scarcity of very low ${\rm B/T}$ bulges in state-of-the-art cosmological simulations compared to a local sample of observed galaxies with $L \backsim L_{\star}$. The discussion is based on a comparison between the hot kinematical state of particles in simulated galactic components and the observed values of the rotational degree of the components of nearby galaxies. It was also shown, however, that when taking into account ${\rm B/T_{V}}$ ratios, which are derived from surface brightness profiles of simulated galaxies in the V-band (G19), the resulting ${\rm B/T_{V}}$ distribution in simulated galaxies is comparable to the observed distribution of {\rm B/T}. 
We find that the median of the {\rm B/T} distribution of our simulated galaxies, $\tilde{\mu}_{\rm B/T-tng50} = 0.09$, is even lower than the median value of the observed sample of \citet{Peebles2020}, $\tilde{\mu}_{\rm B/T-obs} = 0.16 $.  However, our selection criteria explicitly select highly triaxial galaxies, which can bias our sample to galaxies with lower photometric bulge luminosities. To provide a quantitatively meaningful comparison with observations, similar selection criteria for both the simulated and observational samples, as well as a larger observational sample would be needed. It is worth noting the lack of pure discs with a nuclear cluster (like, e.g., M101) in our  sample of simulated late type galaxies.
Although the connection between dynamical and surface brightness decompositions is not straightforward \citep[See][]{Abadi2003}, the point highlighted by \citet{Peebles2020} persists: bulgeless galaxies with very cold discs are not yet easy to achieve with the current set-up in state-of-the-art cosmological simulations.  \citet{Peebles2020} proposes a second order deviation from gaussianity to the density fluctuations in the cosmological model to address this problem. However, according to the review by \citet{Lagos2018} and the aforementioned work by \citet{Bois2010}, there is still room to improve the dynamical and mass resolution, as well as the simplistic models of the interstellar medium of our current simulations. \\

\subsection{Mergers and the ex--situ fraction of stars in the bulge}

The mergers of galaxy pairs with comparable sizes are long known to behave in an inelastic fashion and produce remnants resembling elliptical galaxies \citep{Toomre1977, Barnes1988, Hernquist1992, Hernquist1993}. Whilst the merger scenario seems satisfactory for the formation of high-S\'ersic bulges in galaxies with close to 1:1 mergers, it must be re-examined when talking about MW/M31-like galaxies or $L_{\star}$ galaxies, in light of the results of this work \citep[see also][]{Bell2017}. Only mergers that reach the galactic center and contribute to the ex--situ component of bulges seem to produce a noticeable effect on the shape of the final surface brightness profile of our galaxies.  
We confirm in this work the dominance of a single satellite (or a low number of them) in the ex--situ component of bulges (see Sec.~\ref{sec:insitu-acc}). This then disfavours the scenario of bulge formation via the accretion of many low mass satellites. 
These considerations do not ignore the importance of mergers in the formation and evolution of bulges, but seek only to discuss and reconsider the correlation of the number and kind of mergers with the formation of high-S\'ersic bulges. Indeed, high-S\'ersic bulges have more commonly higher fractions of ex--situ stars than low-S\'ersic bulges, as can be seen in the top panel of Fig.~\ref{fig:sers-bot-accreted-fractionsTNG50}. 

Finally, mergers still explain a large fraction of the stars formed in starburst events that end-up in bulges and may be responsible, in some cases, for triggering bar formation. This may, in turn, explain the formation of another large fraction of the stars that
contribute to the formation of photometric bulges. \\ 

\subsection{Low-S\'ersic bulge formation due to bars and prevention of bar formation due to concentrated bulges}

We showed in Sec.~\ref{sec:barinfluence} that low-S\'ersic bulges exhibit a higher fraction
of bars at all redshifts. This can be interpreted as a causality.  Bars grow, basically, due to transfer of angular momentum and are known drivers of photometric bulge growth, via two channels: i) bars can contribute to photometric bulge growth via the inward pull of gas due to torques that later form stars in the bar itself , near the center of the galaxy or in nuclear rings, as is predicted from simulations \citep{SandersHuntley1976, Athanassoula1992, Kim2012} and confirmed with observations \citep{Phillips1996, Sakamoto1999, Fraser-McKelvie2020, Wang2020} and  ii) bringing stars already formed to inner regions, after reshaping their orbits. Stellar particles captured in resonances can loose orbital angular momentum due to the bar and spiral arms, and suffer a reshaping of their orbits that make them populate the bulge region \citep{SelwoodBinney2002, Minchev2010}. Another mechanism where stars are driven to the central regions can be explained by the invariant manifolds \citep{Romero-Gomez2006, Athanassoula2009}. These act as channels in the ends of bars where stars can be brought from outside co--rotation to the inner regions.

In this line of reasoning, bars would contribute statistically more to the formation of photometric bulges with lower S\'ersic indices.   
However, the excess of bars in galaxies with low S\'ersic indices can also be a consequence of concentrated bulges preventing the formation of bars. The notion of a strong concentration of mass in the central region of galaxies preventing the disc to become unstable
was already presented by \citet{Toomre1981} using linear perturbation theory. It was
argued that the presence of a strong mass concentration could stop the feedback during the ``swing amplifier and feedback loop'' process \citep[see also Chapter 6.3 from][]{BinneyTremaine2008}.  More recently, \citet{Saha2018} and \citet{Kataria2018} used isolated N-body simulations, and showed that cold discs prevent the formation of a bar in the presence of highly concentrated bulges.
Naturally, both the photometric bulge growth via star formation mechanisms associated with bars and the prevention of bar formation, or its strengthening due to a concentrated bulge, can be concomitant processes
in a single galaxy or act with different strengths in different galaxies. This way, both processes are complementary to produce the trend shown in Fig.~\ref{fig:barfrac-z}.  To quantify to what degree these two combined processes lead to different bar fractions in high- and low-S\'ersic photometric bulges, we would need a higher temporal resolution to follow the orbits of individual stars in galaxies and track their formation sites more accurately, in addition to having a larger number of high mass resolution simulated galaxies to find a statistically meaningful sample of them with early central mass concentrations. 

A third, methodological, fact that could contribute to the difference in the bar fractions in high- and low-S\'ersic photometric bulges relies on the way we are messurring Sersic indices, which are obtained from a two-component photometric decomposition. 
As mentioned in Sec.~\ref{sec:2components}, although the excess of light coming from bars is extracted when it is
evident from the SBPs, some contamination that is absorbed by either of the two components remains. Short, weak bars, would contaminate more the S\'ersic component whereas strong, long bars, known to have a nearly exponential profile, would add more contamination to the disc component. Adding a third component to model the bar may yield different results, at the expense of more free parameters and uncertainties in the fitting procedure. In any case, it is beyond of the scope of this paper to compare both procedures.

\begin{figure}
\includegraphics[scale=0.54]{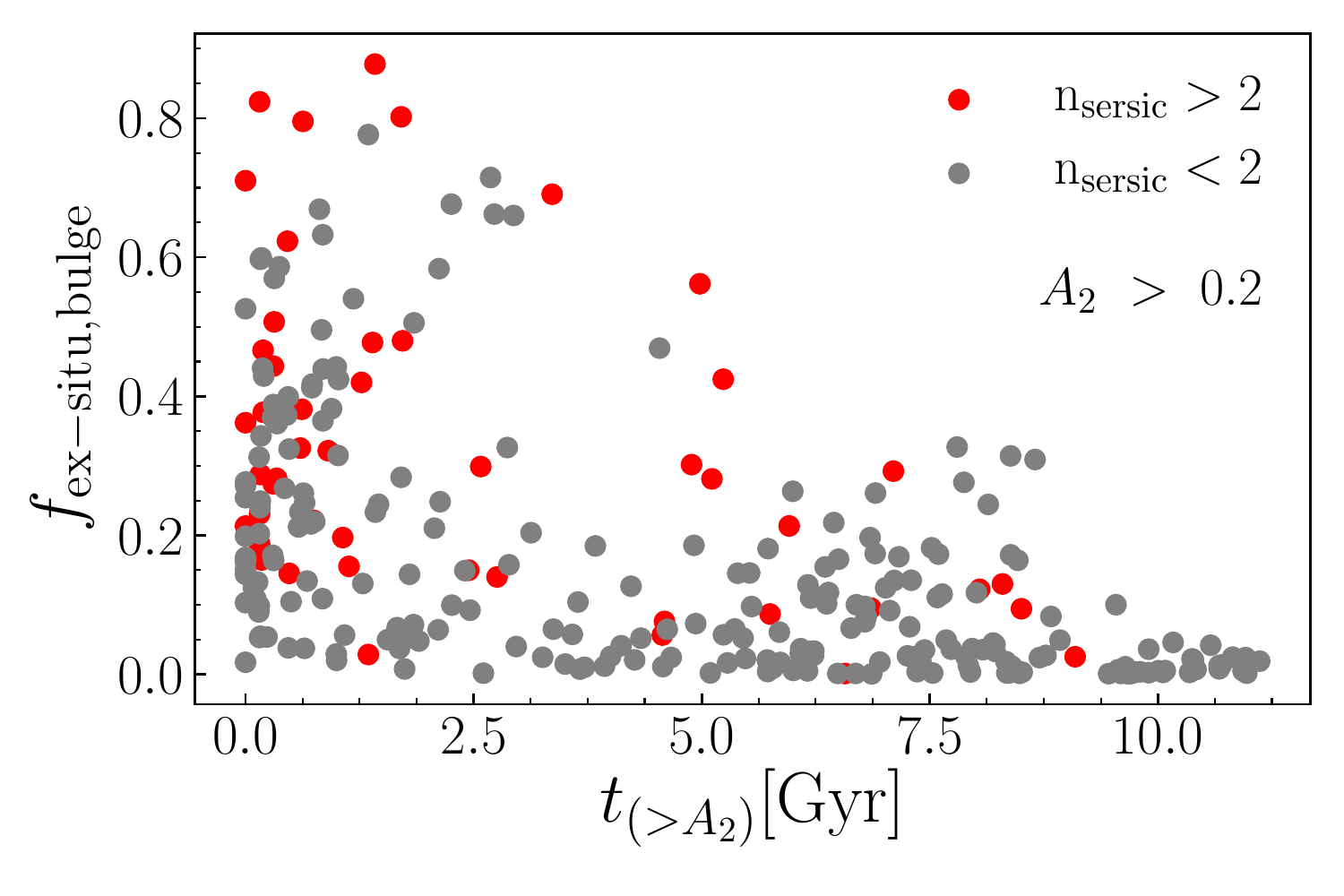}

\caption{Fraction of ex--situ stellar particles in bulges of simulated TNG50 MW/M31-like galaxies as
a function of the time during which galaxies experienced a bar feature with $A_2 > 0.2$. Strong bars develop only in galaxies with ex--situ fractions $f_{\rm acc,bulge} \lesssim 0.2$ }

\label{fig:facc-vs-taA2}
\end{figure}

An interesting trend also found in Sec~\ref{sec:barinfluence} was that of the duration of the bar feature above a given threshold with S\'ersic index, {\rm B/T} and the ex--situ bulge fraction. Although there is no clear correlation between these properties, there is a group of galaxies that develop strong bars, have low ex--situ bulge fractions and present mainly low S\'ersic indices (see Fig.~\ref{fig:time-over-A2}).  These simulated galaxies develop in some cases large massive bulges nonetheless. The lack of large fractions of ex--situ stars, the presence of a strong bar during most of their existence, and the development of large bulges in some cases, suggest the possibility that many of those photometric bulges were formed by star formation mechanisms driven by the bar. This is in line with the results shown in \citet{Fragkoudi2020}, who found, using the Auriga simulations, that bars are more likely to form in galaxies which have low fraction of ex-situ stars in their bulges. They furthermore split their barred sample into galaxies with and without boxy/peanut bulges, and showed that those that have a boxy/peanut bulge will have the lowest fraction of ex-situ stars (see Fig.14 of their article).
In Fig.~\ref{fig:facc-vs-taA2} we show the fraction of ex--situ stars in the bulge, as defined in Sec.~\ref{sec:insitu-acc} as a function of duration of the bar, with a bar strength  $A_2 > 0.2$. Clearly, strong bars mainly develop and prevail in the simulated galaxies with low to moderate ex--situ fractions. This could indicate that either strong bars do not form if bulges already
have accreted a large quantity of stars in a merger, as discussed earlier, or that the enhanced star formation produced by strong bars driving gas to the central regions increases the in--situ bulge fraction. A group of galaxies show a relatively high ex--situ fraction ($f_{\rm ex--situ}> 0.2$) and a strong bar duration  $t_{>A_2} > 3 {\rm  Gyr}$. These systems, where both scenarios combine can be classified as composite.

\section{Summary and Conclusions}
\label{sec:conclusions}

In this work, we have studied the origin of different types of bulges formed in a sample of
MW/M31-like galaxies drawn from the state-of-the-art hydrodynamical cosmological simulation TNG50. Bulges are parametrized in terms of their S\'ersic index and B/T ratios, which are derived from two-component photometric
decompositions of their surface brightness profiles in the V band. In general, there is no trivial correlation between the concentration of the S\'ersic profile that characterizes photometric bulges in simulated MW/M31-like galaxies and the aspects explored in this work, namely, the environment where galaxies lie, the total number of mergers, the properties of the significant mergers they experience, and the presence of a bar. Instead, it is a combination of processes that
contribute to form a bulge with a higher or lower S\'ersic index in a MW/M31-like galaxy.
The presence of strong bars and the occurrence of late mergers play the
more significant role in shaping the inner regions of the surface brightness profiles, whereas bars are shown to be a fundamental driver of bulge mass growth in all types of bulges. 

Our specific findings and conclusions are summarized and listed below: 
\begin{itemize}
\item We found that, of our 287 MW/M31-like galaxies selected using the criteria defined in Sec.~\ref{sec:mw-like-gal}, $17.1\%$ have S\'ersic index $n > 2$ (Fig.~\ref{fig:profiles}).    

\item The S\'ersic index of MW/M31-like galaxies in TNG50 are not affected
  by the environment. Two different methods, one considering the number of neighbours
  inside a fixed sphere of $738.2~{\rm kpc}$ radius and the other that takes into account the
  overdensity of galaxies up to the k{\it-th} neighbour, yield no measureable dependence on environment at $z=0$. {\rm B/T} ratios show a mild increase towards higher overdensities of galaxies. (Fig.~\ref{fig:sers-bot-environmental}).   

\item Galaxies with high-S\'ersic index bulges show, on average, a higher fraction of ex--situ stellar particles in the kinematically selected bulge than galaxies with low-S\'ersic index bulges, by 19 percentage points (Fig.~\ref{fig:sers-bot-accreted-fractionsTNG50}). 

\item Mergers are ubiquitous in most simulated galaxies. The number and total mass accreted via mergers are not correlated with the level of concentration of bulges reflected by their S\'ersic index (Fig.\ref{fig:nummerger}). The last considerable merger ($m_{\rm sat}/m_{\rm host} > 0.1$) occurs on average at later times in simulated galaxies with high-S\'ersic index (Fig.~\ref{fig:lastconsmer}), although a large number of them with low-S\'ersic index also suffer a late massive merger.   

\item Bulge stellar particles in our sample of MW/M31-like galaxies, selected by means of kinematic considerations, are mostly formed in--situ, as shown by previous works (Fig.~\ref{fig:accreted-fractions-Au-TNG50}). Also confirming previous results, a single satellite is commonly responsible for building $50\%$  of the ex--situ component of bulges and a low number of satellites is enough to account for $90\%$  (Fig.~\ref{fig:90-of-acc}).

 \item The bar fraction in our sample of simulated MW/M31-like galaxies shows a reasonable agreement with the bar fraction measured at $z=0$ and do not decrease towards higher redshifts, as observational estimates show for different samples of galaxies (Fig.~\ref{fig:barfrac-sheth}). Simulated galaxies with high-S\'ersic index bulges consistently show a lower bar fraction than those with low-S\'ersic index bulges at all redshifts, pointing to a strong influence of bars in the formation of low-S\'ersic bulges and the effect of concentrated bulges in preventing bar formation. Simulated galaxies with higher ${\rm B/T}$ ratios present a higher
 fraction of bars at all redshifts, indicating that bars are a prominent channel of photometric bulge growth in our sample of simulated galaxies (Fig.~\ref{fig:barfrac-z}).\\
 
Upcoming theoretical attempts to describe and understand the subtleties of the formation of different types of galactic bulges in galaxies like our own and M31, in a cosmological context, should pursue the use of simulations with a challenging combination of characteristics: high dynamical --and mass-- resolution,  to ensure a realistic dynamical behaviour in central regions of simulated galaxies; large cosmological volumes, to capture the observed diversity of bulges, and more realistic ISM models.  
  
\end{itemize}

\section*{Acknowledgements}

I.G. acknowledges financial support from CONICYT Programa de Astronom\'ia, Fondo ALMA-CONICYT 2017 31170048, {\it Consejo Nacional de Investigaciones Cient\'{\i}ficas y T\'ecnicas} (CONICET, PIP-0387), {\it Agencia Nacional de Promoci\'on de la Investigaci\'on, el Desarrollo Tecnol\'ogico y la Innovaci\'on} (Agencia I+D+i, PICT-2018-03743), and {\it Universidad Nacional de La Plata} (G11-150), Argentina. AM acknowledges support from  FONDECYT Regular 1212046.
FAG acknowledges support from  FONDECYT Regular  1211370. IG, AM and FAG acknowledge funding from the Max Planck Society through a “Partner Group” grant. DN acknowledges funding from the Deutsche Forschungsgemeinschaft (DFG) through an Emmy Noether Research Group (grant number NE 2441/1-1).
AP acknowledges support by the Deutsche Forschungsgemeinschaft (DFG, German Research Foundation) -- Project-ID 138713538 -- SFB 881 (``The Milky Way System'', subproject A01)
FM acknowledges support through the Program "Rita Levi Montalcini" of the Italian MUR. The primary TNG simulations were carried out with compute time granted by the Gauss Centre for Supercomputing (GCS) under Large-Scale Projects GCS-ILLU and GCS-DWAR on the GCS share of the supercomputer Hazel Hen at the High Performance Computing Center Stuttgart (HLRS). The public package {\tt PY-SPHVIEWER} \citep{BenitezLlambay2015} was used to produce Fig.~\ref{fig:galaxies} of this paper.

\section*{Data Availability}

The data used in this work is accesible via the IllustrisTNG public database\footnote{https://www.tng-project.org/data}. The Data from the Auriga simulations used for comparison is available upon reasonable request. 

\bibliographystyle{mnras}
\bibliography{references}

\begin{thebibliography}{}
\makeatletter
\relax
\def\mn@urlcharsother{\let\do\@makeother \do\$\do\&\do\#\do\^\do\_\do\%\do\~}
\def\mn@doi{\begingroup\mn@urlcharsother \@ifnextchar [ {\mn@doi@}
  {\mn@doi@[]}}
\def\mn@doi@[#1]#2{\def\@tempa{#1}\ifx\@tempa\@empty \href
  {http://dx.doi.org/#2} {doi:#2}\else \href {http://dx.doi.org/#2} {#1}\fi
  \endgroup}
\def\mn@eprint#1#2{\mn@eprint@#1:#2::\@nil}
\def\mn@eprint@arXiv#1{\href {http://arxiv.org/abs/#1} {{\tt arXiv:#1}}}
\def\mn@eprint@dblp#1{\href {http://dblp.uni-trier.de/rec/bibtex/#1.xml}
  {dblp:#1}}
\def\mn@eprint@#1:#2:#3:#4\@nil{\def\@tempa {#1}\def\@tempb {#2}\def\@tempc
  {#3}\ifx \@tempc \@empty \let \@tempc \@tempb \let \@tempb \@tempa \fi \ifx
  \@tempb \@empty \def\@tempb {arXiv}\fi \@ifundefined
  {mn@eprint@\@tempb}{\@tempb:\@tempc}{\expandafter \expandafter \csname
  mn@eprint@\@tempb\endcsname \expandafter{\@tempc}}}

\bibitem[\protect\citeauthoryear{{Abadi}, {Navarro}, {Steinmetz}  \&
  {Eke}}{{Abadi} et~al.}{2003}]{Abadi2003}
{Abadi} M.~G.,  {Navarro} J.~F.,  {Steinmetz} M.,   {Eke} V.~R.,  2003, \mn@doi
  [\apj] {10.1086/375512}, \href
  {https://ui.adsabs.harvard.edu/#abs/2003ApJ...591..499A} {591, 499}

\bibitem[\protect\citeauthoryear{{Andredakis}, {Peletier}  \&
  {Balcells}}{{Andredakis} et~al.}{1995}]{Andredakis1995}
{Andredakis} Y.~C.,  {Peletier} R.~F.,   {Balcells} M.,  1995, \mn@doi [\mnras]
  {10.1093/mnras/275.3.874}, \href
  {https://ui.adsabs.harvard.edu/abs/1995MNRAS.275..874A} {275, 874}

\bibitem[\protect\citeauthoryear{{Athanassoula}}{{Athanassoula}}{1992}]{Athanassoula1992}
{Athanassoula} E.,  1992, \mn@doi [\mnras] {10.1093/mnras/259.2.345}, \href
  {https://ui.adsabs.harvard.edu/abs/1992MNRAS.259..345A} {259, 345}

\bibitem[\protect\citeauthoryear{{Athanassoula}}{{Athanassoula}}{2005}]{Athanassoula2005}
{Athanassoula} E.,  2005, \mn@doi [\mnras] {10.1111/j.1365-2966.2005.08872.x},
  \href {https://ui.adsabs.harvard.edu/abs/2005MNRAS.358.1477A} {358, 1477}

\bibitem[\protect\citeauthoryear{{Athanassoula}, {Romero-G{\'o}mez}, {Bosma}
  \& {Masdemont}}{{Athanassoula} et~al.}{2009}]{Athanassoula2009}
{Athanassoula} E.,  {Romero-G{\'o}mez} M.,  {Bosma} A.,   {Masdemont} J.~J.,
  2009, \mn@doi [\mnras] {10.1111/j.1365-2966.2009.15583.x}, \href
  {https://ui.adsabs.harvard.edu/abs/2009MNRAS.400.1706A} {400, 1706}

\bibitem[\protect\citeauthoryear{{Barnes}}{{Barnes}}{1988}]{Barnes1988}
{Barnes} J.~E.,  1988, \mn@doi [\apj] {10.1086/166593}, \href
  {https://ui.adsabs.harvard.edu/abs/1988ApJ...331..699B} {331, 699}

\bibitem[\protect\citeauthoryear{{Bell}, {Monachesi}, {Harmsen}, {de Jong},
  {Bailin}, {Radburn-Smith}, {D'Souza}  \& {Holwerda}}{{Bell}
  et~al.}{2017}]{Bell2017}
{Bell} E.~F.,  {Monachesi} A.,  {Harmsen} B.,  {de Jong} R.~S.,  {Bailin} J.,
  {Radburn-Smith} D.~J.,  {D'Souza} R.,   {Holwerda} B.~W.,  2017, \mn@doi
  [\apj] {10.3847/2041-8213/aa6158}, \href
  {https://ui.adsabs.harvard.edu/#abs/2017ApJ...837L...8B} {837, L8}

\bibitem[\protect\citeauthoryear{Benitez-Llambay}{Benitez-Llambay}{2015}]{BenitezLlambay2015}
Benitez-Llambay A.,  2015, py-sphviewer: Py-SPHViewer v1.0.0,
  \mn@doi{10.5281/zenodo.21703}, \url {http://dx.doi.org/10.5281/zenodo.21703}

\bibitem[\protect\citeauthoryear{{Binney} \& {Tremaine}}{{Binney} \&
  {Tremaine}}{2008}]{BinneyTremaine2008}
{Binney} J.,  {Tremaine} S.,  2008, {Galactic Dynamics: Second Edition}

\bibitem[\protect\citeauthoryear{{Bla{\~n}a D{\'\i}az}, {Wegg}, {Gerhard},
  {Erwin}, {Portail}, {Opitsch}, {Saglia}  \& {Bender}}{{Bla{\~n}a D{\'\i}az}
  et~al.}{2017}]{Blana-Diaz2016}
{Bla{\~n}a D{\'\i}az} M.,  {Wegg} C.,  {Gerhard} O.,  {Erwin} P.,  {Portail}
  M.,  {Opitsch} M.,  {Saglia} R.,   {Bender} R.,  2017, \mn@doi [\mnras]
  {10.1093/mnras/stw3294}, \href
  {https://ui.adsabs.harvard.edu/abs/2017MNRAS.466.4279B} {466, 4279}

\bibitem[\protect\citeauthoryear{{Bla{\~n}a D{\'\i}az} et~al.,}{{Bla{\~n}a
  D{\'\i}az} et~al.}{2018}]{Blana-Diaz2018}
{Bla{\~n}a D{\'\i}az} M.,  et~al., 2018, \mn@doi [\mnras]
  {10.1093/mnras/sty2311}, \href
  {https://ui.adsabs.harvard.edu/abs/2018MNRAS.481.3210B} {481, 3210}

\bibitem[\protect\citeauthoryear{{Bland-Hawthorn} \&
  {Gerhard}}{{Bland-Hawthorn} \& {Gerhard}}{2016}]{Bland-HawthornGerhard2016}
{Bland-Hawthorn} J.,  {Gerhard} O.,  2016, \mn@doi [\araa]
  {10.1146/annurev-astro-081915-023441}, \href
  {https://ui.adsabs.harvard.edu/abs/2016ARA&A..54..529B} {54, 529}

\bibitem[\protect\citeauthoryear{{Blanton} \& {Moustakas}}{{Blanton} \&
  {Moustakas}}{2009}]{BlantonMoustakas2009}
{Blanton} M.~R.,  {Moustakas} J.,  2009, \mn@doi [\araa]
  {10.1146/annurev-astro-082708-101734}, \href
  {https://ui.adsabs.harvard.edu/abs/2009ARA&A..47..159B} {47, 159}

\bibitem[\protect\citeauthoryear{{Bl{\'a}zquez-Calero}
  et~al.,}{{Bl{\'a}zquez-Calero} et~al.}{2020}]{Blazquez-Calero2020}
{Bl{\'a}zquez-Calero} G.,  et~al., 2020, \mn@doi [\mnras]
  {10.1093/mnras/stz3125}, \href
  {https://ui.adsabs.harvard.edu/abs/2020MNRAS.491.1800B} {491, 1800}

\bibitem[\protect\citeauthoryear{{Boardman} et~al.,}{{Boardman}
  et~al.}{2020}]{Boardman2020}
{Boardman} N.,  et~al., 2020, \mn@doi [\mnras] {10.1093/mnras/staa2731}, \href
  {https://ui.adsabs.harvard.edu/abs/2020MNRAS.498.4943B} {498, 4943}

\bibitem[\protect\citeauthoryear{{Bois} et~al.,}{{Bois}
  et~al.}{2010}]{Bois2010}
{Bois} M.,  et~al., 2010, \mn@doi [\mnras] {10.1111/j.1365-2966.2010.16885.x},
  \href {https://ui.adsabs.harvard.edu/abs/2010MNRAS.406.2405B} {406, 2405}

\bibitem[\protect\citeauthoryear{{Brooks} \& {Christensen}}{{Brooks} \&
  {Christensen}}{2016}]{BrooksChristensen2015}
{Brooks} A.,  {Christensen} C.,  2016, in {Laurikainen} E.,  {Peletier} R.,
  {Gadotti} D.,  eds, ~ Vol. 418, Galactic Bulges. p.~317,
  \mn@doi{10.1007/978-3-319-19378-6_12}

\bibitem[\protect\citeauthoryear{{Bruzual} \& {Charlot}}{{Bruzual} \&
  {Charlot}}{2003}]{BruzualCharlot2003}
{Bruzual} G.,  {Charlot} S.,  2003, \mn@doi [\mnras]
  {10.1046/j.1365-8711.2003.06897.x}, \href
  {https://ui.adsabs.harvard.edu/#abs/2003MNRAS.344.1000B} {344, 1000}

\bibitem[\protect\citeauthoryear{{Buck}, {Ness}, {Macci{\`o}}, {Obreja}  \&
  {Dutton}}{{Buck} et~al.}{2018}]{Buck2018a}
{Buck} T.,  {Ness} M.~K.,  {Macci{\`o}} A.~V.,  {Obreja} A.,   {Dutton} A.~A.,
  2018, \mn@doi [\apj] {10.3847/1538-4357/aac890}, \href
  {https://ui.adsabs.harvard.edu/#abs/2018ApJ...861...88B} {861, 88}

\bibitem[\protect\citeauthoryear{{Ceverino}, {Dekel}  \& {Bournaud}}{{Ceverino}
  et~al.}{2010}]{Ceverino2010}
{Ceverino} D.,  {Dekel} A.,   {Bournaud} F.,  2010, \mn@doi [\mnras]
  {10.1111/j.1365-2966.2010.16433.x}, \href
  {https://ui.adsabs.harvard.edu/#abs/2010MNRAS.404.2151C} {404, 2151}

\bibitem[\protect\citeauthoryear{{Croton} et~al.,}{{Croton}
  et~al.}{2006}]{Croton2006}
{Croton} D.~J.,  et~al., 2006, \mn@doi [\mnras]
  {10.1111/j.1365-2966.2005.09675.x}, \href
  {https://ui.adsabs.harvard.edu/abs/2006MNRAS.365...11C} {365, 11}

\bibitem[\protect\citeauthoryear{{Dekel}, {Sari}  \& {Ceverino}}{{Dekel}
  et~al.}{2009}]{Dekel2009}
{Dekel} A.,  {Sari} R.,   {Ceverino} D.,  2009, \mn@doi [\apj]
  {10.1088/0004-637X/703/1/785}, \href
  {https://ui.adsabs.harvard.edu/abs/2009ApJ...703..785D} {703, 785}

\bibitem[\protect\citeauthoryear{{Du}, {Ho}, {Debattista}, {Pillepich},
  {Nelson}, {Zhao}  \& {Hernquist}}{{Du} et~al.}{2020}]{Du2020}
{Du} M.,  {Ho} L.~C.,  {Debattista} V.~P.,  {Pillepich} A.,  {Nelson} D.,
  {Zhao} D.,   {Hernquist} L.,  2020, \mn@doi [\apj]
  {10.3847/1538-4357/ab8fa8}, \href
  {https://ui.adsabs.harvard.edu/abs/2020ApJ...895..139D} {895, 139}

\bibitem[\protect\citeauthoryear{{Du}, {Ho}, {Debattista}, {Pillepich},
  {Nelson}, {Hernquist}  \& {Weinberger}}{{Du} et~al.}{2021}]{Du2021}
{Du} M.,  {Ho} L.~C.,  {Debattista} V.~P.,  {Pillepich} A.,  {Nelson} D.,
  {Hernquist} L.,   {Weinberger} R.,  2021, \mn@doi [\apj]
  {10.3847/1538-4357/ac0e98}, \href
  {https://ui.adsabs.harvard.edu/abs/2021ApJ...919..135D} {919, 135}

\bibitem[\protect\citeauthoryear{{Dubois} et~al.,}{{Dubois}
  et~al.}{2020}]{Dubois2020}
{Dubois} Y.,  et~al., 2020, arXiv e-prints, \href
  {https://ui.adsabs.harvard.edu/abs/2020arXiv200910578D} {p. arXiv:2009.10578}

\bibitem[\protect\citeauthoryear{{Elmegreen}}{{Elmegreen}}{1995}]{Elmegreen1995}
{Elmegreen} B.~G.,  1995, \mn@doi [\mnras] {10.1093/mnras/275.4.944}, \href
  {https://ui.adsabs.harvard.edu/abs/1995MNRAS.275..944E} {275, 944}

\bibitem[\protect\citeauthoryear{{Elmegreen}, {Elmegreen}, {Fernandez}  \&
  {Lemonias}}{{Elmegreen} et~al.}{2009}]{Elmegreen2009}
{Elmegreen} B.~G.,  {Elmegreen} D.~M.,  {Fernandez} M.~X.,   {Lemonias} J.~J.,
  2009, \mn@doi [\apj] {10.1088/0004-637X/692/1/12}, \href
  {https://ui.adsabs.harvard.edu/#abs/2009ApJ...692...12E} {692, 12}

\bibitem[\protect\citeauthoryear{{Engler} et~al.,}{{Engler}
  et~al.}{2020}]{Engler2020}
{Engler} C.,  et~al., 2020, arXiv e-prints, \href
  {https://ui.adsabs.harvard.edu/abs/2020arXiv200211119E} {p. arXiv:2002.11119}

\bibitem[\protect\citeauthoryear{{Erwin} et~al.,}{{Erwin}
  et~al.}{2015}]{Erwin2015}
{Erwin} P.,  et~al., 2015, \mn@doi [\mnras] {10.1093/mnras/stu2376}, \href
  {https://ui.adsabs.harvard.edu/abs/2015MNRAS.446.4039E} {446, 4039}

\bibitem[\protect\citeauthoryear{{Fisher} \& {Drory}}{{Fisher} \&
  {Drory}}{2008}]{FisherDrory2008}
{Fisher} D.~B.,  {Drory} N.,  2008, \mn@doi [\aj]
  {10.1088/0004-6256/136/2/773}, \href
  {https://ui.adsabs.harvard.edu/#abs/2008AJ....136..773F} {136, 773}

\bibitem[\protect\citeauthoryear{{Fisher} \& {Drory}}{{Fisher} \&
  {Drory}}{2010}]{FisherDrory2010}
{Fisher} D.~B.,  {Drory} N.,  2010, \mn@doi [\apj]
  {10.1088/0004-637X/716/2/942}, \href
  {https://ui.adsabs.harvard.edu/abs/2010ApJ...716..942F} {716, 942}

\bibitem[\protect\citeauthoryear{{Fisher} \& {Drory}}{{Fisher} \&
  {Drory}}{2011}]{FisherDrory2011}
{Fisher} D.~B.,  {Drory} N.,  2011, \mn@doi [\apj]
  {10.1088/2041-8205/733/2/L47}, \href
  {https://ui.adsabs.harvard.edu/#abs/2011ApJ...733L..47F} {733, L47}

\bibitem[\protect\citeauthoryear{{Fisher} \& {Drory}}{{Fisher} \&
  {Drory}}{2016}]{FisherDrory2016}
{Fisher} D.~B.,  {Drory} N.,  2016, in {Laurikainen} E.,  {Peletier} R.,
  {Gadotti} D.,  eds,  Astrophysics and Space Science Library Vol. 418,
  Galactic Bulges. p.~41 (\mn@eprint {arXiv} {1512.02230}),
  \mn@doi{10.1007/978-3-319-19378-6_3}

\bibitem[\protect\citeauthoryear{{Flynn}, {Holmberg}, {Portinari}, {Fuchs}  \&
  {Jahrei{\ss}}}{{Flynn} et~al.}{2006}]{Flynn2006}
{Flynn} C.,  {Holmberg} J.,  {Portinari} L.,  {Fuchs} B.,   {Jahrei{\ss}} H.,
  2006, \mn@doi [\mnras] {10.1111/j.1365-2966.2006.10911.x}, \href
  {https://ui.adsabs.harvard.edu/abs/2006MNRAS.372.1149F} {372, 1149}

\bibitem[\protect\citeauthoryear{{Fragkoudi} et~al.,}{{Fragkoudi}
  et~al.}{2020}]{Fragkoudi2020}
{Fragkoudi} F.,  et~al., 2020, \mn@doi [\mnras] {10.1093/mnras/staa1104}, \href
  {https://ui.adsabs.harvard.edu/abs/2020MNRAS.494.5936F} {494, 5936}

\bibitem[\protect\citeauthoryear{{Fraser-McKelvie} et~al.,}{{Fraser-McKelvie}
  et~al.}{2020}]{Fraser-McKelvie2020}
{Fraser-McKelvie} A.,  et~al., 2020, \mn@doi [\mnras] {10.1093/mnras/staa1416},
  \href {https://ui.adsabs.harvard.edu/abs/2020MNRAS.495.4158F} {495, 4158}

\bibitem[\protect\citeauthoryear{{Freeman}}{{Freeman}}{1970}]{Freeman1970}
{Freeman} K.~C.,  1970, \mn@doi [\apj] {10.1086/150474}, \href
  {https://ui.adsabs.harvard.edu/abs/1970ApJ...160..811F} {160, 811}

\bibitem[\protect\citeauthoryear{{Gadotti}}{{Gadotti}}{2009}]{Gadotti2009}
{Gadotti} D.~A.,  2009, \mn@doi [\mnras] {10.1111/j.1365-2966.2008.14257.x},
  \href {https://ui.adsabs.harvard.edu/#abs/2009MNRAS.393.1531G} {393, 1531}

\bibitem[\protect\citeauthoryear{{Gadotti}}{{Gadotti}}{2012}]{Gadotti2012}
{Gadotti} D.~A.,  2012, arXiv e-prints, \href
  {https://ui.adsabs.harvard.edu/abs/2012arXiv1208.2295G} {p. arXiv:1208.2295}

\bibitem[\protect\citeauthoryear{{Gargiulo}, {Cora}, {Vega-Mart{\'\i}nez},
  {Gonzalez}, {Zoccali}, {Gonz{\'a}lez}, {Ruiz}  \& {Padilla}}{{Gargiulo}
  et~al.}{2017}]{Gargiulo2017}
{Gargiulo} I.~D.,  {Cora} S.~A.,  {Vega-Mart{\'\i}nez} C.~A.,  {Gonzalez}
  O.~A.,  {Zoccali} M.,  {Gonz{\'a}lez} R.,  {Ruiz} A.~N.,   {Padilla} N.~D.,
  2017, \mn@doi [\mnras] {10.1093/mnras/stx2188}, \href
  {https://ui.adsabs.harvard.edu/#abs/2017MNRAS.472.4133G} {472, 4133}

\bibitem[\protect\citeauthoryear{{Gargiulo} et~al.,}{{Gargiulo}
  et~al.}{2019}]{Gargiulo2019}
{Gargiulo} I.~D.,  et~al., 2019, \mn@doi [\mnras] {10.1093/mnras/stz2536},
  \href {https://ui.adsabs.harvard.edu/abs/2019MNRAS.489.5742G} {489, 5742}

\bibitem[\protect\citeauthoryear{{Genel} et~al.,}{{Genel}
  et~al.}{2014}]{Genel2014}
{Genel} S.,  et~al., 2014, \mn@doi [\mnras] {10.1093/mnras/stu1654}, \href
  {https://ui.adsabs.harvard.edu/abs/2014MNRAS.445..175G} {445, 175}

\bibitem[\protect\citeauthoryear{{G{\'o}mez}, {White}, {Grand}, {Marinacci},
  {Springel}  \& {Pakmor}}{{G{\'o}mez} et~al.}{2017}]{Gomez2017}
{G{\'o}mez} F.~A.,  {White} S. D.~M.,  {Grand} R. J.~J.,  {Marinacci} F.,
  {Springel} V.,   {Pakmor} R.,  2017, \mn@doi [\mnras]
  {10.1093/mnras/stw2957}, \href
  {https://ui.adsabs.harvard.edu/#abs/2017MNRAS.465.3446G} {465, 3446}

\bibitem[\protect\citeauthoryear{{Graham} \& {Driver}}{{Graham} \&
  {Driver}}{2005}]{Graham2005}
{Graham} A.~W.,  {Driver} S.~P.,  2005, \mn@doi [\pasa] {10.1071/AS05001},
  \href {https://ui.adsabs.harvard.edu/abs/2005PASA...22..118G} {22, 118}

\bibitem[\protect\citeauthoryear{{Grand}, {Springel}, {G{\'o}mez}, {Marinacci},
  {Pakmor}, {Campbell}  \& {Jenkins}}{{Grand} et~al.}{2016}]{Grand2016}
{Grand} R. J.~J.,  {Springel} V.,  {G{\'o}mez} F.~A.,  {Marinacci} F.,
  {Pakmor} R.,  {Campbell} D. J.~R.,   {Jenkins} A.,  2016, \mn@doi [\mnras]
  {10.1093/mnras/stw601}, \href
  {https://ui.adsabs.harvard.edu/#abs/2016MNRAS.459..199G} {459, 199}

\bibitem[\protect\citeauthoryear{{Grand} et~al.,}{{Grand}
  et~al.}{2017}]{Grand2017}
{Grand} R. J.~J.,  et~al., 2017, \mn@doi [\mnras] {10.1093/mnras/stx071}, \href
  {https://ui.adsabs.harvard.edu/#abs/2017MNRAS.467..179G} {467, 179}

\bibitem[\protect\citeauthoryear{{Grand} et~al.,}{{Grand}
  et~al.}{2021}]{Grand2021}
{Grand} R. J.~J.,  et~al., 2021, \mn@doi [\mnras] {10.1093/mnras/stab2492},
  \href {https://ui.adsabs.harvard.edu/abs/2021MNRAS.507.4953G} {507, 4953}

\bibitem[\protect\citeauthoryear{{Guedes}, {Mayer}, {Carollo}  \&
  {Madau}}{{Guedes} et~al.}{2013}]{Guedes2013}
{Guedes} J.,  {Mayer} L.,  {Carollo} M.,   {Madau} P.,  2013, \mn@doi [\apj]
  {10.1088/0004-637X/772/1/36}, \href
  {https://ui.adsabs.harvard.edu/#abs/2013ApJ...772...36G} {772, 36}

\bibitem[\protect\citeauthoryear{{Hernquist}}{{Hernquist}}{1992}]{Hernquist1992}
{Hernquist} L.,  1992, \mn@doi [\apj] {10.1086/172009}, \href
  {https://ui.adsabs.harvard.edu/abs/1992ApJ...400..460H} {400, 460}

\bibitem[\protect\citeauthoryear{{Hernquist}}{{Hernquist}}{1993}]{Hernquist1993}
{Hernquist} L.,  1993, \mn@doi [\apj] {10.1086/172686}, \href
  {https://ui.adsabs.harvard.edu/abs/1993ApJ...409..548H} {409, 548}

\bibitem[\protect\citeauthoryear{{Hopkins}, {Cox}, {Younger}  \&
  {Hernquist}}{{Hopkins} et~al.}{2009}]{Hopkins2009}
{Hopkins} P.~F.,  {Cox} T.~J.,  {Younger} J.~D.,   {Hernquist} L.,  2009,
  \mn@doi [\apj] {10.1088/0004-637X/691/2/1168}, \href
  {http://adsabs.harvard.edu/abs/2009ApJ...691.1168H} {691, 1168}

\bibitem[\protect\citeauthoryear{{Kataria} \& {Das}}{{Kataria} \&
  {Das}}{2018}]{Kataria2018}
{Kataria} S.~K.,  {Das} M.,  2018, \mn@doi [\mnras] {10.1093/mnras/stx3279},
  \href {https://ui.adsabs.harvard.edu/abs/2018MNRAS.475.1653K} {475, 1653}

\bibitem[\protect\citeauthoryear{{Kent}}{{Kent}}{1985}]{Kent1985}
{Kent} S.~M.,  1985, \mn@doi [\apjs] {10.1086/191066}, \href
  {https://ui.adsabs.harvard.edu/abs/1985ApJS...59..115K} {59, 115}

\bibitem[\protect\citeauthoryear{{Kim}, {Seo}, {Stone}, {Yoon}  \&
  {Teuben}}{{Kim} et~al.}{2012}]{Kim2012}
{Kim} W.-T.,  {Seo} W.-Y.,  {Stone} J.~M.,  {Yoon} D.,   {Teuben} P.~J.,  2012,
  \mn@doi [\apj] {10.1088/0004-637X/747/1/60}, \href
  {https://ui.adsabs.harvard.edu/#abs/2012ApJ...747...60K} {747, 60}

\bibitem[\protect\citeauthoryear{{Koekemoer}, {Faber}, {Ferguson}  \& et
  al.}{{Koekemoer} et~al.}{2011}]{Koekemoer2011}
{Koekemoer} A.~M.,  {Faber} S.~M.,  {Ferguson}  et al. 2011, \mn@doi [\apjs]
  {10.1088/0067-0049/197/2/36}, \href
  {https://ui.adsabs.harvard.edu/abs/2011ApJS..197...36K} {197, 36}

\bibitem[\protect\citeauthoryear{{Kormendy}}{{Kormendy}}{1977}]{Kormendy1977}
{Kormendy} J.,  1977, \mn@doi [\apj] {10.1086/155687}, \href
  {https://ui.adsabs.harvard.edu/#abs/1977ApJ...218..333K} {218, 333}

\bibitem[\protect\citeauthoryear{{Kormendy} \& {Kennicutt}}{{Kormendy} \&
  {Kennicutt}}{2004}]{KormendyKennicutt2004}
{Kormendy} J.,  {Kennicutt} Robert~C. J.,  2004, \mn@doi [Annual Review of
  Astronomy and Astrophysics] {10.1146/annurev.astro.42.053102.134024}, \href
  {https://ui.adsabs.harvard.edu/#abs/2004ARA&A..42..603K} {42, 603}

\bibitem[\protect\citeauthoryear{{Kunder} et~al.,}{{Kunder}
  et~al.}{2016}]{Kunder2016}
{Kunder} A.,  et~al., 2016, \mn@doi [\apjl] {10.3847/2041-8205/821/2/L25},
  \href {https://ui.adsabs.harvard.edu/abs/2016ApJ...821L..25K} {821, L25}

\bibitem[\protect\citeauthoryear{{Kunder} et~al.,}{{Kunder}
  et~al.}{2020}]{Kunder2020}
{Kunder} A.,  et~al., 2020, \mn@doi [\aj] {10.3847/1538-3881/ab8d35}, \href
  {https://ui.adsabs.harvard.edu/abs/2020AJ....159..270K} {159, 270}

\bibitem[\protect\citeauthoryear{{Lagos}}{{Lagos}}{2018}]{Lagos2018}
{Lagos} C. d.~P.,  2018, arXiv e-prints, \href
  {https://ui.adsabs.harvard.edu/abs/2018arXiv181013074L} {p. arXiv:1810.13074}

\bibitem[\protect\citeauthoryear{{Laurikainen}, {Salo}  \&
  {Buta}}{{Laurikainen} et~al.}{2005}]{Laurikainen2005}
{Laurikainen} E.,  {Salo} H.,   {Buta} R.,  2005, \mn@doi [\mnras]
  {10.1111/j.1365-2966.2005.09404.x}, \href
  {https://ui.adsabs.harvard.edu/abs/2005MNRAS.362.1319L} {362, 1319}

\bibitem[\protect\citeauthoryear{{Licquia} \& {Newman}}{{Licquia} \&
  {Newman}}{2015}]{LicquiaNewman2015}
{Licquia} T.~C.,  {Newman} J.~A.,  2015, \mn@doi [\apj]
  {10.1088/0004-637X/806/1/96}, \href
  {https://ui.adsabs.harvard.edu/#abs/2015ApJ...806...96L} {806, 96}

\bibitem[\protect\citeauthoryear{{Lintott} et~al.,}{{Lintott}
  et~al.}{2008}]{Lintott2008}
{Lintott} C.~J.,  et~al., 2008, \mn@doi [\mnras]
  {10.1111/j.1365-2966.2008.13689.x}, \href
  {https://ui.adsabs.harvard.edu/abs/2008MNRAS.389.1179L} {389, 1179}

\bibitem[\protect\citeauthoryear{{Lovell}, {Frenk}, {Eke}, {Jenkins}, {Gao}  \&
  {Theuns}}{{Lovell} et~al.}{2014}]{Lovell2014}
{Lovell} M.~R.,  {Frenk} C.~S.,  {Eke} V.~R.,  {Jenkins} A.,  {Gao} L.,
  {Theuns} T.,  2014, \mn@doi [\mnras] {10.1093/mnras/stt2431}, \href
  {https://ui.adsabs.harvard.edu/abs/2014MNRAS.439..300L} {439, 300}

\bibitem[\protect\citeauthoryear{{Lovell} et~al.,}{{Lovell}
  et~al.}{2018}]{Lovell2018}
{Lovell} M.~R.,  et~al., 2018, \mn@doi [\mnras] {10.1093/mnras/sty2339}, \href
  {https://ui.adsabs.harvard.edu/abs/2018MNRAS.481.1950L} {481, 1950}

\bibitem[\protect\citeauthoryear{{Luo} et~al.,}{{Luo} et~al.}{2020}]{Luo2020}
{Luo} Y.,  et~al., 2020, \mn@doi [\mnras] {10.1093/mnras/staa328}, \href
  {https://ui.adsabs.harvard.edu/abs/2020MNRAS.493.1686L} {493, 1686}

\bibitem[\protect\citeauthoryear{{Marchesini}, {Stefanon}, {Brammer}  \&
  {Whitaker}}{{Marchesini} et~al.}{2012}]{Marchesini2012}
{Marchesini} D.,  {Stefanon} M.,  {Brammer} G.~B.,   {Whitaker} K.~E.,  2012,
  \mn@doi [\apj] {10.1088/0004-637X/748/2/126}, \href
  {https://ui.adsabs.harvard.edu/abs/2012ApJ...748..126M} {748, 126}

\bibitem[\protect\citeauthoryear{{Marinacci} et~al.,}{{Marinacci}
  et~al.}{2018}]{Marinacci2018}
{Marinacci} F.,  et~al., 2018, \mn@doi [\mnras] {10.1093/mnras/sty2206}, \href
  {https://ui.adsabs.harvard.edu/abs/2018MNRAS.480.5113M} {480, 5113}

\bibitem[\protect\citeauthoryear{{Martig}, {Bournaud}, {Croton}, {Dekel}  \&
  {Teyssier}}{{Martig} et~al.}{2012}]{Martig2012}
{Martig} M.,  {Bournaud} F.,  {Croton} D.~J.,  {Dekel} A.,   {Teyssier} R.,
  2012, \mn@doi [\apj] {10.1088/0004-637X/756/1/26}, \href
  {https://ui.adsabs.harvard.edu/abs/2012ApJ...756...26M} {756, 26}

\bibitem[\protect\citeauthoryear{{Melvin} et~al.,}{{Melvin}
  et~al.}{2014}]{Melvin2014}
{Melvin} T.,  et~al., 2014, \mn@doi [\mnras] {10.1093/mnras/stt2397}, \href
  {https://ui.adsabs.harvard.edu/abs/2014MNRAS.438.2882M} {438, 2882}

\bibitem[\protect\citeauthoryear{{M{\'e}ndez-Abreu}, {Debattista}, {Corsini}
  \& {Aguerri}}{{M{\'e}ndez-Abreu} et~al.}{2014}]{Mendez-Abreu2014}
{M{\'e}ndez-Abreu} J.,  {Debattista} V.~P.,  {Corsini} E.~M.,   {Aguerri}
  J.~A.~L.,  2014, \mn@doi [\aap] {10.1051/0004-6361/201423955}, \href
  {https://ui.adsabs.harvard.edu/abs/2014A&A...572A..25M} {572, A25}

\bibitem[\protect\citeauthoryear{{Men{\'e}ndez-Delmestre}, {Sheth},
  {Schinnerer}, {Jarrett}  \& {Scoville}}{{Men{\'e}ndez-Delmestre}
  et~al.}{2007}]{MenendezDelmestre2007}
{Men{\'e}ndez-Delmestre} K.,  {Sheth} K.,  {Schinnerer} E.,  {Jarrett} T.~H.,
  {Scoville} N.~Z.,  2007, \mn@doi [\apj] {10.1086/511025}, \href
  {https://ui.adsabs.harvard.edu/abs/2007ApJ...657..790M} {657, 790}

\bibitem[\protect\citeauthoryear{{Minchev} \& {Famaey}}{{Minchev} \&
  {Famaey}}{2010}]{Minchev2010}
{Minchev} I.,  {Famaey} B.,  2010, \mn@doi [\apj]
  {10.1088/0004-637X/722/1/112}, \href
  {http://adsabs.harvard.edu/abs/2010ApJ...722..112M} {722, 112}

\bibitem[\protect\citeauthoryear{{Monachesi}, {Bell}, {Radburn-Smith},
  {Bailin}, {de Jong}, {Holwerda}, {Streich}  \& {Silverstein}}{{Monachesi}
  et~al.}{2016}]{Monachesi2016}
{Monachesi} A.,  {Bell} E.~F.,  {Radburn-Smith} D.~J.,  {Bailin} J.,  {de Jong}
  R.~S.,  {Holwerda} B.,  {Streich} D.,   {Silverstein} G.,  2016, \mn@doi
  [\mnras] {10.1093/mnras/stv2987}, \href
  {https://ui.adsabs.harvard.edu/#abs/2016MNRAS.457.1419M} {457, 1419}

\bibitem[\protect\citeauthoryear{{Monachesi} et~al.,}{{Monachesi}
  et~al.}{2019}]{Monachesi2019}
{Monachesi} A.,  et~al., 2019, \mn@doi [\mnras] {10.1093/mnras/stz538}, \href
  {https://ui.adsabs.harvard.edu/abs/2019MNRAS.485.2589M} {485, 2589}

\bibitem[\protect\citeauthoryear{{Mould}}{{Mould}}{2013}]{Mould2013}
{Mould} J.,  2013, \mn@doi [\pasa] {10.1017/pas.2013.004}, \href
  {https://ui.adsabs.harvard.edu/abs/2013PASA...30...27M} {30, e027}

\bibitem[\protect\citeauthoryear{{Naiman} et~al.,}{{Naiman}
  et~al.}{2018}]{Naiman2018}
{Naiman} J.~P.,  et~al., 2018, \mn@doi [\mnras] {10.1093/mnras/sty618}, \href
  {https://ui.adsabs.harvard.edu/abs/2018MNRAS.477.1206N} {477, 1206}

\bibitem[\protect\citeauthoryear{{Nelson} et~al.,}{{Nelson}
  et~al.}{2015}]{Nelson2015}
{Nelson} D.,  et~al., 2015, \mn@doi [Astronomy and Computing]
  {10.1016/j.ascom.2015.09.003}, \href
  {https://ui.adsabs.harvard.edu/abs/2015A&C....13...12N} {13, 12}

\bibitem[\protect\citeauthoryear{{Nelson} et~al.,}{{Nelson}
  et~al.}{2018}]{Nelson2018}
{Nelson} D.,  et~al., 2018, \mn@doi [\mnras] {10.1093/mnras/stx3040}, \href
  {https://ui.adsabs.harvard.edu/abs/2018MNRAS.475..624N} {475, 624}

\bibitem[\protect\citeauthoryear{{Nelson} et~al.,}{{Nelson}
  et~al.}{2019a}]{Nelson2019a}
{Nelson} D.,  et~al., 2019a, arXiv e-prints, \href
  {https://ui.adsabs.harvard.edu/abs/2019arXiv190205554N} {p. arXiv:1902.05554}

\bibitem[\protect\citeauthoryear{{Nelson} et~al.,}{{Nelson}
  et~al.}{2019b}]{Nelson2019b}
{Nelson} D.,  et~al., 2019b, \mn@doi [Computational Astrophysics and Cosmology]
  {10.1186/s40668-019-0028-x}, \href
  {https://ui.adsabs.harvard.edu/abs/2019ComAC...6....2N} {6, 2}

\bibitem[\protect\citeauthoryear{{Nelson} et~al.,}{{Nelson}
  et~al.}{2021}]{Nelson2021}
{Nelson} E.~J.,  et~al., 2021, \mn@doi [\mnras] {10.1093/mnras/stab2131}, \href
  {https://ui.adsabs.harvard.edu/abs/2021MNRAS.508..219N} {508, 219}

\bibitem[\protect\citeauthoryear{{Okamoto}}{{Okamoto}}{2013}]{Okamoto2013}
{Okamoto} T.,  2013, \mn@doi [\mnras] {10.1093/mnras/sts067}, \href
  {http://adsabs.harvard.edu/abs/2013MNRAS.428..718O} {428, 718}

\bibitem[\protect\citeauthoryear{{Pakmor}, {Marinacci}  \& {Springel}}{{Pakmor}
  et~al.}{2014}]{Pakmor2014}
{Pakmor} R.,  {Marinacci} F.,   {Springel} V.,  2014, \mn@doi [\apjl]
  {10.1088/2041-8205/783/1/L20}, \href
  {http://esoads.eso.org/abs/2014ApJ...783L..20P} {783, L20}

\bibitem[\protect\citeauthoryear{{Peebles}}{{Peebles}}{2020a}]{Peeblesbook2020}
{Peebles} P.~J.~E.,  2020a, {Cosmology's Century: An Inside History of our
  Modern Understanding of the Universe}

\bibitem[\protect\citeauthoryear{{Peebles}}{{Peebles}}{2020b}]{Peebles2020}
{Peebles} P.~J.~E.,  2020b, \mn@doi [\mnras] {10.1093/mnras/staa2649}, \href
  {https://ui.adsabs.harvard.edu/abs/2020MNRAS.498.4386P} {498, 4386}

\bibitem[\protect\citeauthoryear{{Peebles} \& {Nusser}}{{Peebles} \&
  {Nusser}}{2010}]{PeeblesNusser2010}
{Peebles} P.~J.~E.,  {Nusser} A.,  2010, \mn@doi [\nat] {10.1038/nature09101},
  \href {https://ui.adsabs.harvard.edu/abs/2010Natur.465..565P} {465, 565}

\bibitem[\protect\citeauthoryear{{Perez}, {Valenzuela}, {Tissera}  \&
  {Michel-Dansac}}{{Perez} et~al.}{2013}]{Perez2013}
{Perez} J.,  {Valenzuela} O.,  {Tissera} P.~B.,   {Michel-Dansac} L.,  2013,
  \mn@doi [\mnras] {10.1093/mnras/stt1563}, \href
  {https://ui.adsabs.harvard.edu/#abs/2013MNRAS.436..259P} {436, 259}

\bibitem[\protect\citeauthoryear{{Phillips}}{{Phillips}}{1996}]{Phillips1996}
{Phillips} A.~C.,  1996, in {Buta} R.,  {Crocker} D.~A.,   {Elmegreen} B.~G.,
  eds,  Astronomical Society of the Pacific Conference Series Vol. 91, IAU
  Colloq. 157: Barred Galaxies. p.~44

\bibitem[\protect\citeauthoryear{{Pillepich}, {Madau}  \& {Mayer}}{{Pillepich}
  et~al.}{2015}]{Pillepich2015}
{Pillepich} A.,  {Madau} P.,   {Mayer} L.,  2015, \mn@doi [\apj]
  {10.1088/0004-637X/799/2/184}, \href
  {https://ui.adsabs.harvard.edu/\#abs/2015ApJ...799..184P} {799, 184}

\bibitem[\protect\citeauthoryear{{Pillepich} et~al.,}{{Pillepich}
  et~al.}{2018a}]{Pillepich2018b}
{Pillepich} A.,  et~al., 2018a, \mn@doi [\mnras] {10.1093/mnras/stx2656}, \href
  {https://ui.adsabs.harvard.edu/abs/2018MNRAS.473.4077P} {473, 4077}

\bibitem[\protect\citeauthoryear{{Pillepich} et~al.,}{{Pillepich}
  et~al.}{2018b}]{Pillepich2018a}
{Pillepich} A.,  et~al., 2018b, \mn@doi [\mnras] {10.1093/mnras/stx3112}, \href
  {https://ui.adsabs.harvard.edu/abs/2018MNRAS.475..648P} {475, 648}

\bibitem[\protect\citeauthoryear{{Pillepich} et~al.,}{{Pillepich}
  et~al.}{2019}]{Pillepich2019}
{Pillepich} A.,  et~al., 2019, arXiv e-prints, \href
  {https://ui.adsabs.harvard.edu/abs/2019arXiv190205553P} {p. arXiv:1902.05553}

\bibitem[\protect\citeauthoryear{{Pillepich}, {Nelson}, {Truong}, {Weinberger},
  {Martin-Navarro}, {Springel}, {Faber}  \& {Hernquist}}{{Pillepich}
  et~al.}{2021}]{Pillepich2021b}
{Pillepich} A.,  {Nelson} D.,  {Truong} N.,  {Weinberger} R.,  {Martin-Navarro}
  I.,  {Springel} V.,  {Faber} S.~M.,   {Hernquist} L.,  2021, \mn@doi [\mnras]
  {10.1093/mnras/stab2779}, \href
  {https://ui.adsabs.harvard.edu/abs/2021MNRAS.tmp.2620P} {}

\bibitem[\protect\citeauthoryear{{Pulsoni}, {Gerhard}, {Arnaboldi},
  {Pillepich}, {Nelson}, {Hernquist}  \& {Springel}}{{Pulsoni}
  et~al.}{2020}]{Pulsoni2020}
{Pulsoni} C.,  {Gerhard} O.,  {Arnaboldi} M.,  {Pillepich} A.,  {Nelson} D.,
  {Hernquist} L.,   {Springel} V.,  2020, \mn@doi [\aap]
  {10.1051/0004-6361/202038253}, \href
  {https://ui.adsabs.harvard.edu/abs/2020A&A...641A..60P} {641, A60}

\bibitem[\protect\citeauthoryear{{Rodriguez-Gomez} et~al.,}{{Rodriguez-Gomez}
  et~al.}{2015}]{Rodriguez-Gomez2015}
{Rodriguez-Gomez} V.,  et~al., 2015, \mn@doi [\mnras] {10.1093/mnras/stv264},
  \href {https://ui.adsabs.harvard.edu/abs/2015MNRAS.449...49R} {449, 49}

\bibitem[\protect\citeauthoryear{{Rodriguez-Gomez} et~al.,}{{Rodriguez-Gomez}
  et~al.}{2016}]{Rodriguez-Gomez2016}
{Rodriguez-Gomez} V.,  et~al., 2016, \mn@doi [\mnras] {10.1093/mnras/stw456},
  \href {https://ui.adsabs.harvard.edu/abs/2016MNRAS.458.2371R} {458, 2371}

\bibitem[\protect\citeauthoryear{{Romero-G{\'o}mez}, {Masdemont},
  {Athanassoula}  \& {Garc{\'\i}a-G{\'o}mez}}{{Romero-G{\'o}mez}
  et~al.}{2006}]{Romero-Gomez2006}
{Romero-G{\'o}mez} M.,  {Masdemont} J.~J.,  {Athanassoula} E.,
  {Garc{\'\i}a-G{\'o}mez} C.,  2006, \mn@doi [\aap]
  {10.1051/0004-6361:20054653}, \href
  {https://ui.adsabs.harvard.edu/abs/2006A&A...453...39R} {453, 39}

\bibitem[\protect\citeauthoryear{{Rosas-Guevara} et~al.,}{{Rosas-Guevara}
  et~al.}{2021}]{Rosas-Guevara2021}
{Rosas-Guevara} Y.,  et~al., 2021, arXiv e-prints, \href
  {https://ui.adsabs.harvard.edu/abs/2021arXiv211004537R} {p. arXiv:2110.04537}

\bibitem[\protect\citeauthoryear{{Saglia} et~al.,}{{Saglia}
  et~al.}{2010}]{Saglia2010}
{Saglia} R.~P.,  et~al., 2010, \mn@doi [\aap] {10.1051/0004-6361/200912805},
  \href {https://ui.adsabs.harvard.edu/abs/2010A&A...509A..61S} {509, A61}

\bibitem[\protect\citeauthoryear{{Saha} \& {Elmegreen}}{{Saha} \&
  {Elmegreen}}{2018}]{Saha2018}
{Saha} K.,  {Elmegreen} B.,  2018, \mn@doi [\apj] {10.3847/1538-4357/aabacd},
  \href {https://ui.adsabs.harvard.edu/abs/2018ApJ...858...24S} {858, 24}

\bibitem[\protect\citeauthoryear{{Sakamoto}, {Okumura}, {Ishizuki}  \&
  {Scoville}}{{Sakamoto} et~al.}{1999}]{Sakamoto1999}
{Sakamoto} K.,  {Okumura} S.~K.,  {Ishizuki} S.,   {Scoville} N.~Z.,  1999,
  \mn@doi [\apj] {10.1086/307910}, \href
  {https://ui.adsabs.harvard.edu/abs/1999ApJ...525..691S} {525, 691}

\bibitem[\protect\citeauthoryear{{Sanders} \& {Huntley}}{{Sanders} \&
  {Huntley}}{1976}]{SandersHuntley1976}
{Sanders} R.~H.,  {Huntley} J.~M.,  1976, \mn@doi [\apj] {10.1086/154692},
  \href {https://ui.adsabs.harvard.edu/#abs/1976ApJ...209...53S} {209, 53}

\bibitem[\protect\citeauthoryear{{Scoville} et~al.,}{{Scoville}
  et~al.}{2007}]{Scoville2007}
{Scoville} N.,  et~al., 2007, \mn@doi [\apjs] {10.1086/516585}, \href
  {https://ui.adsabs.harvard.edu/abs/2007ApJS..172....1S} {172, 1}

\bibitem[\protect\citeauthoryear{{Sellwood} \& {Binney}}{{Sellwood} \&
  {Binney}}{2002}]{SelwoodBinney2002}
{Sellwood} J.~A.,  {Binney} J.~J.,  2002, \mn@doi [\mnras]
  {10.1046/j.1365-8711.2002.05806.x}, \href
  {http://adsabs.harvard.edu/abs/2002MNRAS.336..785S} {336, 785}

\bibitem[\protect\citeauthoryear{{S{\'e}rsic}}{{S{\'e}rsic}}{1963}]{Sersic1963}
{S{\'e}rsic} J.~L.,  1963, Boletin de la Asociacion Argentina de Astronomia La
  Plata Argentina, \href
  {https://ui.adsabs.harvard.edu/abs/1963BAAA....6...41S} {6, 41}

\bibitem[\protect\citeauthoryear{{S\'ersic}}{{S\'ersic}}{1968}]{Sersic1968}
{S\'ersic} J.~L.,  1968, {Atlas de Galaxias Australes}

\bibitem[\protect\citeauthoryear{{Shen}, {Rich}, {Kormendy}, {Howard}, {De
  Propris}  \& {Kunder}}{{Shen} et~al.}{2010}]{Shen2010}
{Shen} J.,  {Rich} R.~M.,  {Kormendy} J.,  {Howard} C.~D.,  {De Propris} R.,
  {Kunder} A.,  2010, \mn@doi [\apj] {10.1088/2041-8205/720/1/L72}, \href
  {https://ui.adsabs.harvard.edu/#abs/2010ApJ...720L..72S} {720, L72}

\bibitem[\protect\citeauthoryear{{Sheth} et~al.,}{{Sheth}
  et~al.}{2008}]{Sheth2008}
{Sheth} K.,  et~al., 2008, \mn@doi [\apj] {10.1086/524980}, \href
  {https://ui.adsabs.harvard.edu/abs/2008ApJ...675.1141S} {675, 1141}

\bibitem[\protect\citeauthoryear{{Sick}, {Courteau}, {Cuillandre}, {Dalcanton},
  {de Jong}, {McDonald}, {Simard}  \& {Tully}}{{Sick} et~al.}{2015}]{Sick2015}
{Sick} J.,  {Courteau} S.,  {Cuillandre} J.-C.,  {Dalcanton} J.,  {de Jong} R.,
   {McDonald} M.,  {Simard} D.,   {Tully} R.~B.,  2015, in {Cappellari} M.,
  {Courteau} S.,  eds, ~ Vol. 311, Galaxy Masses as Constraints of Formation
  Models. pp 82--85 (\mn@eprint {arXiv} {1410.0017}),
  \mn@doi{10.1017/S1743921315003440}

\bibitem[\protect\citeauthoryear{{Sijacki}, {Springel}, {Di Matteo}  \&
  {Hernquist}}{{Sijacki} et~al.}{2007}]{Sijacki2007}
{Sijacki} D.,  {Springel} V.,  {Di Matteo} T.,   {Hernquist} L.,  2007, \mn@doi
  [\mnras] {10.1111/j.1365-2966.2007.12153.x}, \href
  {https://ui.adsabs.harvard.edu/abs/2007MNRAS.380..877S} {380, 877}

\bibitem[\protect\citeauthoryear{{Simmons}}{{Simmons}}{2014}]{Simmons2014}
{Simmons} B.~D. e.~a.,  2014, \mn@doi [\mnras] {10.1093/mnras/stu1817}, \href
  {https://ui.adsabs.harvard.edu/abs/2014MNRAS.445.3466S} {445, 3466}

\bibitem[\protect\citeauthoryear{{Sparre} \& {Springel}}{{Sparre} \&
  {Springel}}{2016}]{SparreSpringel2016}
{Sparre} M.,  {Springel} V.,  2016, \mn@doi [\mnras] {10.1093/mnras/stw1793},
  \href {https://ui.adsabs.harvard.edu/abs/2016MNRAS.462.2418S} {462, 2418}

\bibitem[\protect\citeauthoryear{{Springel}}{{Springel}}{2010}]{Springel2010}
{Springel} V.,  2010, \mn@doi [\mnras] {10.1111/j.1365-2966.2009.15715.x},
  \href {http://adsabs.harvard.edu/abs/2010MNRAS.401..791S} {401, 791}

\bibitem[\protect\citeauthoryear{{Springel} \& {Hernquist}}{{Springel} \&
  {Hernquist}}{2003}]{SpringelHernquist2003}
{Springel} V.,  {Hernquist} L.,  2003, \mn@doi [\mnras]
  {10.1046/j.1365-8711.2003.06206.x}, \href
  {https://ui.adsabs.harvard.edu/\#abs/2003MNRAS.339..289S} {339, 289}

\bibitem[\protect\citeauthoryear{{Springel} et~al.,}{{Springel}
  et~al.}{2018}]{Springel2018}
{Springel} V.,  et~al., 2018, \mn@doi [\mnras] {10.1093/mnras/stx3304}, \href
  {https://ui.adsabs.harvard.edu/abs/2018MNRAS.475..676S} {475, 676}

\bibitem[\protect\citeauthoryear{{Tacchella} et~al.,}{{Tacchella}
  et~al.}{2019}]{Tacchella2019}
{Tacchella} S.,  et~al., 2019, \mn@doi [\mnras] {10.1093/mnras/stz1657}, \href
  {https://ui.adsabs.harvard.edu/abs/2019MNRAS.487.5416T} {487, 5416}

\bibitem[\protect\citeauthoryear{{Tamm}, {Tempel}, {Tenjes}, {Tihhonova}  \&
  {Tuvikene}}{{Tamm} et~al.}{2012}]{Tamm2012}
{Tamm} A.,  {Tempel} E.,  {Tenjes} P.,  {Tihhonova} O.,   {Tuvikene} T.,  2012,
  \mn@doi [\aap] {10.1051/0004-6361/201220065}, \href
  {https://ui.adsabs.harvard.edu/abs/2012A&A...546A...4T} {546, A4}

\bibitem[\protect\citeauthoryear{{Tissera}, {White}  \&
  {Scannapieco}}{{Tissera} et~al.}{2012}]{Tissera2012}
{Tissera} P.~B.,  {White} S. D.~M.,   {Scannapieco} C.,  2012, \mn@doi [\mnras]
  {10.1111/j.1365-2966.2011.20028.x}, \href
  {https://ui.adsabs.harvard.edu/\#abs/2012MNRAS.420..255T} {420, 255}

\bibitem[\protect\citeauthoryear{{Tissera}, {Beers}, {Carollo}  \&
  {Scannapieco}}{{Tissera} et~al.}{2014}]{Tissera2014}
{Tissera} P.~B.,  {Beers} T.~C.,  {Carollo} D.,   {Scannapieco} C.,  2014,
  \mn@doi [\mnras] {10.1093/mnras/stu181}, \href
  {https://ui.adsabs.harvard.edu/abs/2014MNRAS.439.3128T} {439, 3128}

\bibitem[\protect\citeauthoryear{{Tissera}, {Machado}, {Carollo}, {Minniti},
  {Beers}, {Zoccali}  \& {Meza}}{{Tissera} et~al.}{2018}]{Tissera2018}
{Tissera} P.~B.,  {Machado} R. E.~G.,  {Carollo} D.,  {Minniti} D.,  {Beers}
  T.~C.,  {Zoccali} M.,   {Meza} A.,  2018, \mn@doi [\mnras]
  {10.1093/mnras/stx2431}, \href
  {https://ui.adsabs.harvard.edu/#abs/2018MNRAS.473.1656T} {473, 1656}

\bibitem[\protect\citeauthoryear{{Toomre}}{{Toomre}}{1977}]{Toomre1977}
{Toomre} A.,  1977, in {Tinsley} B.~M.,  {Larson} Richard B.~Gehret D.~C.,
  eds, Evolution of Galaxies and Stellar Populations. p.~401

\bibitem[\protect\citeauthoryear{{Toomre}}{{Toomre}}{1981}]{Toomre1981}
{Toomre} A.,  1981, in {Fall} S.~M.,  {Lynden-Bell} D.,  eds, Structure and
  Evolution of Normal Galaxies. pp 111--136

\bibitem[\protect\citeauthoryear{{Vogelsberger} et~al.,}{{Vogelsberger}
  et~al.}{2014a}]{Vogelsberger2014b}
{Vogelsberger} M.,  et~al., 2014a, \mn@doi [\mnras] {10.1093/mnras/stu1536},
  \href {https://ui.adsabs.harvard.edu/abs/2014MNRAS.444.1518V} {444, 1518}

\bibitem[\protect\citeauthoryear{{Vogelsberger} et~al.,}{{Vogelsberger}
  et~al.}{2014b}]{Vogelsberger2014a}
{Vogelsberger} M.,  et~al., 2014b, \mn@doi [\nat] {10.1038/nature13316}, \href
  {https://ui.adsabs.harvard.edu/abs/2014Natur.509..177V} {509, 177}

\bibitem[\protect\citeauthoryear{{Wang}, {Athanassoula}, {Yu}, {Wolf}, {Shao},
  {Gao}  \& {Randriamampand ry}}{{Wang} et~al.}{2020}]{Wang2020}
{Wang} J.,  {Athanassoula} E.,  {Yu} S.-Y.,  {Wolf} C.,  {Shao} L.,  {Gao} H.,
   {Randriamampand ry} T.~H.,  2020, \mn@doi [\apj] {10.3847/1538-4357/ab7fad},
  \href {https://ui.adsabs.harvard.edu/abs/2020ApJ...893...19W} {893, 19}

\bibitem[\protect\citeauthoryear{{Weinberger} et~al.,}{{Weinberger}
  et~al.}{2017}]{Weinberger2017}
{Weinberger} R.,  et~al., 2017, \mn@doi [\mnras] {10.1093/mnras/stw2944}, \href
  {https://ui.adsabs.harvard.edu/abs/2017MNRAS.465.3291W} {465, 3291}

\bibitem[\protect\citeauthoryear{{Weinzirl}, {Jogee}, {Khochfar}, {Burkert}  \&
  {Kormendy}}{{Weinzirl} et~al.}{2009}]{Weinzirl2009}
{Weinzirl} T.,  {Jogee} S.,  {Khochfar} S.,  {Burkert} A.,   {Kormendy} J.,
  2009, \mn@doi [\apj] {10.1088/0004-637X/696/1/411}, \href
  {https://ui.adsabs.harvard.edu/abs/2009ApJ...696..411W} {696, 411}

\bibitem[\protect\citeauthoryear{{de Vaucouleurs}}{{de
  Vaucouleurs}}{1948}]{DeVaucouleurs1948}
{de Vaucouleurs} G.,  1948, Annales d'Astrophysique, \href
  {https://ui.adsabs.harvard.edu/abs/1948AnAp...11..247D} {11, 247}

\makeatother
\end{thebibliography}

\appendix
\section{Resolution convergence}
\label{sec:appendix-res}

In this appendix, we conduct an analysis of the resolution convergence of the SBPs used to obtain the parameters of the two-component fits described in Sec.~\ref{sec:2components}. For that purpose, we compare the values of surface brightness in radial bins of the SBPs, in the sample of  MW/M31--like galaxies in TNG50--1 (labelled as TNG50 throughout this article) with those of galaxies residing in the analogous DM haloes in the sibling simulations of lower resolution TNG50--2 and TNG50--3 (see the illustrisTNG project webpage for a complete description of the simulations). We exclude from our analysis the TNG50--4 simulation, which has a resolution level that is a factor $512$ lower, and a softening length a factor of $8$ times larger.  A thorough resolution convergence analysis of the TNG50 simulation in quantities like the disc sizes, disc scale heights and kinematic measures can be found in \citet{Pillepich2019}.  

In order to compare the SBPs on a galaxy-galaxy basis me must find the analogue galaxies of our original sample in TNG50--1, in the lower resolution realizations. For that purpose, galaxies were matched between simulations using the database produced in Lovell et al. (in prep.) with the matching algorithm described in \citet[][]{Lovell2014, Lovell2018}. In this implementation, analogue objects are identified using the initial conditions. DM haloes that originated from the same Lagrangian patch in the initial conditions are considered analogues accross simulations. However, a perfect match does not always occur. DM haloes are tagged with a score, based on a quality--of--match statistic that indicates the certainty of the match. Scores close to 1 indicate a perfect match, while scores lower than $0.5$ indicate that there is no match. Almost all galaxies of our sample of 287 MW/M31--like galaxies in TNG50--1 have a match in TNG50--2 and TNG50--3,  with a score higher than $0.8$. Only three of them have no counterpart in the lower resolution runs.     

\begin{figure}
\includegraphics[scale=0.42]{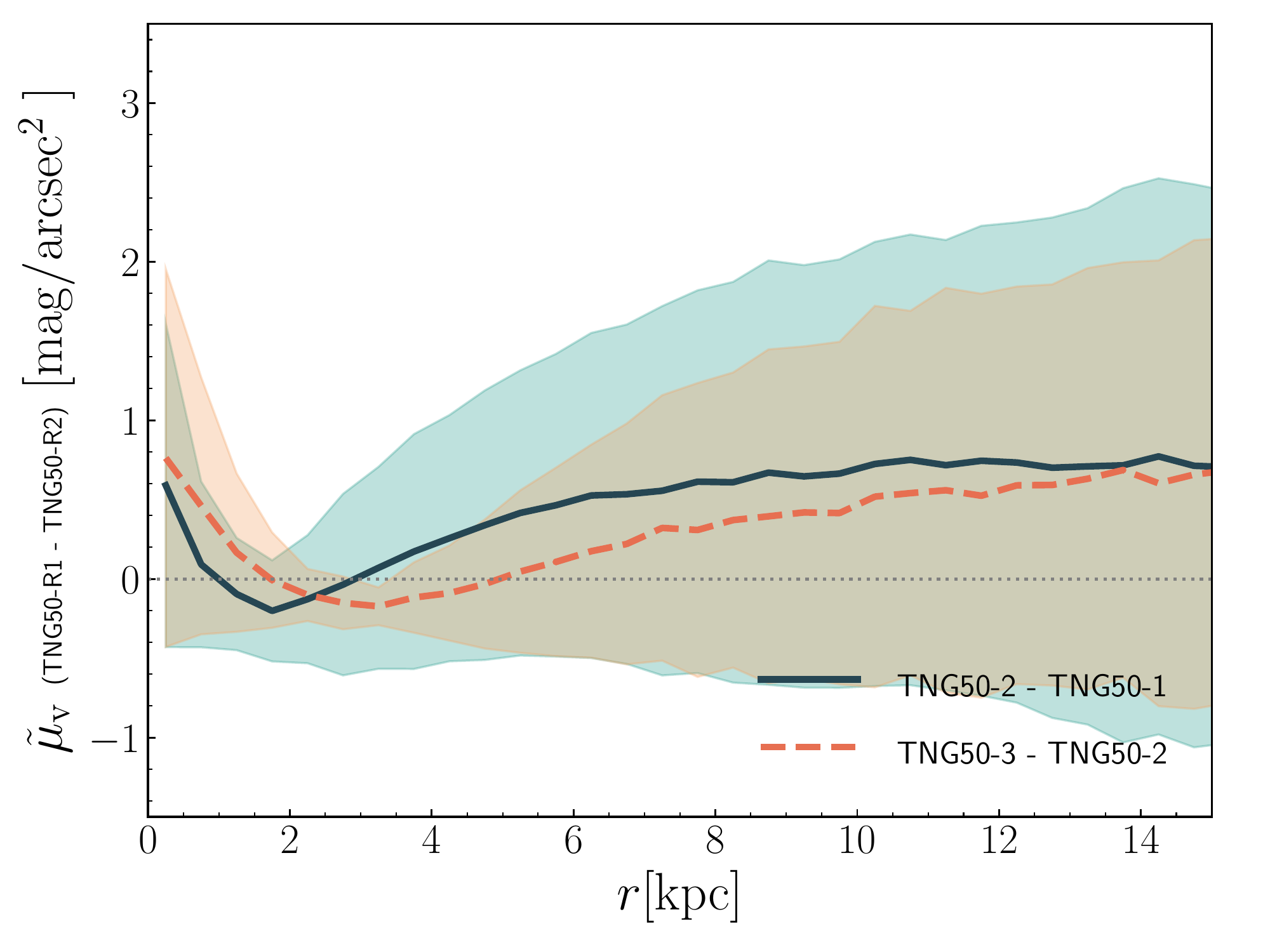}

\caption{Binned median values of the difference of surface brightness as a function of radius for  our sample of simulated galaxies in different resolution hierarchies. The solid dark green line shows the median of the difference between analogues in the TNG50--2 simulation and galaxies in our sample in TNG50--1 and the green filled region shows the corresponding interquartile ranges. The dashed orange line and filled orange region shows the same, but for TNG50--3 and TNG50--2. A dotted grey line indicates the zero difference value as visual aid. The SBPs in central regions of MW/M31--like galaxies are not fully converged in the highest resolution run in TNG50.}

\label{fig:resconv}
\end{figure}

Fig.~\ref{fig:resconv} shows the median difference of surface brightness in each radial bin of the SBPs between galaxies in TNG50--1 and their analogue galaxies in TNG50--2 and between these analogues in TNG50--2 and TNG50--3. We can see that in the central region, the median difference is positive in both cases, meaning that surface brightness profiles in TNG50--1 have a statistical excess in surface brightness there with respect to TNG50--2. Accordingly, TNG50--2 analogues show an excess of surface brightness in central regions with respect to analogues in TNG50--3. The difference between TNG50--2 and TNG50--3 is close to 0.8  ${\rm mag~arcsec}^{-2}$, while between TNG50--1 and TNG50--2 the median difference is close to 0.6 ${\rm mag ~arcsec}^{-2}$.  At radii in the range [1,3] {\rm kpc}, the median difference becomes slightly negative between TNG50--2 and TNG50--1, probably due to the presence of brighter bars in TNG50--1. The median difference between TNG50--3 and TNG50--2 analogues becomes negative in the range [2,5] {\rm kpc} .  Moving to the outer regions, the median difference in the SBP becomes more substantial, as the surface brightnesses of discs diminishes.

\label{lastpage} 
\end{document}